\RequirePackage{graphicx}
\RequirePackage{xspace}

\documentclass[%
 superscriptaddress,
reprint,
showpacs,preprintnumbers,
nofootinbib,
 amsmath,amssymb,
aps,
prd,
floatfix,
]{revtex4-1}
\usepackage{color} %% TEMP: only for editing
\usepackage{hyperref}
\usepackage{subfigure}

\newcommand{\pt}{p_\textrm{T}}

\newcommand{\dpt}{\delta \pt}
\newcommand{\dptv}{\delta\vec{p}_\textrm{T}}
\newcommand{\dalphat}{\delta\alpha_\textrm{T}}
\newcommand{\dphit}{\delta\phi_\textrm{T}}
\newcommand{\nucleon}{\textrm{N}}
\newcommand{\pni}{\vec{p}_{\nucleon}}
\newcommand{\ptni}{\vec{p}_\textrm{T}^{\,\nucleon}}
\newcommand{\Dptv}{\Delta\vec{p}_\textrm{T}}
\renewcommand{\BibitemShut}[1]{}

\begin{document}

%\preprint{}

\title{Characterisation of nuclear effects in muon-neutrino scattering on hydrocarbon with a measurement of final-state kinematics and correlations in charged-current pionless interactions at T2K}
%\input{author20180119}
%\input{author20180214}
%%%%%%%%%%%%%%%%%%%%%%%%%%%%%%%%%%%%%%%%%%%%%%%%%%%%%%%%%%%%%%
% T2K author list generated on Wed, 14 Feb 2018 17:58:08 +0900
% setting: extra = 0 revtex = 1 ptep = 0 simple = 0 xml = 0 yearrule = 1 shiftrule = 1
%         author list from archive (starting 2018/01/19 until now)
% Number of authors = 305
%%%%%%%%%%%%%%%%%%%%%%%%%%%%%%%%%%%%%%%%%%%%%%%%%%%%%%%%%%%%%%

\newcommand{\INSTHD}{\affiliation{University Autonoma Madrid, Department of Theoretical Physics, 28049 Madrid, Spain}}
\newcommand{\INSTEE}{\affiliation{University of Bern, Albert Einstein Center for Fundamental Physics, Laboratory for High Energy Physics (LHEP), Bern, Switzerland}}
\newcommand{\INSTFE}{\affiliation{Boston University, Department of Physics, Boston, Massachusetts, U.S.A.}}
\newcommand{\INSTD}{\affiliation{University of British Columbia, Department of Physics and Astronomy, Vancouver, British Columbia, Canada}}
\newcommand{\INSTGA}{\affiliation{University of California, Irvine, Department of Physics and Astronomy, Irvine, California, U.S.A.}}
\newcommand{\INSTI}{\affiliation{IRFU, CEA Saclay, Gif-sur-Yvette, France}}
\newcommand{\INSTGB}{\affiliation{University of Colorado at Boulder, Department of Physics, Boulder, Colorado, U.S.A.}}
\newcommand{\INSTFG}{\affiliation{Colorado State University, Department of Physics, Fort Collins, Colorado, U.S.A.}}
\newcommand{\INSTFH}{\affiliation{Duke University, Department of Physics, Durham, North Carolina, U.S.A.}}
\newcommand{\INSTBA}{\affiliation{Ecole Polytechnique, IN2P3-CNRS, Laboratoire Leprince-Ringuet, Palaiseau, France }}
\newcommand{\INSTEG}{\affiliation{University of Geneva, Section de Physique, DPNC, Geneva, Switzerland}}
\newcommand{\INSTDG}{\affiliation{H. Niewodniczanski Institute of Nuclear Physics PAN, Cracow, Poland}}
\newcommand{\INSTCB}{\affiliation{High Energy Accelerator Research Organization (KEK), Tsukuba, Ibaraki, Japan}}
\newcommand{\INSTED}{\affiliation{Institut de Fisica d'Altes Energies (IFAE), The Barcelona Institute of Science and Technology, Campus UAB, Bellaterra (Barcelona) Spain}}
\newcommand{\INSTEC}{\affiliation{IFIC (CSIC \& University of Valencia), Valencia, Spain}}
\newcommand{\INSTEI}{\affiliation{Imperial College London, Department of Physics, London, United Kingdom}}
\newcommand{\INSTGF}{\affiliation{INFN Sezione di Bari and Universit\`a e Politecnico di Bari, Dipartimento Interuniversitario di Fisica, Bari, Italy}}
\newcommand{\INSTBE}{\affiliation{INFN Sezione di Napoli and Universit\`a di Napoli, Dipartimento di Fisica, Napoli, Italy}}
\newcommand{\INSTBF}{\affiliation{INFN Sezione di Padova and Universit\`a di Padova, Dipartimento di Fisica, Padova, Italy}}
\newcommand{\INSTBD}{\affiliation{INFN Sezione di Roma and Universit\`a di Roma ``La Sapienza'', Roma, Italy}}
\newcommand{\INSTEB}{\affiliation{Institute for Nuclear Research of the Russian Academy of Sciences, Moscow, Russia}}
\newcommand{\INSTHA}{\affiliation{Kavli Institute for the Physics and Mathematics of the Universe (WPI), The University of Tokyo Institutes for Advanced Study, University of Tokyo, Kashiwa, Chiba, Japan}}
\newcommand{\INSTCC}{\affiliation{Kobe University, Kobe, Japan}}
\newcommand{\INSTCD}{\affiliation{Kyoto University, Department of Physics, Kyoto, Japan}}
\newcommand{\INSTEJ}{\affiliation{Lancaster University, Physics Department, Lancaster, United Kingdom}}
\newcommand{\INSTFC}{\affiliation{University of Liverpool, Department of Physics, Liverpool, United Kingdom}}
\newcommand{\INSTFI}{\affiliation{Louisiana State University, Department of Physics and Astronomy, Baton Rouge, Louisiana, U.S.A.}}
\newcommand{\INSTHB}{\affiliation{Michigan State University, Department of Physics and Astronomy,  East Lansing, Michigan, U.S.A.}}
\newcommand{\INSTCE}{\affiliation{Miyagi University of Education, Department of Physics, Sendai, Japan}}
\newcommand{\INSTDF}{\affiliation{National Centre for Nuclear Research, Warsaw, Poland}}
\newcommand{\INSTFJ}{\affiliation{State University of New York at Stony Brook, Department of Physics and Astronomy, Stony Brook, New York, U.S.A.}}
\newcommand{\INSTGJ}{\affiliation{Okayama University, Department of Physics, Okayama, Japan}}
\newcommand{\INSTCF}{\affiliation{Osaka City University, Department of Physics, Osaka, Japan}}
\newcommand{\INSTGG}{\affiliation{Oxford University, Department of Physics, Oxford, United Kingdom}}
\newcommand{\INSTBB}{\affiliation{UPMC, Universit\'e Paris Diderot, CNRS/IN2P3, Laboratoire de Physique Nucl\'eaire et de Hautes Energies (LPNHE), Paris, France}}
\newcommand{\INSTGC}{\affiliation{University of Pittsburgh, Department of Physics and Astronomy, Pittsburgh, Pennsylvania, U.S.A.}}
\newcommand{\INSTFA}{\affiliation{Queen Mary University of London, School of Physics and Astronomy, London, United Kingdom}}
\newcommand{\INSTE}{\affiliation{University of Regina, Department of Physics, Regina, Saskatchewan, Canada}}
\newcommand{\INSTGD}{\affiliation{University of Rochester, Department of Physics and Astronomy, Rochester, New York, U.S.A.}}
\newcommand{\INSTHC}{\affiliation{Royal Holloway University of London, Department of Physics, Egham, Surrey, United Kingdom}}
\newcommand{\INSTBC}{\affiliation{RWTH Aachen University, III. Physikalisches Institut, Aachen, Germany}}
\newcommand{\INSTFB}{\affiliation{University of Sheffield, Department of Physics and Astronomy, Sheffield, United Kingdom}}
\newcommand{\INSTDI}{\affiliation{University of Silesia, Institute of Physics, Katowice, Poland}}
\newcommand{\INSTEH}{\affiliation{STFC, Rutherford Appleton Laboratory, Harwell Oxford,  and  Daresbury Laboratory, Warrington, United Kingdom}}
\newcommand{\INSTCH}{\affiliation{University of Tokyo, Department of Physics, Tokyo, Japan}}
\newcommand{\INSTBJ}{\affiliation{University of Tokyo, Institute for Cosmic Ray Research, Kamioka Observatory, Kamioka, Japan}}
\newcommand{\INSTCG}{\affiliation{University of Tokyo, Institute for Cosmic Ray Research, Research Center for Cosmic Neutrinos, Kashiwa, Japan}}
\newcommand{\INSTHF}{\affiliation{Tokyo Institute of Technology, Department of Physics, Tokyo, Japan}}
\newcommand{\INSTGI}{\affiliation{Tokyo Metropolitan University, Department of Physics, Tokyo, Japan}}
\newcommand{\INSTF}{\affiliation{University of Toronto, Department of Physics, Toronto, Ontario, Canada}}
\newcommand{\INSTB}{\affiliation{TRIUMF, Vancouver, British Columbia, Canada}}
\newcommand{\INSTG}{\affiliation{University of Victoria, Department of Physics and Astronomy, Victoria, British Columbia, Canada}}
\newcommand{\INSTDJ}{\affiliation{University of Warsaw, Faculty of Physics, Warsaw, Poland}}
\newcommand{\INSTDH}{\affiliation{Warsaw University of Technology, Institute of Radioelectronics, Warsaw, Poland}}
\newcommand{\INSTFD}{\affiliation{University of Warwick, Department of Physics, Coventry, United Kingdom}}
\newcommand{\INSTGH}{\affiliation{University of Winnipeg, Department of Physics, Winnipeg, Manitoba, Canada}}
\newcommand{\INSTEA}{\affiliation{Wroclaw University, Faculty of Physics and Astronomy, Wroclaw, Poland}}
\newcommand{\INSTHE}{\affiliation{Yokohama National University, Faculty of Engineering, Yokohama, Japan}}
\newcommand{\INSTH}{\affiliation{York University, Department of Physics and Astronomy, Toronto, Ontario, Canada}}

\INSTHD
\INSTEE
\INSTFE
\INSTD
\INSTGA
\INSTI
\INSTGB
\INSTFG
\INSTFH
\INSTBA
\INSTEG
\INSTDG
\INSTCB
\INSTED
\INSTEC
\INSTEI
\INSTGF
\INSTBE
\INSTBF
\INSTBD
\INSTEB
\INSTHA
\INSTCC
\INSTCD
\INSTEJ
\INSTFC
\INSTFI
\INSTHB
\INSTCE
\INSTDF
\INSTFJ
\INSTGJ
\INSTCF
\INSTGG
\INSTBB
\INSTGC
\INSTFA
\INSTE
\INSTGD
\INSTHC
\INSTBC
\INSTFB
\INSTDI
\INSTEH
\INSTCH
\INSTBJ
\INSTCG
\INSTHF
\INSTGI
\INSTF
\INSTB
\INSTG
\INSTDJ
\INSTDH
\INSTFD
\INSTGH
\INSTEA
\INSTHE
\INSTH

\author{K.\,Abe}\INSTBJ
\author{J.\,Amey}\INSTEI
\author{C.\,Andreopoulos}\INSTEH\INSTFC
\author{L.\,Anthony}\INSTFC
\author{M.\,Antonova}\INSTEC
\author{S.\,Aoki}\INSTCC
\author{A.\,Ariga}\INSTEE
\author{Y.\,Ashida}\INSTCD
\author{Y.\,Azuma}\INSTCF
\author{S.\,Ban}\INSTCD
% ...

\collaboration{The T2K Collaboration}\noaffiliation

%This is the way to include the full author list
%\input{author20160208.tex}



%\author{}
%\thanks{}
%\affiliation{}


\date{\today}
%\linenumbers

\begin{abstract}
This paper reports measurements of final-state proton multiplicity, muon and proton kinematics, and their correlations in charged-current pionless neutrino interactions, measured by the T2K ND280 near detector in its plastic scintillator (C$_8$H$_8$) target. The data were taken between years 2010 and 2013, corresponding to approximately 6$\times10^{20}$ protons on target. Thanks to their exploration of the proton kinematics and of imbalances between the proton and muon kinematics, the results offer a novel probe of the nuclear-medium effects most pertinent to the (sub-)GeV neutrino-nucleus interactions that are used in accelerator-based long-baseline neutrino oscillation measurements.
These results are compared to many neutrino-nucleus interaction models which all fail to describe at least part of the observed phase space.
In case of events without a proton above a detection threshold in the final state, a fully consistent implementation of the local Fermi gas model with multinucleon interactions gives the best description of the data. In the case of at least one proton in the final state the spectral function model agrees well with the data, most notably when measuring the kinematic imbalance between the muon and the proton in the plane transverse to the incoming neutrino. Within the models considered, only the existence of multinucleon interactions are able to describe the extracted cross-section within regions of high transverse kinematic imbalance. The effect of final-state interactions is also discussed. 
\end{abstract}

%\pacs{13.15.+g,25.30.Pt}
%\keywords{}

\maketitle

% Put \label in argument of \section for cross referencing: \section{\label{}}

%\input{Introduction}
\section{Introduction \label{sec:introduction}}

Neutrino interactions with nuclei are the experimental tool exploited to provide evidence of neutrino oscillations~\cite{Fukuda:1998mi,Ahmad:2001an,Ahmad:2002jz,Araki:2004mb,Aliu:2004sq,Michael:2006rx,Abe:2011sj,An:2012eh} and to search for leptonic CP-symmetry violation~\cite{Abe:2017vif,Adamson:2017gxd,Abe:2015zbg,Acciarri:2015uup}. In long-baseline accelerator-based neutrino oscillation experiments, neutrino beams are produced with energies in the range of hundreds of MeV to a few GeV. The produced neutrinos interact then with the bound nucleons of nuclei in the detectors via reactions such as quasi-elastic scattering (QE), resonant production (RES), and deep inelastic scattering (DIS). A precise measurement of the oscillation parameters relies on the understanding of the incoming neutrino beam flux, of the scattering of neutrinos with nucleons, and of the nuclear medium effects in the nucleus. The systematic uncertainties arising from neutrino-nucleus interactions, especially those related to nuclear effects, are currently one of the limiting factors for oscillation measurements~\cite{Alvarez-Ruso:2017oui} in T2K~\cite{Abe:2011ks} and NOvA~\cite{novafirst}, and will become the dominant uncertainties for future long-baseline experiments, such as DUNE~\cite{Acciarri:2015uup} and Hyper-Kamiokande~\cite{Abe:2014oxa}.

Neutrinos of such energies can probe nuclear structure at the nucleon level and therefore an accurate description of the nucleus in terms of nucleonic degrees of freedom is essential. To a first approximation, in the independent particle model (IPM), each nucleon is subject to Fermi motion (FM) and a mean-field potential. It is then common to factorise neutrino-nucleus interactions into an interaction with such a bound nucleon (the impulse approximation), leaving the remaining nucleus in a one-particle-one-hole (1p1h) excitation state, and a separate description of the subsequent final state reinteractions inside the nucleus~\cite{impuseApprox}. Driven by precision measurements of electron-nucleus scattering and first large statistics neutrino-nucleus scattering measurements~\cite{k2kma,AguilarArevalo:2010zc}, various theoretical developments beyond these approximations have been proposed. In the random phase approximation (RPA) approach~\cite{Singh:1992dc,Gil:1997bm,Nieves:2004wx,Nieves:2005rq,Martini:2009uj}, collective excitations approximated as a superposition of 1p1h excitations are calculated. This particular medium effect is parametrised as a correction factor to the interaction cross section as a function of the squared four-momentum transfer $Q^2$. In addition to such long-range correlations, short-range correlations (SRCs) are also captured by the spectral function (SF) approach~\cite{Benhar:1994hw,Benhar:2005dj,Benhar:2006wy,Ankowski:2010yh}, which accounts for nucleon-nucleon correlations beyond the mean-field dynamics. These correlations produce an enhancement in the ground-state nucleon momentum distribution beyond the Fermi momentum, and can lead to two-particle-two-hole (2p2h) excitations of the nucleus (and, more in general, to npnh excitations with n~$>1$). Formalisms developed for electron-nucleus scattering have been adapted to describe neutrino data, proposing that 2p2h contributions, notably due to meson-exchange currents (MEC), might be significant in neutrino-nucleus interactions~\cite{Delorme:1985ps,Marteau:1999jp,Martini:2009uj,Martini:2010ex,Nieves:2011pp,Nieves:2011yp,Martini:2011wp}. 

%; yet their kinematic nature is unclear (see e.g. Refs.~\cite{Niewczas:2015iea, Weinstein:2016inx}).~\footnote{\textcolor{red}{need to explain high momentum tail in dpt from 2p2h in discussion.}}

Among the reactions relevant for GeV energy neutrinos, the charged-current (CC) QE, 
\begin{align}
\nu\mathrm{N}\to\ell\mathrm{N}', \label{eq:ccqe}
\end{align}
is of primary importance for neutrino detection in oscillation experiments, where $\nu$ and $\ell$ are the neutrino and the corresponding charged lepton, $\mathrm{N}$ and $\mathrm{N}'$ are the initial- and final-state nucleons. Embedded in a nucleus, the final-state nucleon propagates through and interacts with the nucleus remnant. These final-state interactions (FSI) could be highly inelastic, causing energy dissipation which can prevent hadrons escaping the nuclear medium or alternatively stimulate additional hadrons to be emitted. As a result, the QE reaction in Eq.~\ref{eq:ccqe} is not directly accessible. What can be measured are the CC interactions without pion in the extra-nucleus final state (CC$0\pi$). This process includes not only other reactions such as pion production, in which the pion is absorbed inside the nucleus, but also 2p2h excitations involving two-nucleon knockout. CC$0\pi$ (sometimes called ``CCQE-like'') interactions have been extensively measured~\cite{Lyubushkin:2008pe,AguilarArevalo:2010zc,Aguilar-Arevalo:2013dva,Fields:2013zhk,Fiorentini:2013ezn,Walton:2014esl,Wolcott:2015hda,Betancourt:2017uso,Abe:2014iza,Abe:2015oar,Abe:2016tmq,Acciarri:2014gev}, yet the unambiguous identification of various nuclear effects has proved difficult. This is primarily because the often measured single-particle final-state kinematics, such as momentum and angular distributions, are determined by both the intrinsic dynamics of Eq.~\ref{eq:ccqe} and by nuclear effects.
% (see, e.g., Fig.~3 of Ref.~\cite{Nieves:2011yp}). 
%%modified from SB%%%%%

This paper reports measurements of muon-neutrino CC$0\pi$ interactions with the T2K beam, which has a peak energy of around 600~MeV. The multi-differential cross section using muon and proton kinematics, their correlations, and the final-state multiplicity of protons (above a threshold energy) are measured. These measurements are performed using the T2K near detector (ND280), on a plastic scintillator (C$_8$H$_8$) target, with approximately $6\times10^{20}$ protons on target (POT). The main aim of such new measurements is to improve the understanding of nuclear effects in neutrino interactions, notably with a view to minimizing the corresponding uncertainties in neutrino oscillation measurements. In oscillation measurements neutrino-interaction models are used to infer the neutrino energy from the final state particles and to extrapolate the near detector constraints to the far detector. To test the correctness of such inference, detailed comparisons of the measured cross sections with the most recent neutrino-nucleus interaction models are reported in this paper.

%This paper reports a measurement of final-state multiplicity of protons (above a threshold energy), muon and proton kinematics, and their correlations in CC$0\pi$ interaction, using the T2K near detector (ND280) and its plastic scintillator (C$_8$H$_8$) target with $\approx6\times10^{20}$ protons on target (POT). %% SD: I reworded this above

The modelling of neutrino energy reconstruction in the CC$0\pi$ sample, exploited for neutrino oscillation measurements, is affected
by large uncertainties due to nuclear effects: even when protons can be in principle detected, the detector response depends on the actual kinematics of the
outgoing protons. In the absence of a robust model prediction on the hadronic final state, a multi-differential measurement of single-particle kinematics and nucleon multiplicity, provides valuable input for the modelling of neutrino energy reconstruction and detector response. Furthermore, measurements of proton kinematics from neutrino-nucleus scattering may be used to infer neutron multiplicity and kinematics in the corresponding antineutrino reaction.
While the single-particle kinematics and the multiplicity measurements provide a comprehensive description of the CC$0\pi$ final state, the measurement of muon-proton correlations in the final state provides a powerful probe of nuclear effects. Considering the dynamics of Eq.~\ref{eq:ccqe} in the case of scattering on a free nucleon $\nu\mathrm{n}\to\ell\mathrm{p}$ in the absence of nuclear effects, the final-state proton kinematics can be uniquely determined by that of the muon. In a CC$0\pi$ measurement the deviation of proton multiplicity and kinematics from what is expected in the simple process of Eq.~\ref{eq:ccqe} originates solely from nuclear effects. Such deviations can be characterised using so-called transverse kinematic imbalances (introduced for the first time in Ref.~\cite{Lu:2015tcr}) and proton inferred kinematics, which are measured in the analyses presented here.

%such deviations a can be formulated in terms of transverse kinematic imbalances (introduced for the first time in~\cite{Lu:2015tcr}) and inferred kinematics, which will be discussed in the following sections.
%%I moved this to abstract
%This paper reports a measurement of final-state proton multiplicity, muon and proton kinematics, and their correlations in CC$0\pi$ interaction, measured by the T2K ND280 near detector with its plastic scintillator (C$_8$H$_8$) target. The data was taken in the neutrino Runs 1--4 between years 2010 and 2013, corresponding to 7.48$\times10^{20}$ protons on target (POT). This measurement indicates that ... \textcolor{red}{Summarize the major conclusions of the paper here. Eg, 2p2h kinematics by dpt, MA insensitive results}

This paper is organised as follows. %\textcolor{red}{tbc}.
%SB
After a short description of the T2K experiment in Sec.~\ref{sec:T2K}, the measurements presented in this paper and the new variables are introduced in Sec.~\ref{sec:anaStrategy}. Sec.~\ref{sec:anaDescription} describes the analysis procedure, including the simulations used, the event selection and the method for cross section evaluation. Following this the results are reported for each of three analyses: one using proton and muon kinematics, another using transverse kinematic imbalances and a third using proton inferred kinematics. The interpretation of the results is discussed in Sec.~\ref{sec:discussion}, followed by conclusions in Sec.~\ref{sec:conclusions}.

\section{The T2K experiment \label{sec:T2K}}
%Very brief introduction on T2K

%Describe the beam line and INGRID

%Describe ND280

%{\it Need to explain that for analysis described in this paper we are using fully active scintilator target FGD (not the FGD2), and that we study tracks which travers TPC downstream of the target, or are contained in the FGD. We should not use names FGD1, TPC2 which are not clear for people from outside collaboration.%%}

The Tokai-to-Kamioka (T2K) experiment~\cite{Abe:2011ks} is an accelerator-based long-baseline experiment which measures neutrino oscillations in a $\nu_{\mu}$ ($\bar{\nu}_{\mu}$) beam~\cite{Abe:2017vif}.
%T2K provides precise measurements of atmospheric neutrino oscillation parameters $\Delta m^2_{32}$ and $\theta_{23}$ by
%studying the $\nu_{\mu} \rightarrow \nu_e$ and $\bar{\nu}_\mu  \rightarrow \bar{\nu}_e$) transitions~\cite{Abe:2017bay}. It was the first experiment to observe the appearance of $\nu_e$ due to non-zero mixing angle $\theta_{13}$~\cite{Abe:2013hdq} and recently also provided first hints of leptonic CP violation~\cite{Abe:2017vif}.
The T2K neutrino beam is produced by the Japan Proton Accelerator Research Complex.
A 30~GeV proton beam collides with a graphite target producing positive and negative pions and kaons which are focused and charge-selected by three horn magnets. The positive (negative) hadrons decay to produce a flux highly dominated by $\nu_{\mu}$ ($\bar{\nu}_{\mu}$)~\cite{t2kflux}.
  
\begin{figure}
\begin{center}
 \includegraphics[width=9cm]{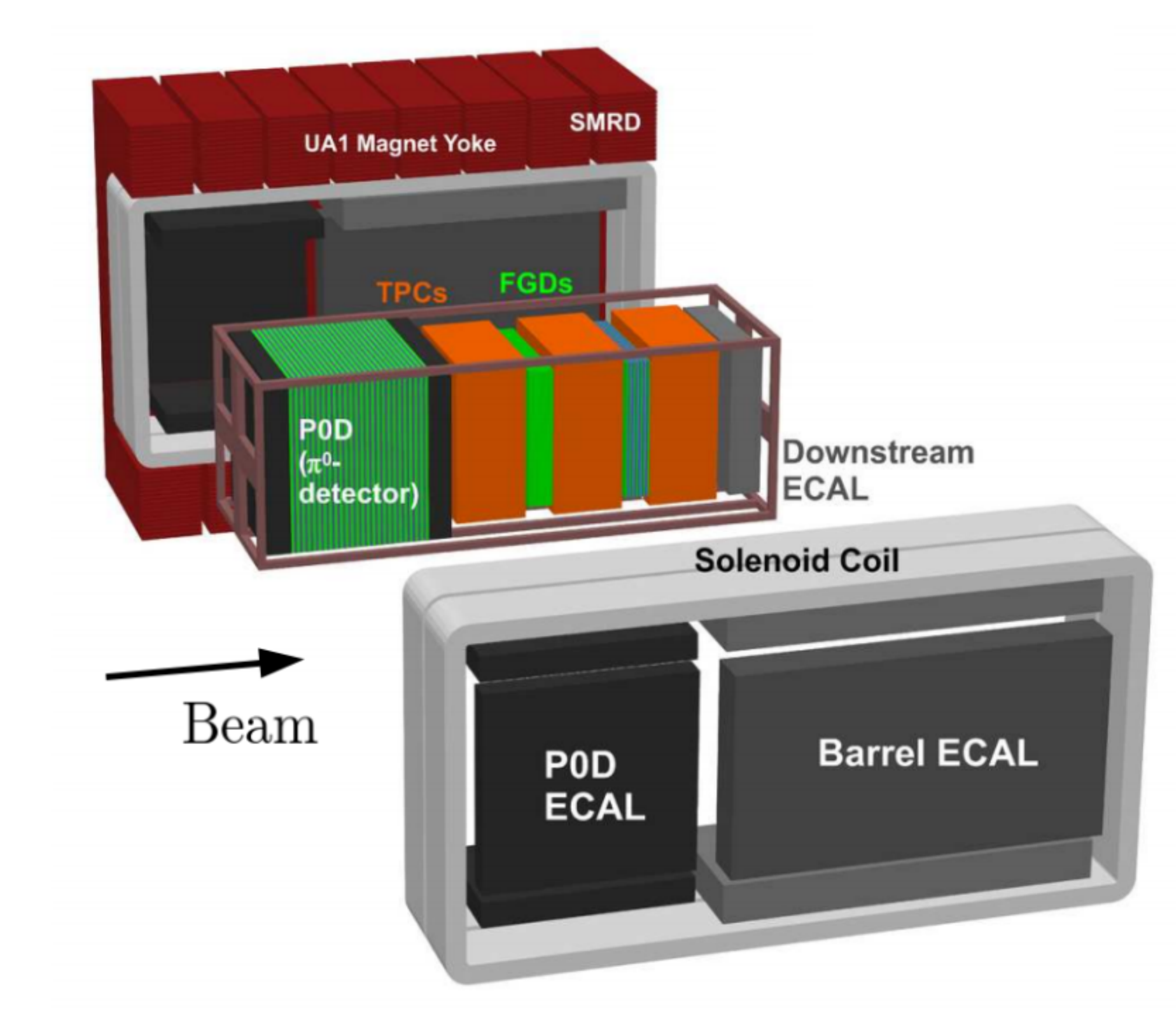}
 \end{center}
\caption{An exploded view of the ND280 off-axis near detector labelling each sub-detector. Adapted from~\cite{Abe:2011ks}.}
\label{fig:nd280}
\end{figure}

The Super-Kamiokande far detector is located 295~km away from the production point and sits $2.5^o$ off the beam axis. T2K is further equipped with two near detectors: ND280 and INGRID. INGRID~\cite{ingrid} is designed to monitor the direction of the neutrino beam whilst ND280 is dedicated to the study of the un-oscillated spectrum of neutrinos at 280~m from the production target and is the detector used by the analyses presented here. ND280 is positioned off-axis so that it has the same peak neutrino energy as Super-Kamiokande. Such configuration ensures a narrow energy spectrum of the beam centred around 600~MeV, in correspondence with the oscillation maximum. It also suppresses the intrinsic $\nu_e$ and the non-QE interactions, which are primarily produced by the high-energy tail of the neutrino flux.
ND280 is composed of an upstream $\pi^0$ detector (P0D)~\cite{Assylbekov:2011sh} and a central tracker region, described below, surrounded by an electromagnetic calorimeter (ECal)~\cite{ecal}, consisting of interleaved layers of lead and scintillator, which itself is all contained within a magnet, providing a 0.2 T dipole field. The magnet is instrumented with the side range muon detector~\cite{smrd}. A schematic of ND280 is shown in Fig.~\ref{fig:nd280}.

The primary component of ND280 used in the analyses presented here is the central tracker region, comprising of three time projection chambers (TPCs)~\cite{Abgrall:2010hi} and two fine grained detectors (FGD1 and FGD2)~\cite{Amaudruz:2012agx}. The FGDs are both instrumented with finely segmented scintillating bars which provide both charged particle tracking as well as a target mass for neutrino interactions and, whilst FGD1 is fully active, FGD2 also contains inactive water layers. In these analyses only FGD1 is used as a hydrocarbon (C$_8$H$_8$) target. Events leaving the FGDs can be tracked into the TPCs, which provide high-resolution tracking and thereby allow the curvature of charged particles to be used to make accurate measurements of their momenta (the TPCs provide an inverse momentum resolution of 10\% at 1 GeV). This can then be combined with measurements of particle energy loss for charged particle identification (PID). If charged particles stop before leaving the FGD1, their momentum is determined by their length. In this case the PID is performed using both track length and the total energy-deposition. Muons and pions can also be identified by searching for delayed signal at the track end due to the Michel electron from the decay of muons (including muons from pion decay).

\section{Measurement strategy \label{sec:anaStrategy}}

\subsection{Observables \label{sec:obsvables}} 

This paper presents three different analyses which study the kinematics of the outgoing muon and protons in charged-current events without pions in the final state (CC0$\pi$). Each of these analyses measure differential cross sections as a function of different observables and with a slightly different selection, optimized to the observables being measured.

The first `multi-differential' analysis measures the differential cross-section as a function of the momentum and angle of the particles in the final state. 
%Since the efficiency of detecting muons and protons
%in ND280 is not flat as a function of the particles angle and momentum, and given the very poor theoretical knowledge of the relative
%kinematics between muon and protons in $CC0\pi$ events, the measurement is performed in a multidimensional way, as a function of
%muon and proton momentum and angle. 
This approach minimises the dependency of the result on the input neutrino-nucleus scattering simulations, as will be described later, and provides the most complete information to characterise the final state. 
%This approach does not rely on strong assumptions regarding the relative kinematics
%between muon and proton in the process of correcting for the detection efficiencies. 
Such results can therefore be compared with present and future models of CC0$\pi$ processes, even if their direct interpretation in terms of different nuclear effects is not straightforward. This multi-dimensional analysis simultaneously measures the cross section of events with and without detected protons in the final state, allowing a complete description of CC0$\pi$ events and, due to improved constraints on systematic uncertainties, surpasses the accuracy of results previously reported by the T2K collaboration in Ref.~\cite{Abe:2016tmq}. Since this analysis classifies events based on the number of reconstructed protons, it is also able to measure a cross-section as a function of the multiplicity of protons above detection threshold. The other two analyses require the presence of at least one proton and, in the case where multiple protons are reconstructed, only the most energetic one is used to form the measured observables.

\begin{figure*}
\begin{center}
 \includegraphics[width=12cm]{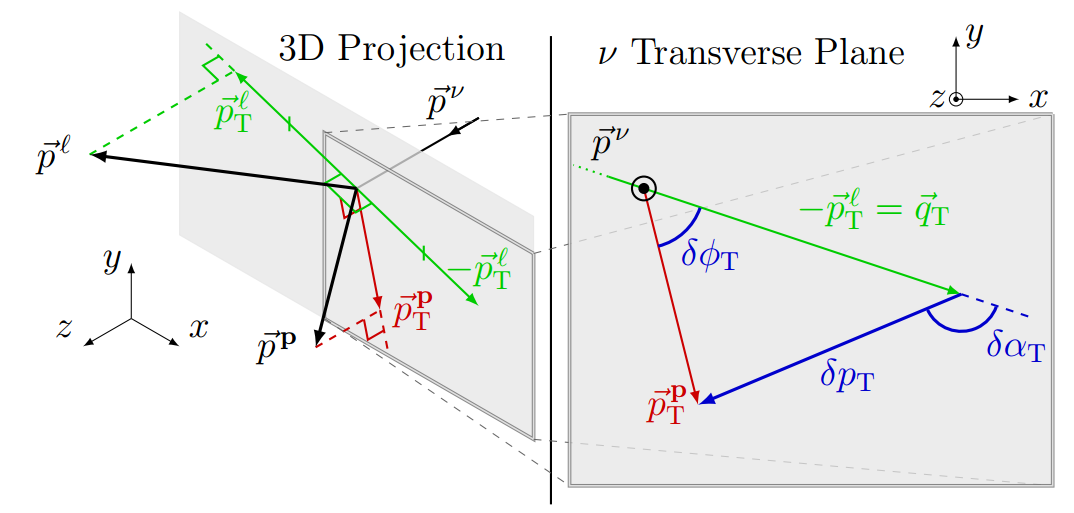}
 \end{center}
\caption{Schematic view of the definition of the Single Transverse Variables: $\dpt$, $\dalphat$ and $\dphit$. The left side shows an incoming neutrino interacting and producing a lepton ($\ell$) and a proton $\mathrm{p}$, whose momenta are projected onto the plane transverse to the neutrino ($\nu$). The right side then shows the momenta in this transverse plane and how the STV are formed from considering the imbalance within it. Taken from Ref.~\cite{Pickering:2016fiq}}
\label{fig:STVdiagram}
\end{figure*}

The second `single transverse variables (STV)' analysis measures the cross-section of CC0$\pi$ events with (at least) one proton in the final state as a function of the STV, which are defined in Ref.~\cite{Lu:2015tcr}. The MINERvA experiment are also measuring transverse kinematic imbalances with a $\sim 3$ GeV peak neutrino beam energy~\cite{Lu:2018stk}. These variables are built specifically to characterise, and minimise the degeneracy between, the nuclear effects most pertinent to long-baseline oscillation experiments. In particular, the STV facilitate the possible identification of: Fermi motion of the initial state nucleon, final state re-interactions of the nucleons in the nucleus and multinucleon interactions (2p2h). 
As shown in Fig.~\ref{fig:STVdiagram}, the STV are defined by projecting the lepton and proton momentum on the plane perpendicular to the neutrino direction. In the absence of any nuclear effects, the proton and muon momenta are equal and opposite in this plane and therefore the measured difference between their projections is a direct probe of nuclear effects in QE events:
\begin{align}
\dptv&=\ptni-\Dptv, \label{eq:dptconvolution}
\end{align}
where $\ptni$ is the initial state nucleon transverse momentum and $\Dptv$ is the modification due to final state effects.
$\dptv$ can be fully characterized in terms of the vector magnitude ($\dpt$) and the two angles ($\dalphat$ and $\dphit$):
\begin{align}
\dpt &= |\overrightarrow{p}_T^l + \overrightarrow{p}_T^p|, \\
\delta \alpha_T &= \arccos \frac{-\overrightarrow{p}_T^l \cdot \delta \overrightarrow{p}_T}{p_T^l \dpt}, \\
\delta \phi_T &= \arccos \frac{-\overrightarrow{p}_T^l \cdot \overrightarrow{p}_T^p}{p_T^l p_T^p},
\end{align}
where ${p}_T^l$ and ${p}_T^p$ are, respectively, the projections of the momentum of the outgoing lepton and proton on the transverse plane. Different nuclear effects alter the distributions of such STV in different and predictable ways. Measurements of the STV therefore have a unique sensitivity to identify nuclear effects, as will be exploited in Sec.\ref{sec:discussion}. This allows cross sections extracted using these observables to act as a powerful tool to tune and distinguish nuclear models. Furthermore, in case of disagreement the STV distributions provide useful hints on the possible causes of the discrepancies. 

%added by Xianguo ->>
%\textcolor{red}{Please describe the convolution nature of $\dptv$, cf. Eq.~7 in~\cite{Lu:2015tcr}. This will be used later in Discussion. Thanks! Xianguo}
%\begin{align}
%\dptv&=\ptni-\Dptv \label{eq:dptconvolution}
%\end{align}
%added by Xianguo <<-
%SB: Done!

The third `inferred kinematics' analysis utilises a similar kinematic imbalance to the STV analysis to probe nuclear effects in CC0$\pi$ interactions by comparing the measured proton momentum and angle with the proton kinematics which can be inferred from the measured muon kinematics in the simplified QE hypothesis. Such inferred proton kinematics are estimated as follows:
\begin{equation}
\label{eq:inferred2}
E_\nu = \frac{m_p^2-m_\mu^2+2E_\mu(m_n-E_b)-(m_n-E_b)^2}{2[(m_n-E_b)-E_\mu+p_\mu cos\theta_\mu]},\\
\end{equation}
\begin{align}
\label{eq:inferred}
E^{inferred}_p &= E_\nu-E_\mu+m_p, \nonumber\\ 
\overrightarrow{p}^{inferred}_p &= (-p_\mu^x,-p_\mu^y,-p_\mu^z+E_\nu), 
\end{align}
where the z axis corresponds to the neutrino direction, $n$, $p$, $\mu$ and $\nu$ denote the neutron, proton, muon and neutrino and $E_b$ is the nuclear binding energy. The value of $E_b$ used in the definition of these variables is 25 MeV for carbon, but this may be different from the event-by-event ``physical'' value of $E_b$.
The cross section for events with a muon and (at least) one proton in the final state is then measured as a function of three observables:
\begin{align}
\label{eq:inferred3}
 \Delta p_p  &=  |\overrightarrow{p}_p^{measured}|-|\overrightarrow{p}_p^{inferred}|, \nonumber\\
 \Delta\theta_p  &=  \theta_p^{measured}-\theta_p^{inferred}, \nonumber \\
 |\Delta \textbf{p}|  &=  |\overrightarrow{p}_p^{measured}-\overrightarrow{p}_p^{inferred}|.
\end{align}
%These observables are built such as to enhance nuclear effects which manifest themselves as deviations from zero imbalance but, in contrast to the STV, the information along the neutrino direction is also included. 
%A muon and a proton in the final state define five independent degrees of freedom, as there is rotation symmetry around the neutrino direction. Once integrated over the muon momentum and angle, only three degrees of freedom are left. Therefore there is a relation between the STV and the variables of Eq.\ref{eq:inferred3}, but such relation is not trivial because  the latter depends directly on the neutrino energy distribution (and the momentum of the initial state nucleon along the neutrino direction), while this is projected away in the STV.
These observables are built such as to enhance nuclear effects which manifest themselves as deviations from zero imbalance.
The STV depend only on transverse components of muon and proton momentum vectors with respect to the neutrino direction, while the variables of Eq.\ref{eq:inferred3} depend also on the longitudinal components of both vectors. As such, there is no trivial relation between the two sets of variables such that each gives complimentary information about the nuclear effects involved in neutrino interactions.
As can be seen in Eq.\ref{eq:inferred}, the definition of the inferred proton kinematics relies on the same QE formula as is used in the estimation of neutrino energy in oscillations measurements at T2K. Therefore the observed deviations from the expected proton inferred kinematic imbalance provide hints of the biases that may be caused from the mismodeling of nuclear effects in neutrino oscillations measurements at T2K. The measurement of the differential cross section as a function of these proton inferred kinematic variables is performed separately in bins of muon kinematics. This can highlight the possible mismodeling of nuclear effects in different regions of the muon kinematic phase-space and is also essential in order to mitigate the model dependence in the efficiency corrections (this will be further discussed in Sec.~\ref{sec:modeldep}). Once de-convoluted from detector effects, this analysis measures how the true particle kinematics deviate from their inferred values under a QE approximation.

%The measured cross section as a function of these observables, after correcting the muon and proton kinematics for detector effects, is compared to the prediction from the reference T2K model of neutrino interactions, where the true muon and proton kinematics at generator level will be used to build the same variables. %SD: Reworded above

\subsection{Minimisation of input-model dependence \label{sec:modeldep}} 

In all three analyses extensive precautions are taken to ensure that the results are minimally dependent on the signal model used in the reference T2K simulation (this model is detailed in Sec.\ref{sec:anaDescription}). This is particularly important for these analyses since the predictive power of available interaction models for the outgoing proton kinematics, and the relative kinematics between muon and protons, is poor. One crucial way to minimise such model-dependence is to ensure that the analyses' signal definition is only reliant on observables which are experimentally accessible at ND280. As such, the signal is defined as all events with no pions in the final state (CC0$\pi$) without correcting for FSI pion absorption. Moreover, for the analyses which integrate over large regions of kinematic phase space or do not estimate the efficiency as a function of all relevant kinematic variables, it is also absolutely necessary to apply phase space restrictions in the signal definition in order to avoid model dependence in the efficiency correction. The phase space restrictions used in the analyses presented here are shown in Tab.~\ref{tab:phaseSpace}. Since the efficiency of detecting muons and protons in ND280 is not flat as function of the particles angle and momentum, the efficiency correction should be made as a function of the momentum and angle of both the outgoing particles. The relative angle between the outgoing particles is also important but, due to the magnetic field and the very good spatial resolution of the TPCs, this has only a second-order effect on the efficiency. The multi-differential analysis performs a complete multi-dimensional efficiency correction and therefore only a loose phase space restriction on the proton momentum is applied. The STV analysis may, in principle, be the most affected by this issue since each bin of the STV integrates over all possible muon and proton kinematics. As a consequence, the STV measurements use the most stringent restrictions in the signal phase space, selecting only regions of flat and/or well understood efficiency. Finally, the inferred kinematics analysis performs a measurement binned in muon momentum and angle and thus it requires only restrictions on the proton phase space. It should be noted that the restrictions listed in Tab.~\ref{tab:phaseSpace} are applied in the signal identification at generator level, therefore the multiplicity of the protons is defined counting only protons above the thresholds in the table. The final measurements do not correct for protons which cannot be detected efficiently and therefore the same restrictions have to be applied to any model in order to compare with the results presented in this paper.

To further alleviate model-dependence, the measured differential cross sections are flux-integrated, normalising all the bins of the measured variables to the same flux:
\begin{equation}
\frac{d\sigma}{dx_i}=\frac{N_i^{CC0\pi}}{\epsilon_i\Phi N^{FV}_{nucleons}\Delta x_i},
\label{eqn:xsec}
\end{equation} 
where $N_i^{CC0\pi}$ is the measured number of signal events in the $i$-th bin, $\epsilon_i$ 
is the efficiency in that bin, $\Phi$ is the overall flux integral,
$N^{FV}_{nucleons}$ is the number of nucleons in the fiducial volume and $x$ is the measured variable.

\begin{center} 
\begin{table}[h!]
\footnotesize
\begin{tabular}{ |l|c|c|c|c| } 
 \hline
Analysis & $p_p$ & $cos\theta_p$ & $p_\mu$ & $cos\theta_\mu$  \\
 \hline
\hline
Multi-dimensional 0p & $<500$~MeV & - & - & - \\
& (or no proton) & & & \\
Multi-dimensional 1p & $>500$~MeV & - & - & - \\
\hline
STV &  0.45-1~GeV & $>0.4$ & $>250$~MeV & $>-0.6$ \\
\hline
Inferred kinematics & $>450$~MeV & $>0.4$ & - & - \\
\hline
\end{tabular}
\caption{Signal phase space restrictions for the three analyses. The cuts apply to the proton with the highest momentum.}
 \label{tab:phaseSpace}
\end{table}
\end{center}

The analyses can be further affected by model-dependent assumptions in the process of correcting for detector effects. The multi-dimensional and the STV analyses use a binned likelihood fit, similar to that used for `Analysis I' in Ref.~\cite{Abe:2016tmq}.
The results of this method, when unregularised, are completely independent on the nominal model used to create the reference templates for the signal. The STV analysis also provides results after applying a regularisation method which has been tuned and thoroughly tested in order to minimize the dependence on the signal model. The third analysis exploits the D'Agostini unfolding procedure~\cite{dagostini1, dagostini2}, also described for 'Analysis II' in Ref.~\cite{Abe:2016tmq}.
%
%Moreover it relies on purity corrections, evaluated from the simulation of the T2K reference model, in order to correct for background. This approach
%is known to be affected by some model-dependence on the signal. The other two analyses constrain the background from a fit to 
%dedicated control regions in data thus minimizing the systematic uncertainties related with background modeling. 
% 

To additionally reduce model-dependence, and to minimise systematic uncertainties related to background modelling, each analysis employs dedicated control regions to achieve a data-driven background estimation and subtraction. Since the control regions chosen and the background subtraction method differs slightly between analyses, these will be discussed in the details of the strategy for each of the analyses which will be reported in Sec.\ref{sec:anaDescription}.

% Make clear that one option is to change particuar dials (for a few cases) which is a closure test but we change the models and try to find cases to break the method.

Despite the many aforementioned precautions, it is still possible that residual model-dependence can bias analysis results. To ensure this does not happen, a comprehensive set of studies with mock datasets has been performed. A first set of mock datasets is created by modifying systematic parameters of particular interest within the reference model (for example 2p2h normalisation or $M_A^{QE}$). The cross section extraction methods must be able to recover the truth when each mock dataset is treated exactly as real data. However, this only tests that the methods can extract the truth from mock data which are systematic variations of the input model, and so is more of a closure test than a true evaluation of possible bias. For a more rigorous test, alternative Monte Carlo event generators, which employ some entirely different signal and background models, are used to produce mock data. Moreover, some of these mock data is specialised to specifically modify the models of the nuclear effects that the analyses wish to characterise, namely modifying 2p2h shape, Fermi motion and FSI models. Using such mock data as an input, it has been verified that, even in the case of extreme deviations from the input signal model, the cross section extraction machinery for each analysis can recover the truth such that it is always well within the uncertainties on the extracted result and also produces a small $\chi^2$ when the full resultant covariances are considered. Some examples of such studies can be found in~\cite{dolanThesis}.

Finally, it should be noted that the three analyses exploit the same data and rely on similar selections. The systematic uncertainties are also evaluated in similar ways, for instance relying on the same data in control regions. As a consequence, it is a very good approximation to assume all the uncertainties to be fully correlated between the different analyses thus the results of the analysis should not be used together in a joint fit. A full discussion on the interpretation of the results will be reported in Sec.~\ref{sec:discussion}.

\section{Analyses description and results \label{sec:anaDescription}}
\subsection{Simulation \label{sec:simulation}} 
%Describe briefly the simulation of the beam (and mention the tuning)

%Describe in full details the Monte Carlo used as main simulation sample in the three analyses (NEUT)

%Describe in details the alternative Monte Carlo used for fake datasets: GENIE and NuWro

%(For each MC say which models are implemented)

%{\it Need access to information what values of paranetes were set when running MC. $M_a = ...$ \\
%NEUT models are described in TN192
%NuWro explain LFG vs SF ... what other parameters Steven set while running NuWro \\
%GENIE (any TN available?) what models it uses? There is diifferent FSI model as in NEUT and no 2p2h implemanted (thats what is in TN-287) ... or we compare to GENIE with 2p2h?  }

The analysis of the neutrino data relies on simulation in order to correct the measured quantities for flux normalization, for detector effects and to estimate the systematic uncertainties.

The T2K flux simulation is based on the modelling of interactions of protons with a graphite target using the FLUKA 2011 package~\cite{Ferrari:2005zk,Fluka:2014}. The modelling of hadron re-interactions and decays outside the target is performed using GEANT3~\cite{GEANT3} and GCALOR~\cite{GCALOR} software packages. Multiplicities and differential cross sections of produced pions and kaons are tuned based on the NA61/SHINE data~\cite{Abgrall:2011ae,Abgrall:2011ts,Abgrall:2016fs} and on other experiments~\cite{eichten,allaby,e910}, allowing the reduction of the overall flux normalisation uncertainty to 8.5$\%$.

The neutrino interaction cross-section with nuclei in the detector and the kinematics of the outgoing particles are simulated by the T2K neutrino event generator NEUT 5.3.2~\cite{Hayato:2002sd,Hayato:2009}. The final state particles are then propagated through the detector material using GEANT4~\cite{Agostinelli:2002hh}. Various additional neutrino event generators are used in the analyses presented in this paper in order to both test the robustness of the results (as discussed in Sec.~\ref{sec:modeldep}) and to compare the final measurements to different models. To this end NEUT 5.3.2.2, NEUT 5.4.0, GENIE 2.12.4~\cite{Andreopoulos:2009rq}, GENIE 2.8.0, NuWro 11q~\cite{Golan:2012wx} and GIBUU 2016~\cite{Gallmeister:2016dnq} are used. 

NEUT version 5.3.2 utilises the Llewellyn-Smith formalism~\cite{llewellyn-smith} to describe the CCQE neutrino-nucleon cross section and the spectral function (SF) from Ref.~\cite{Benhar:1995} is used as a nuclear model. The axial mass used for quasi-elastic processes ($M_A^{QE}$) is set to 1.21 GeV, based on the Super-Kamiokande measurement of atmospheric neutrinos and the K2K measurement on the accelerator neutrino beam~\cite{k2kma}, while the resonant pion production process is described by the Rein Sehgal model~\cite{rein-sehgal} with the axial mass $M_A^{RES}$ set to 1.21 GeV.
The simulation of multinucleon interactions, when the neutrino interacts with a correlated pair of nucleons, also called 2p2h interactions, is based on the model from Nieves et. al in Ref.~\cite{Nieves:2012}. 

The deep inelastic scattering (DIS), relevant at neutrino energy above 1~GeV, is modeled using the parton distribution function GRV98~\cite{Gluck:1998xa} with corrections by Bodek and Yang~\cite{Bodek:2003wd}. The FSI, describing the transport of the hadrons produced in the elementary neutrino interaction through the nucleus, are simulated using a semi-classical intranuclear cascade model. 

A different version of NEUT (5.3.2.2) is used in the comparison of the final results with the models, which differs from the version used for the main analysis of the data by its different value of $M_A^{QE}=1.03$~GeV and its more realistic, reduced strength of proton FSI. NEUT additionally 5.3.2.2 facilitates the alteration of nucleon FSI strength by varying the mean free path between FSI during the intranuclear cascade. The final results are also compared to a third NEUT version (NEUT 5.4.0) where a fully consistent local Fermi gas (LFG) 1p1h and 2p2h model based on the work of Nieves et. al in Ref.~\cite{Nieves:2012} has been implemented.

GENIE, an alternative neutrino generator exploited in these analyses, uses different values of the axial masses ($M_A^{QE}=0.99$~GeV and $M_A^{RES}=1.12$~GeV) and relies on a different nuclear model for CCQE events: a relativistic Fermi gas (RFG) with Bodek and Ritchie modifications~\cite{Bodek:1981wr}. A parametrized model of FSI is used (known as GENIE's ``hA'' model). Both GENIE 2.8.0 and 2.12.4 are used within the analyses, the latter facilitates the optional inclusion of 2p2h interactions using the so called `empirical' MEC model alongside other improvements to the FSI model. 

The NuWro 11q version is also used in these analyses. It simulates the CCQE process with the Llewellyn-Smith model, assuming an axial mass $M_A^{QE}=1.0$~GeV, and the 2p2h process by the model in Ref.~\cite{Nieves:2012}, similarly to NEUT. Different nuclear models are considered in the comparison to the data: SF, RFG and LFG. For LFG and RFG the effect of Random Phase Approximations (RPA) corrections, as computed in Ref.~\cite{Nieves:2011pp}, is tested. RPA is not applied to SF since the model already partially contains the short- and long-range correlations between the nucleons in the nucleus. Similarly the 2p2h contribution should be different in SF with respect to what has been calculated in Ref.~\cite{Nieves:2011pp} for LFG. However, since a dedicated computation of the 2-body current for the SF is not yet available in simulations, the same 2p2h contribution as in LFG is added on top of the SF in both the NEUT and NuWro simulations. For pion production a single $\Delta$ model by Adler-Rarita-Schwinger is used for the hadronic mass $W<1.6$~GeV with $M_A^{RES}=0.94$~GeV. A smooth transition to deep inelastic processes is made for $W$ between 1.3 and 1.6 GeV. The total cross section for DIS is based on the Bodek and Yang approach, similarly to other generators. Like NEUT the FSI are simulated with a semi-classical cascade model.
%Additionally, hadronization is done by custom made model, which relies on the [] and uses PYTHIS6 for the fragmentation.
%Notice that a much longer discussion is needed for the hadronization part if we want to include it, but this is really not relevant for these analyses

The measurements presented in this paper are also compared to GiBUU 2016 where the Giessen-Boltzmann-Uehling-Uhlenbeck implementation of quantum-kinetic transport theory~\cite{Buss:2011mx} is used.
The nucleons are inserted in a coordinate- and momentum-dependent potential using the LFG momentum distribution. The CCQE process is modeled as in Ref.~\cite{Leitner:2006ww} with $M_A^{QE}=1.03$~GeV.
The 2p2h contribution is simulated by considering only the transverse contributions and translating to neutrino scattering the response measured in electron scattering~\cite{Gallmeister:2016dnq}. In these comparisons the default GiBUU 2016 initial state isospin for 2p2h interactions is used ($\mathcal{T}=1$). The model used for single pion production~\cite{Hernandez:2013jka} mostly differs from the other generators for the inclusion of medium effects on the $\Delta$ resonance. The DIS is simulated with PYTHIA v6.4.

The comparison of the measurements presented in this paper to the various mentioned models is performed in the framework of NUISANCE~\cite{Stowell:2016jfr}.

\subsection{Event selection \label{sec:selection}}

%Describe Signal selection and report efficiency and purity
%In this context explain the phase space restrictions.
%Introduce also the various different topologies (mu only, mu TPC+proton FGD, mu TPC + proton TPC...)

%Describe control region selection

%All along the text describe what is common to the 3 analyses and mention explicitly what is different and why (if we find large differences in the selection we may separate the selection section into the following subsection for each analysis but I do not think that it is the case)

%Just refer to experiment section for PIDs

\begin{figure*}[!htbp] 
\centering
\includegraphics[width=0.9\textwidth]{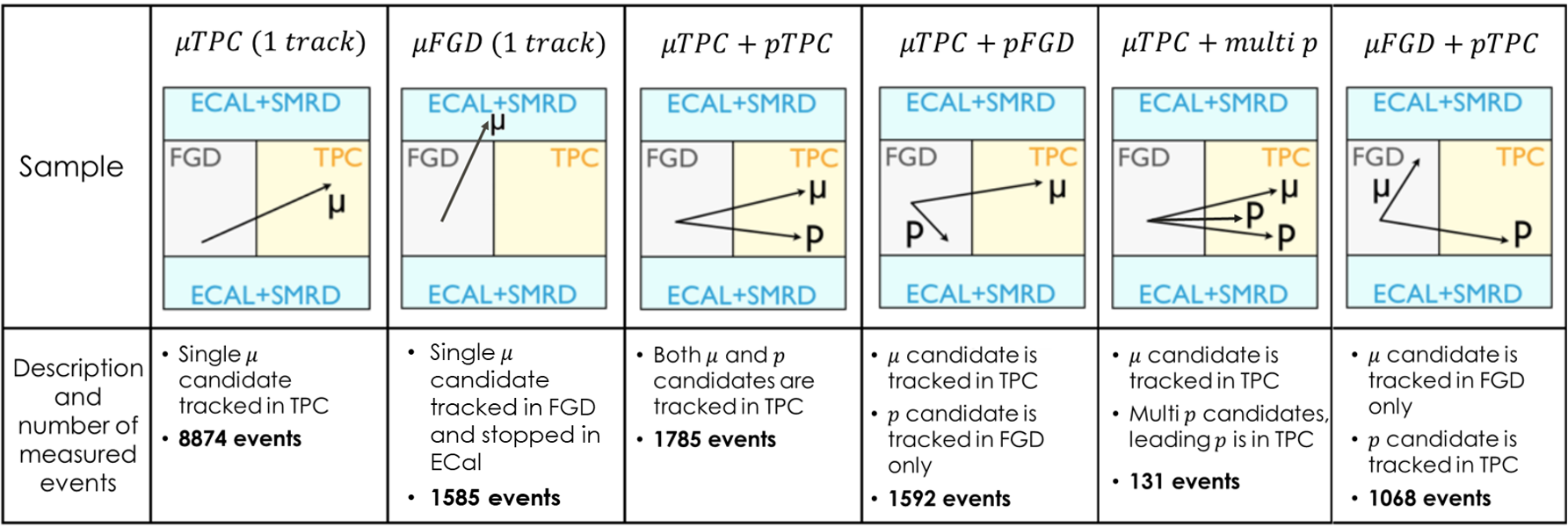}
\caption{A diagram summarising the different signal samples used. The number of events selected in data for each sample is indicated.}
\label{fig:topo}
\end{figure*}

The three analyses presented herein share a common basic event selection, which aims to identify muon neutrino interactions with a hydrocarbon target producing one muon, no pions and any number of protons in the final state. Events are pre-selected by identifying a vertex in the most upstream fine-grained detector (FGD1) associated with either the highest momentum negative track in the central TPC or, if there is no negative track, the highest momentum positive track. If there is no such TPC track the event is rejected. This pre-selection is split depending on the charge of this primary track, as shown in Fig.~\ref{fig:topo}. 

If the primary track is negative then the track is required to be muon-like using the TPC PID. Extra tracks sharing a common vertex with the primary track must either have good quality measurement in the TPC, or be contained in FGD1 such that their kinematics can be reliably determined and they must be identified as proton-like by the TPC or FGD PID respectively. If there is more than one extra track sharing such a common vertex then it is required that at least one of these tracks enter the TPC but each must be identified as proton-like. Following this selection, any events with other tracks that are not muon- or proton-like are rejected. To reject events with low momentum charged or neutral pions, it is required that no Michel electrons (electrons from the decay of the muon that itself is from the pion decay) are tagged within the FGD and that there is no activity in the tracker ECal consistent with a photon. The selected events are then split into samples based on whether there was zero, one or more than one proton-like track and, if so, whether it left a track in the TPC. 

If the primary track is positive (and there are therefore no identified negative TPC tracks) then the selection requires the identification of a single extra FGD track sharing a common vertex position with the primary track. This track must then either stop in FGD1 or in the surrounding ECal and be identified as muon-like by the FGD or ECal PID respectively. In the latter case, time of flight information between FGD1 and the ECal is used to ensure that energy depositions seen in the ECal are related to the same track that traversed the FGD. 

Finally a last sample is selected with a single track travelling through the FGD before stopping in the ECal. This sample uses the measured time of flight between the track ends to verify propagation direction, and the ECal hit topologies to verify whether the track is muon-like. This is a small sample but all concentrated at high angle, therefore it is included only in the multi-differential analysis which measures the cross-section with finer binning in muon angle.

\begin{figure}[!htbp] 
\centering
\includegraphics[width=0.49\textwidth]{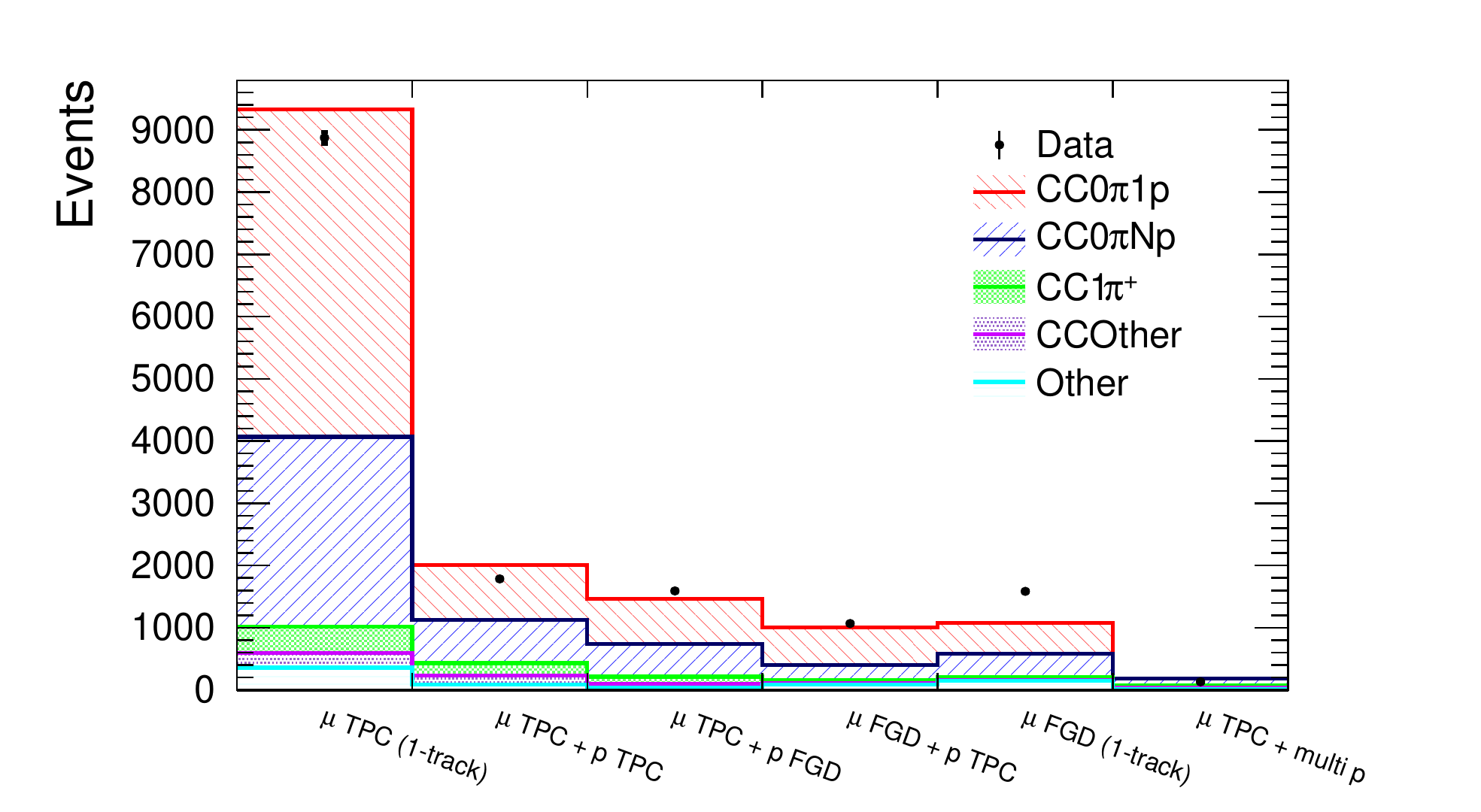}
\caption{The number of selected events in each sample of the event selection within data and the NEUT 5.3.2 simulation. The simulated events are broken down by true interaction topology as predicted by the generator.} 
\label{fig:sigEvtsByBranch}
\end{figure}

Fig.~\ref{fig:topo} summarises the topology and the number of selected events within the six signal samples discussed while the number of selected events in each sample, broken down by true interaction topology, is shown in Fig.~\ref{fig:sigEvtsByBranch}. Other samples are possible but typically with very poor efficiency, resolution and larger detector systematic uncertainties, for example events with a negative primary track and multiple FGD1 contained protons. Since such alternative samples are found to make up a very small number of selected events, less than 30 events in the available data, they are excluded.

%\subsubsection{Efficiency and phase-space constraints}

\begin{figure*}[!htbp]
 \centering
 \includegraphics[width=0.49\textwidth]{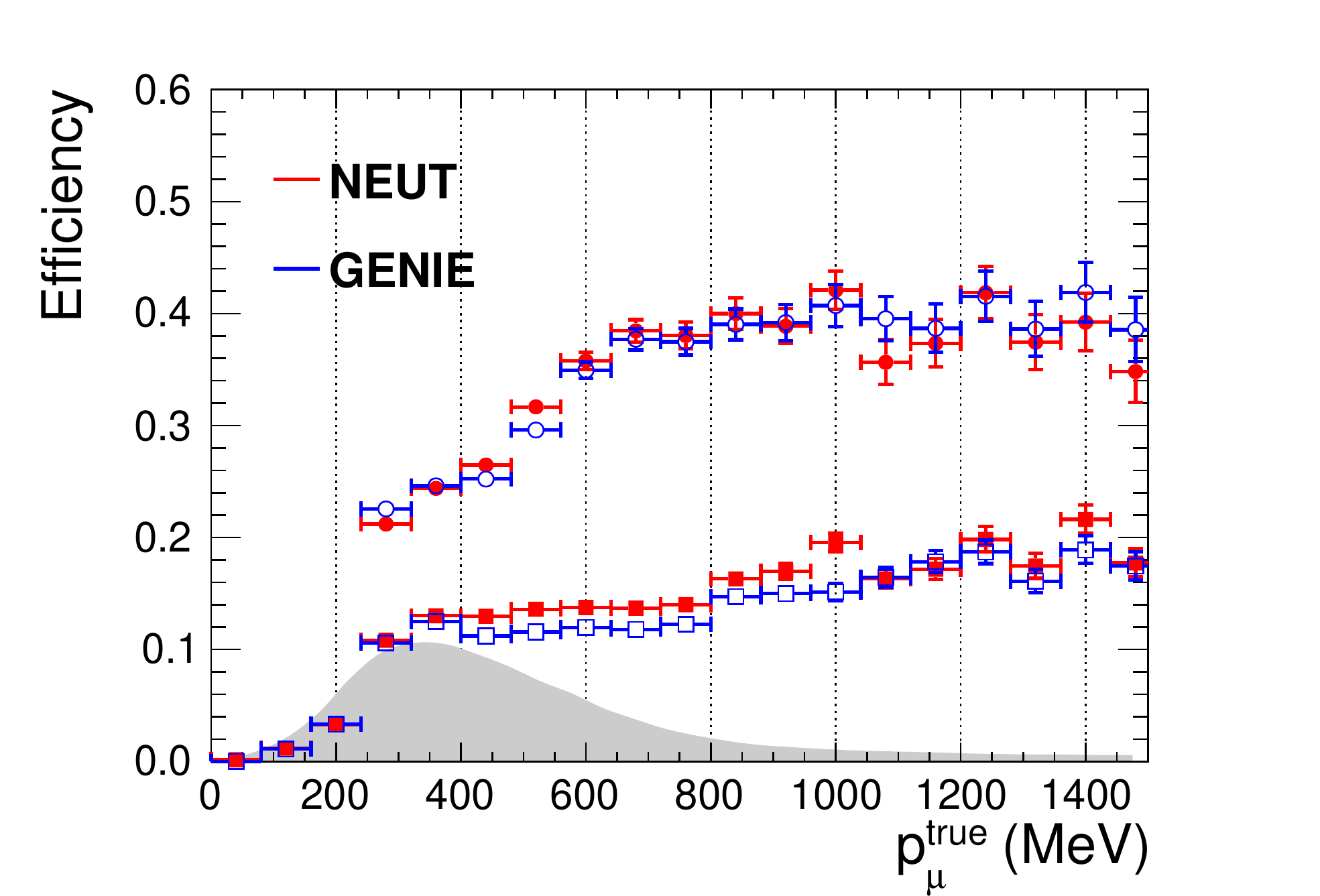}
 \includegraphics[width=0.49\textwidth]{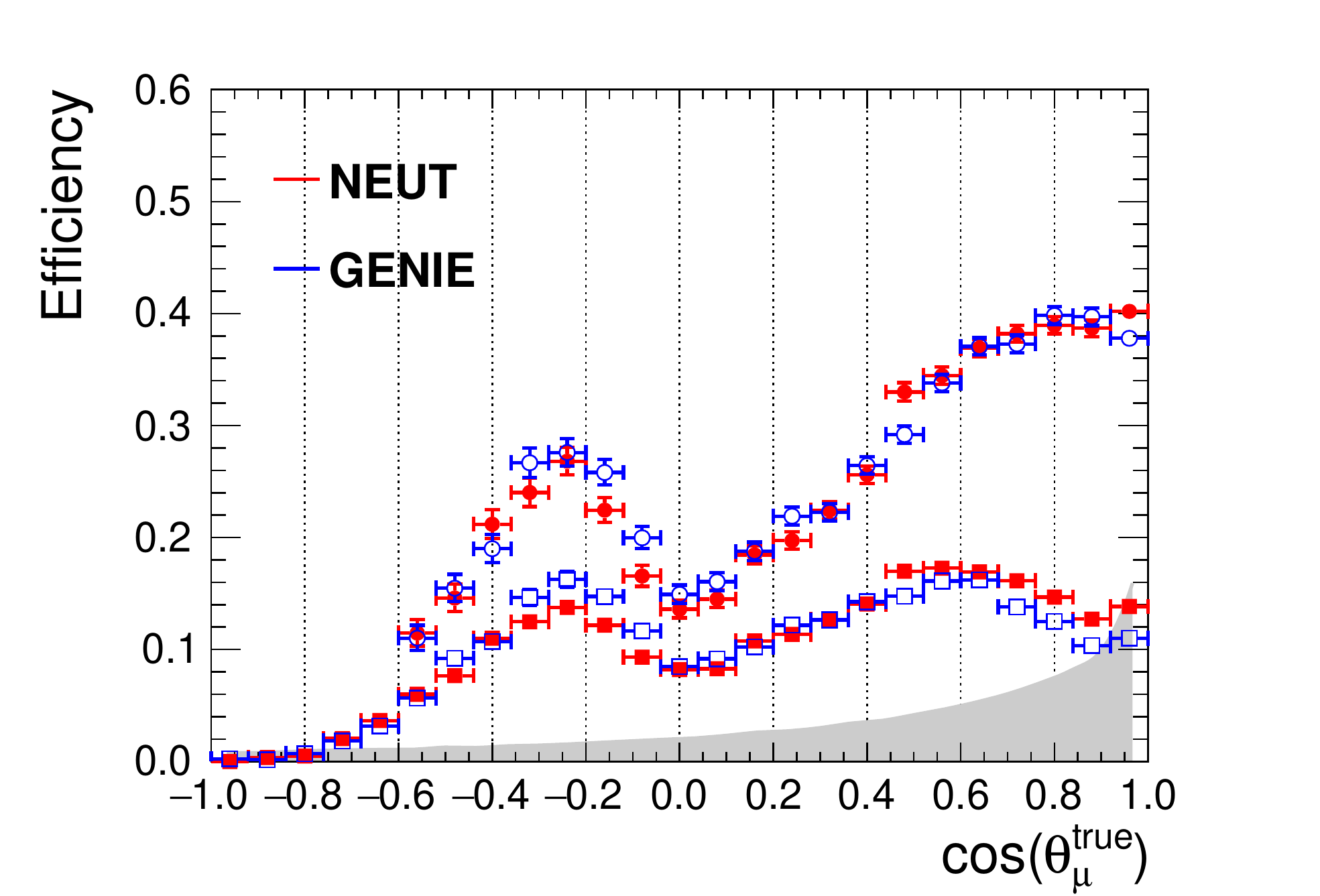}

 \includegraphics[width=0.49\textwidth]{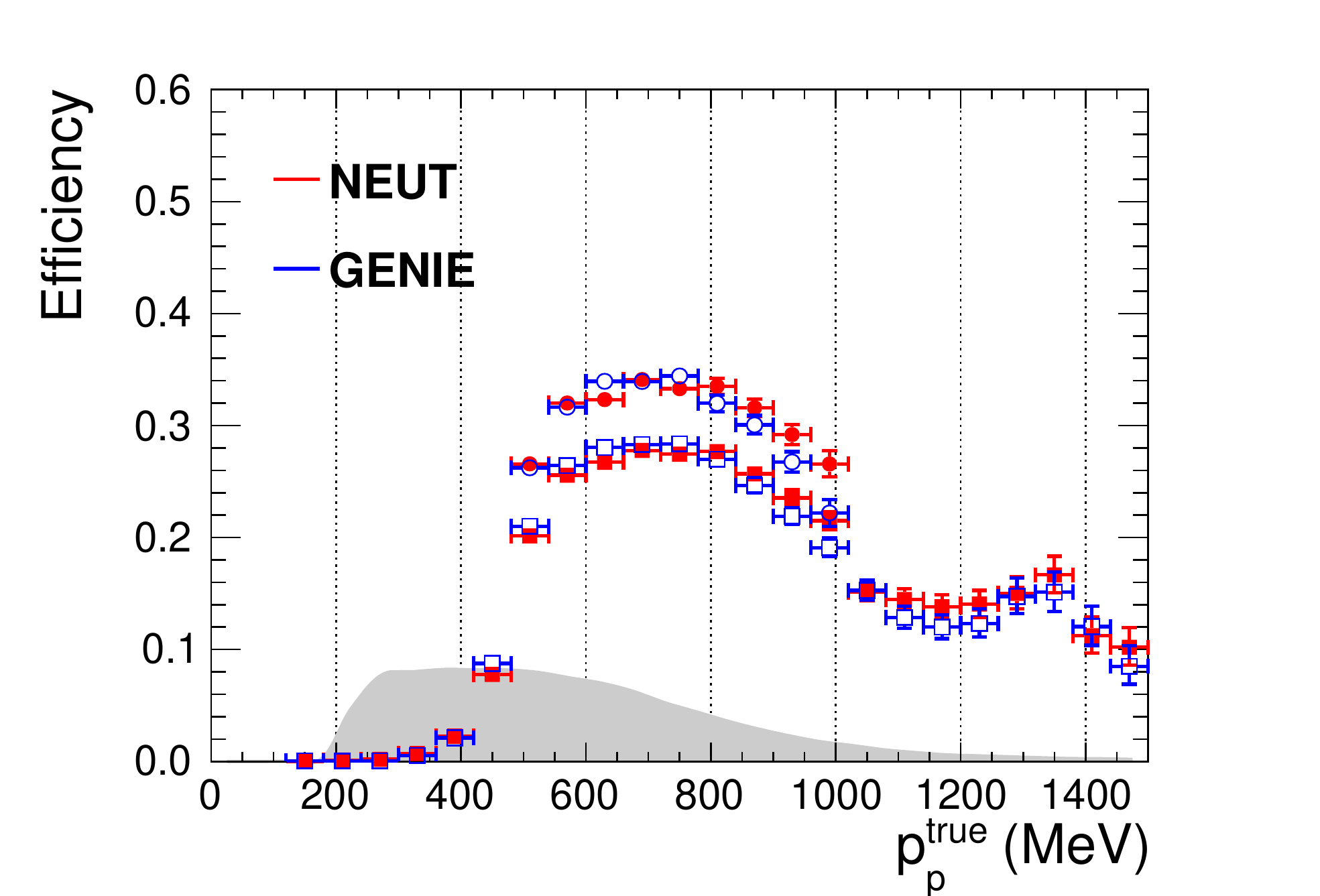}
 \includegraphics[width=0.49\textwidth]{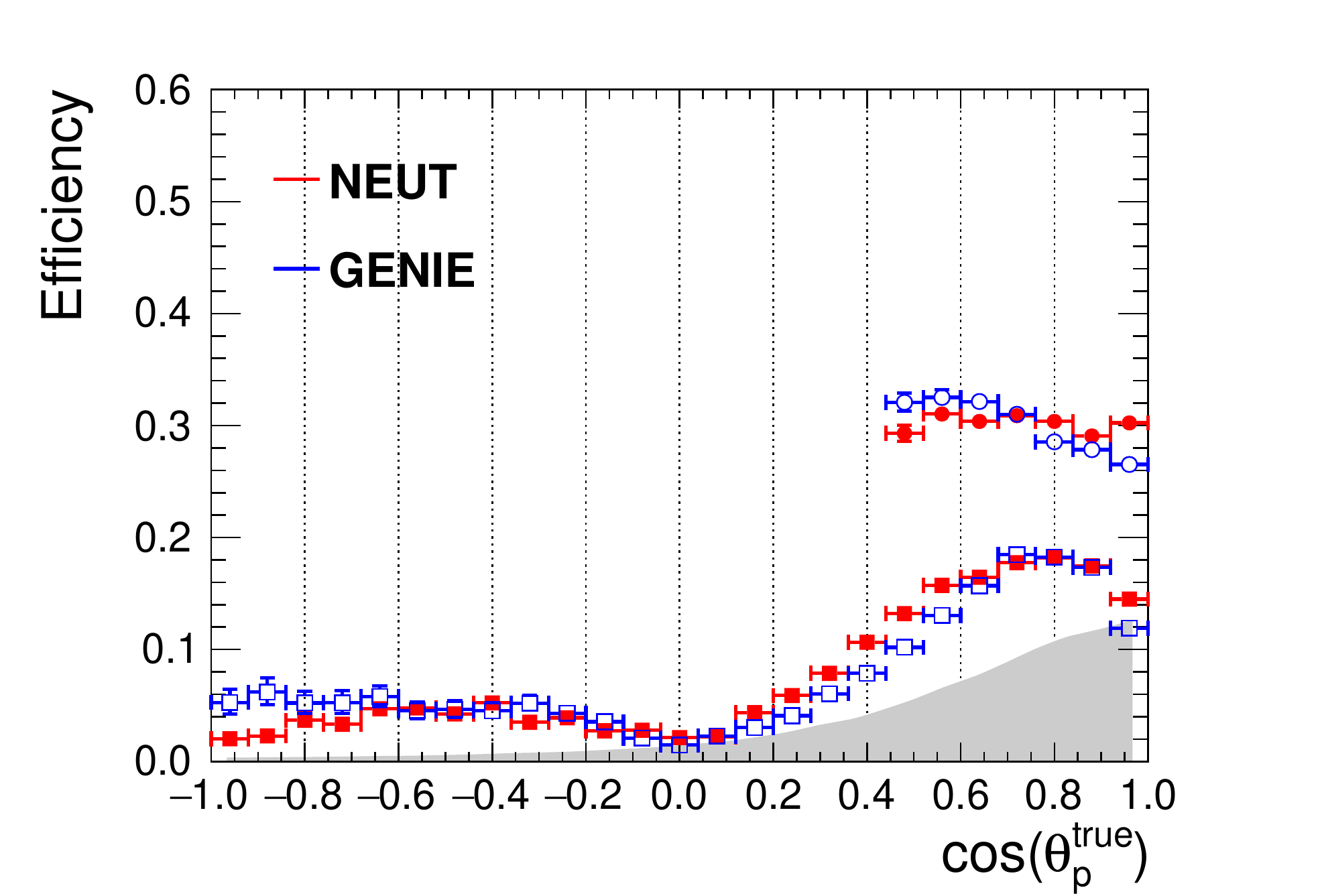}
 \caption[]{ \label{fig:eff1} Efficiencies, after all the selection steps, as function of true muon (upper plots) and proton (lower plots) kinematics as predicted by NEUT 5.3.2 and GENIE 2.8.0. The square points show the efficiency prediction before any phase-space constraints whilst the circular points have had the proton and muon kinematic constraints for the STV analysis in Tab.~\ref{tab:phaseSpace} applied. The grey filled distribution shows the shape of the CC0$\pi$ cross section predicted by NEUT 5.3.2.2.}
\end{figure*}

As discussed in Sec.~\ref{sec:modeldep}, it is important not to attempt to correct for low efficiency in regions of kinematic phase-space that the detector is not sensitive to. This is particularly important when measuring a differential cross section in observables that do not well characterise a detector's acceptance such as the single-transverse and proton inferred kinematic observables. To avoid input-model bias from integrating over regions of changing efficiency, it is necessary to set appropriate limitations on the kinematic phase-space of the final state particles.
%Ideally the efficiency correction should be able to be determined from GEANT4 and the ND280 detector simulation such that it is entirely independent of the neutrino interaction model. For example, if single muons are being measured an event with a particular true momentum and angle should either be seen or not, so the efficiency should be a 2D equivalent of a step function in muon $p$, $cos(\theta)$\footnote{In reality this is not quite the case since factors such as vertex position and secondary interaction rates also play a role in whether an event can be reconstructed, but this is still independent of the neutrino-nucleus interaction model.}. However, if the efficiency curve of just momentum or angle in 1D is considered, then the input simulation interaction model is relied upon to fold in the effects of the other dimension (i.e. to inform the momentum distribution falls into each $cos(\theta)$ bin or vice-versa). 
In the analyses presented here, both muons and protons are identified and therefore ND280's acceptance is reasonably well characterised by the muon and proton momentum and angle\footnote{It should be noted that ND280's acceptance has also a small dependence on other factors, most markedly the vertex position and the angle between the outgoing muon and proton. However, the distribution of the former is not dependent on the interaction model while, as discussed in Sec.~\ref{sec:modeldep},  the impact of the latter is fairly small.}. 
%but, for the reasons discussed above, it is important to note that such projections rely on the model predictions and therefore do not fully characterise the selections acceptance. 
Ideally, the selected phase-space restrictions should leave a flat efficiency within the four dimension regions of muon- and proton-kinematics which will be integrated over in the final measurement. This ensures that the efficiency corrections are independent of the distribution of kinematics which are not measured. 
To determine the phase-space restrictions introduced in Tab.~\ref{tab:phaseSpace}, the efficiency and selected event distributions were studied in various projections of the underlying four-dimensional kinematics in order to find a suitable balance between efficiency flatness and the number of CC0$\pi$+Np events that fall out of the restricted phase space (which are then considered as background). The resultant impact of the phase-space restrictions is shown for both the ND280 NEUT 5.3.2 and GENIE 2.8.0 simulations in Fig.~\ref{fig:eff1}, which shows the efficiency after all the selection steps projected into the relevant kinematic variables, before and after phase-space restrictions are applied. In general it can be seen that the chosen phase space restrictions ensure a more flat efficiency within the regions of kinematic phase space that contribute most to the CC0$\pi$ cross section, particularly in the poorly understood outgoing proton kinematics.
%Once the signal definition is altered the purity and efficiency of the selection also changed and a new background is manifest in 

\begin{figure*}[htbp] 
\centering
\includegraphics[width=0.49\textwidth]{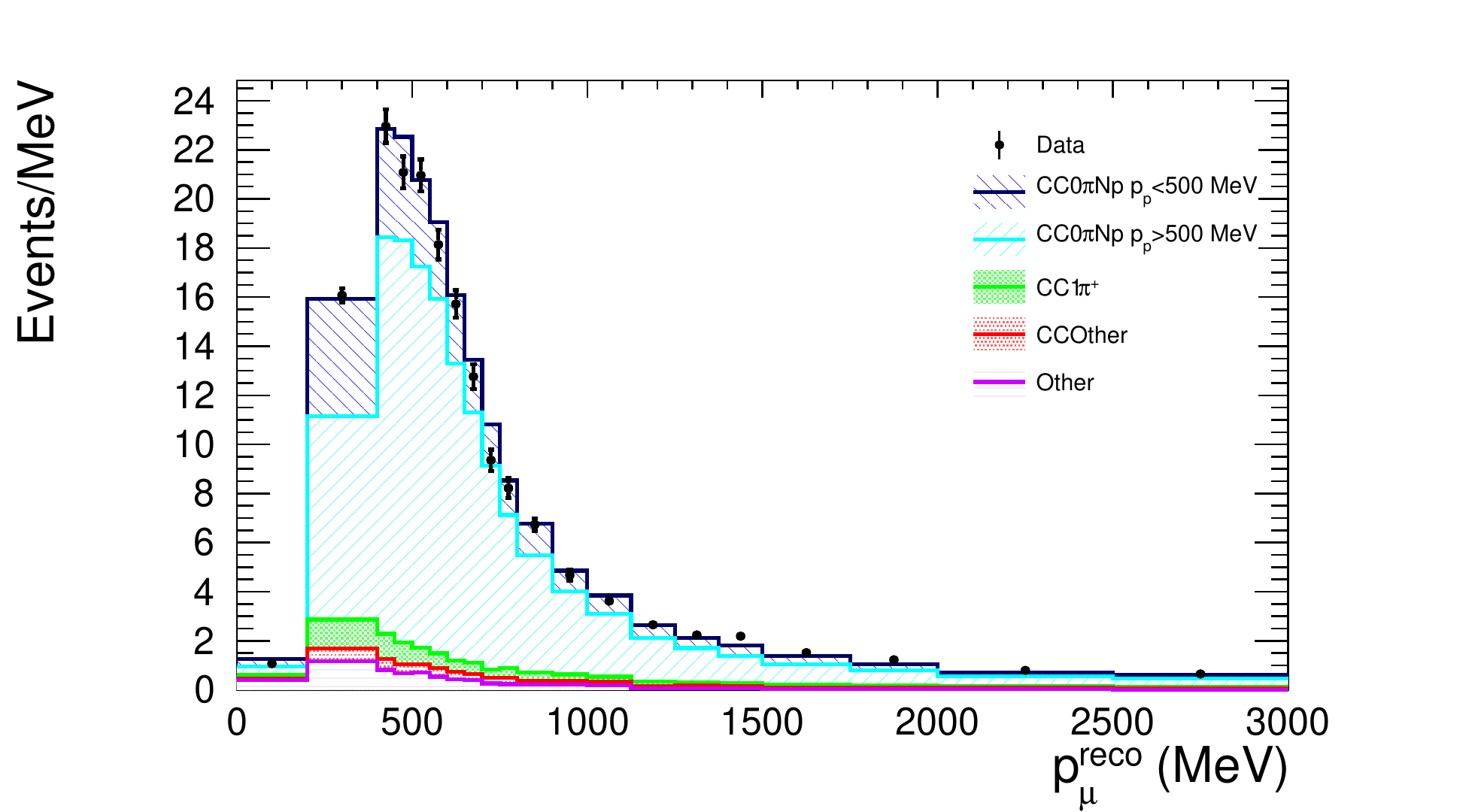}
\includegraphics[width=0.49\textwidth]{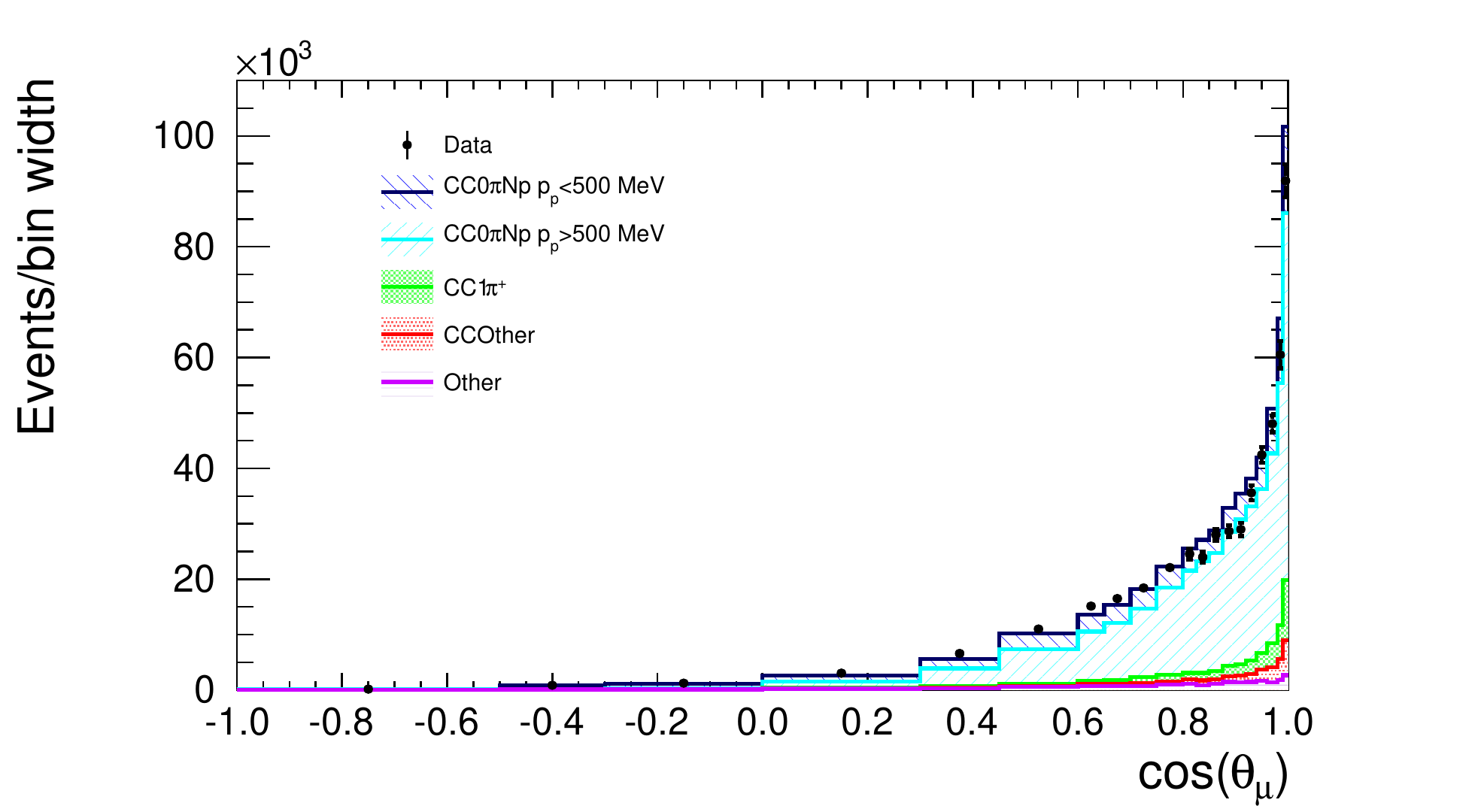}
\includegraphics[width=0.49\textwidth]{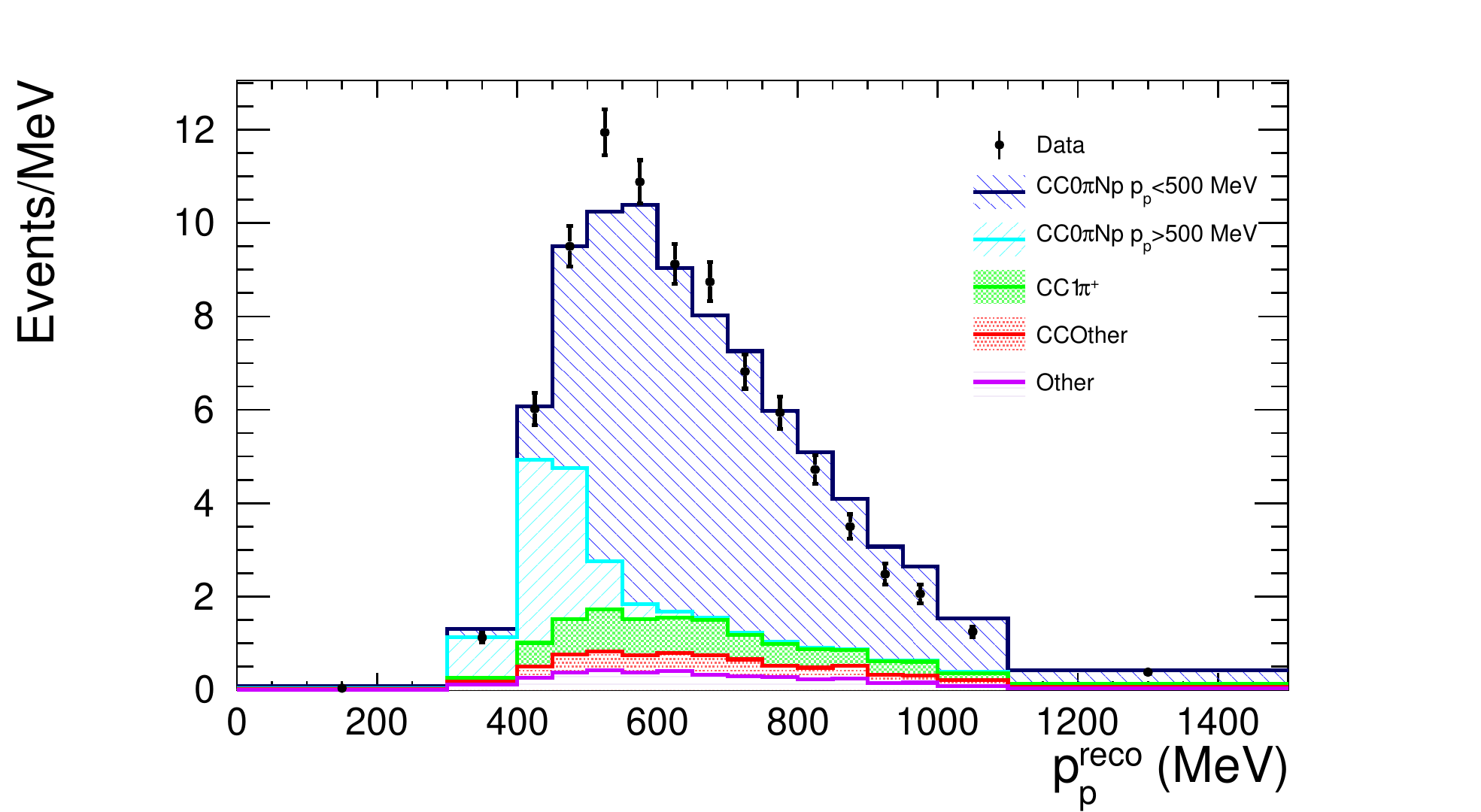}
\includegraphics[width=0.49\textwidth]{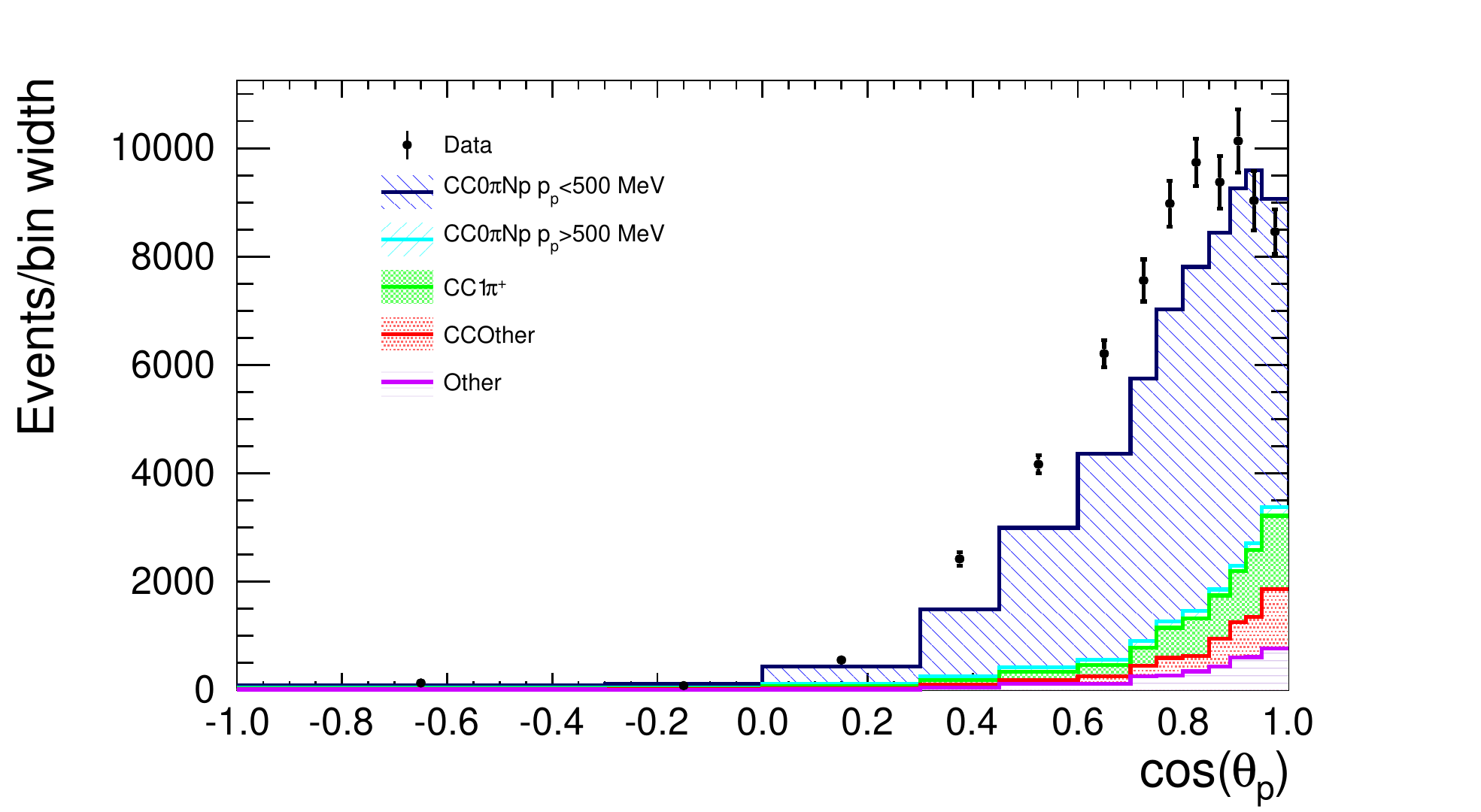}
\caption{The distribution of reconstructed observables used within the multi-differential analyses following the event selection for both NEUT 5.3.2 and data. The muon kinematic plots show events from all samples while the proton kinematics are limited to showing events from the samples which identify a proton. The plots are broken down by interaction topology and the CC0$\pi$ contribution is further split depending on whether the interaction falls within the multi-differential analysis phase space constraints from Tab.~\ref{tab:phaseSpace}. }
\label{fig:selmup}
\end{figure*}

\begin{figure*}[!hp] 
\centering
\includegraphics[width=0.49\textwidth]{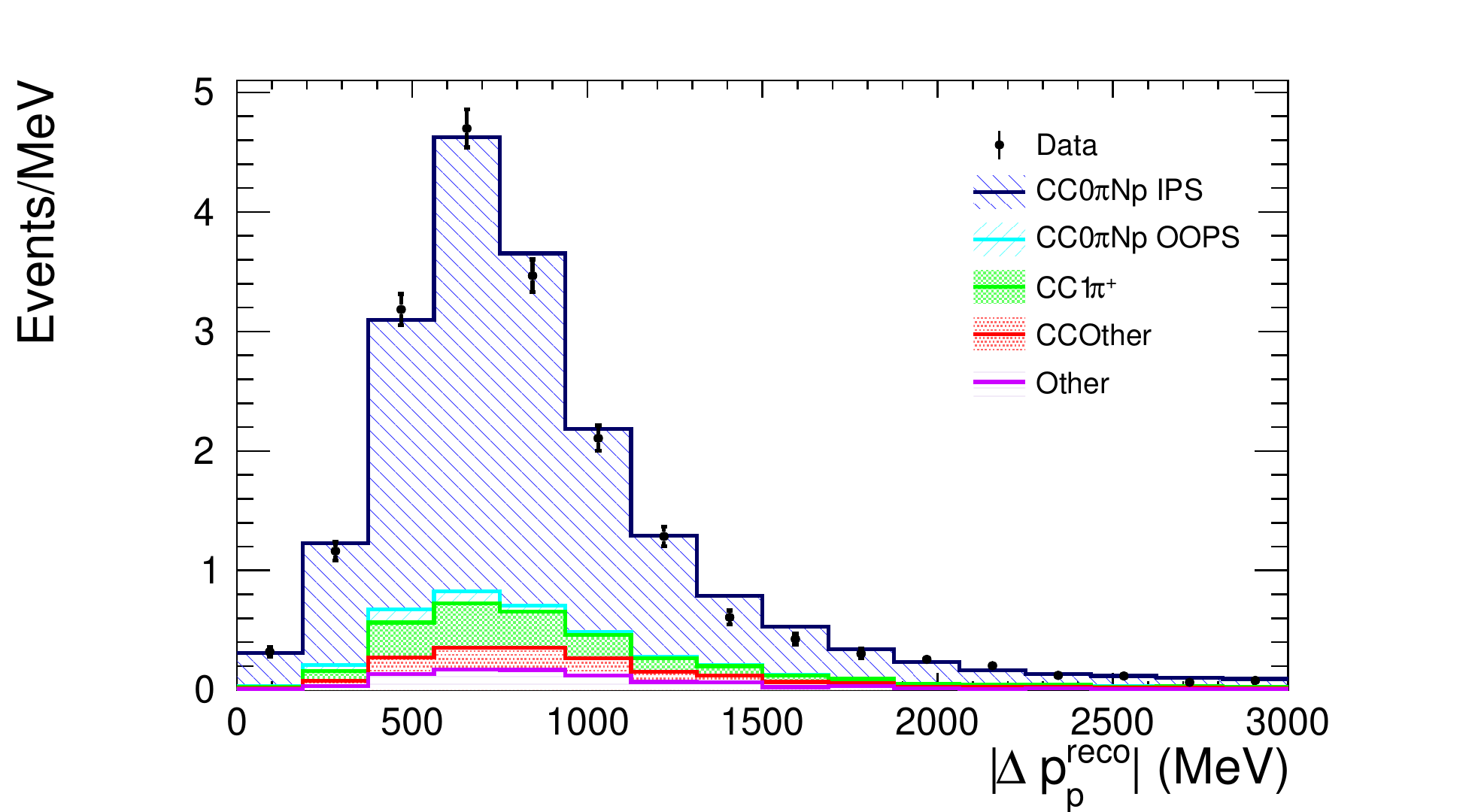}
\includegraphics[width=0.49\textwidth]{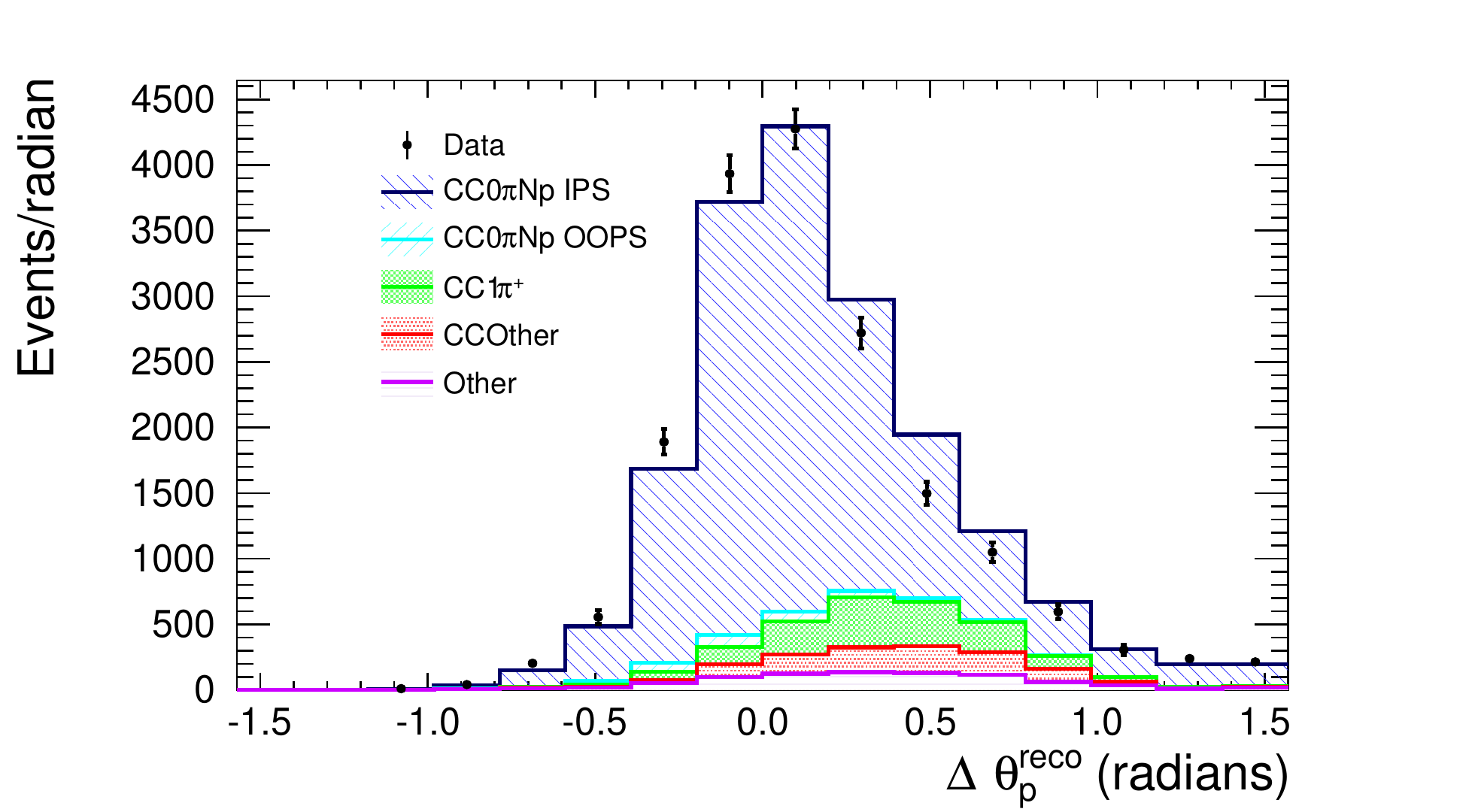}
\includegraphics[width=0.49\textwidth]{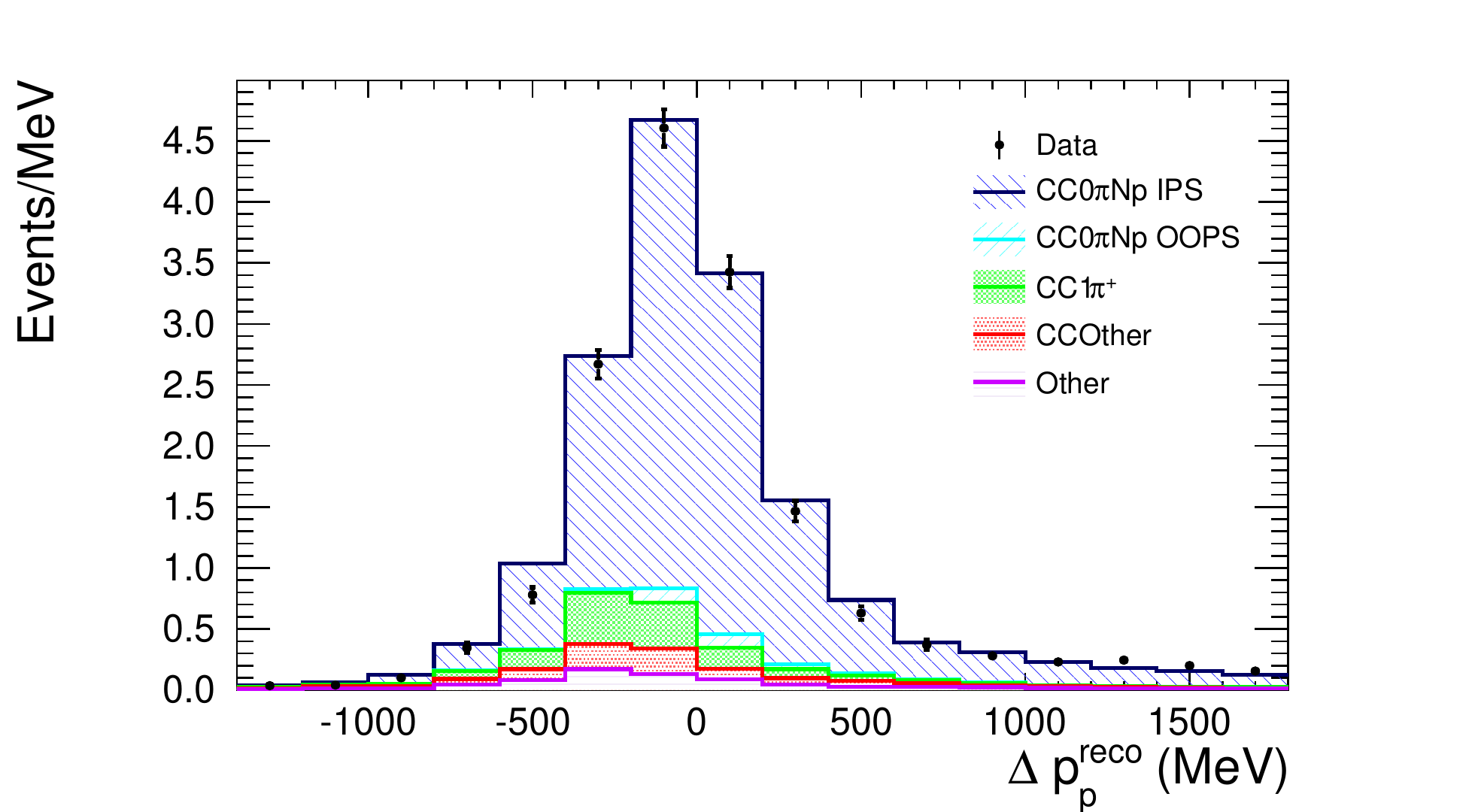}
\includegraphics[width=0.49\textwidth]{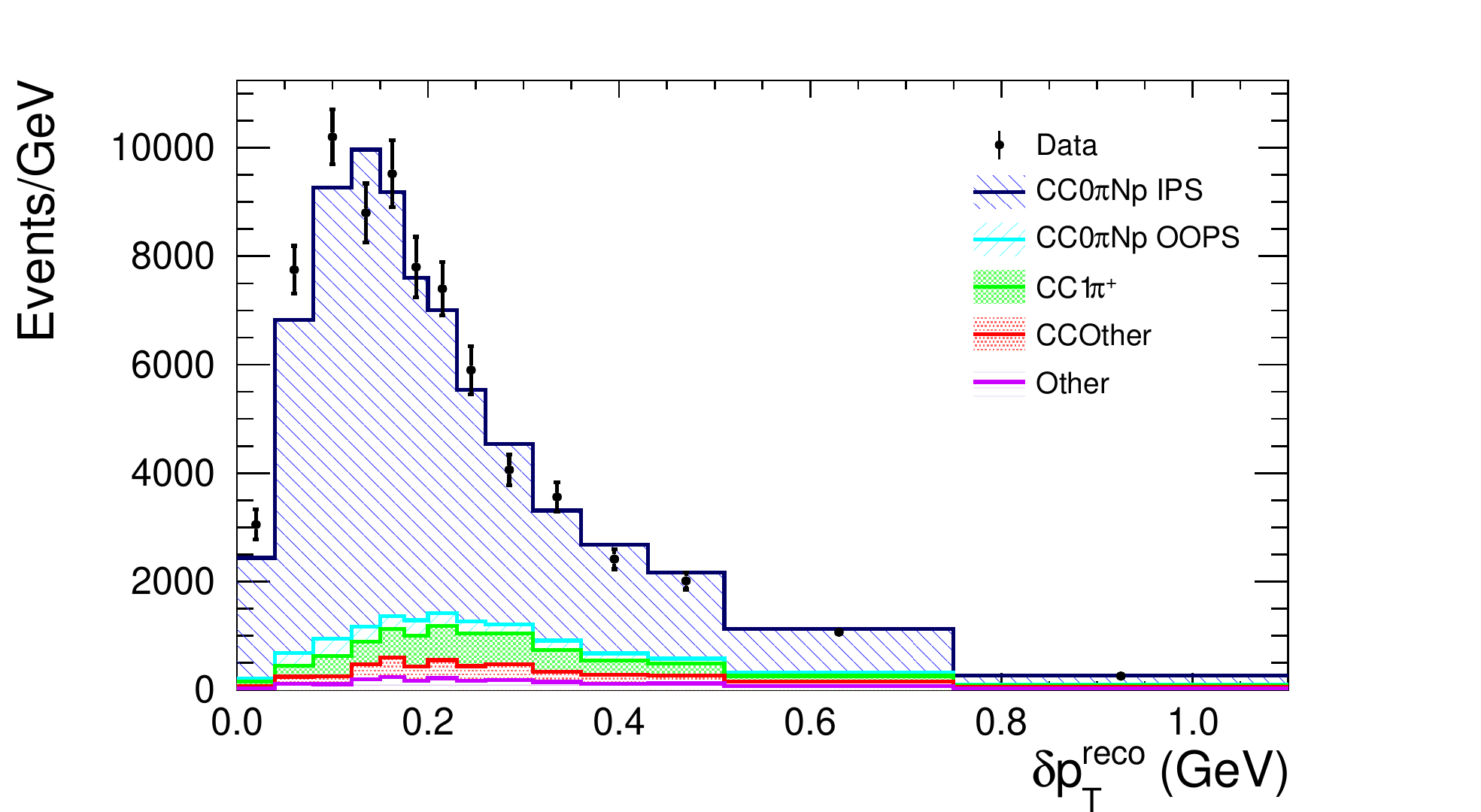}
\includegraphics[width=0.49\textwidth]{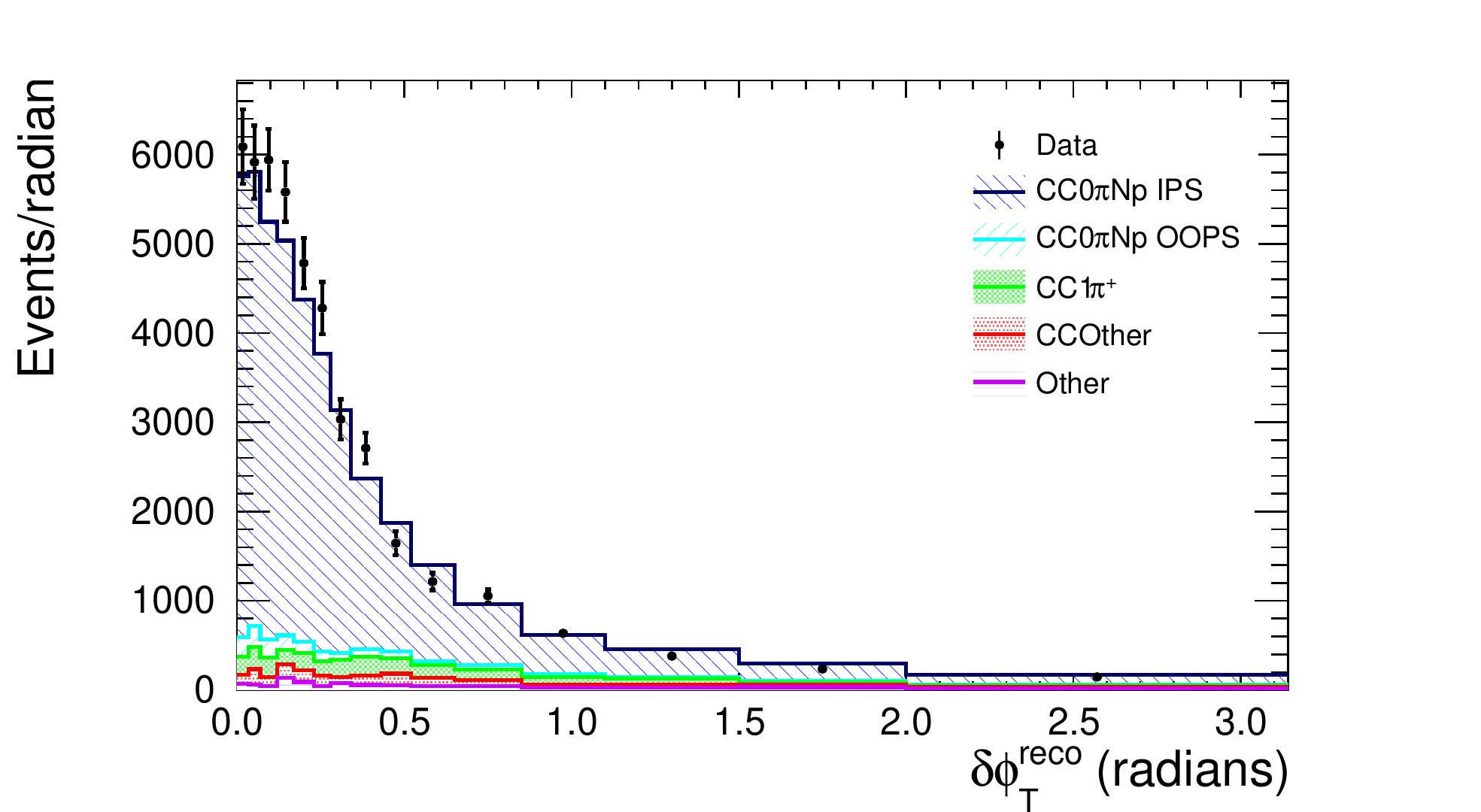}
\includegraphics[width=0.49\textwidth]{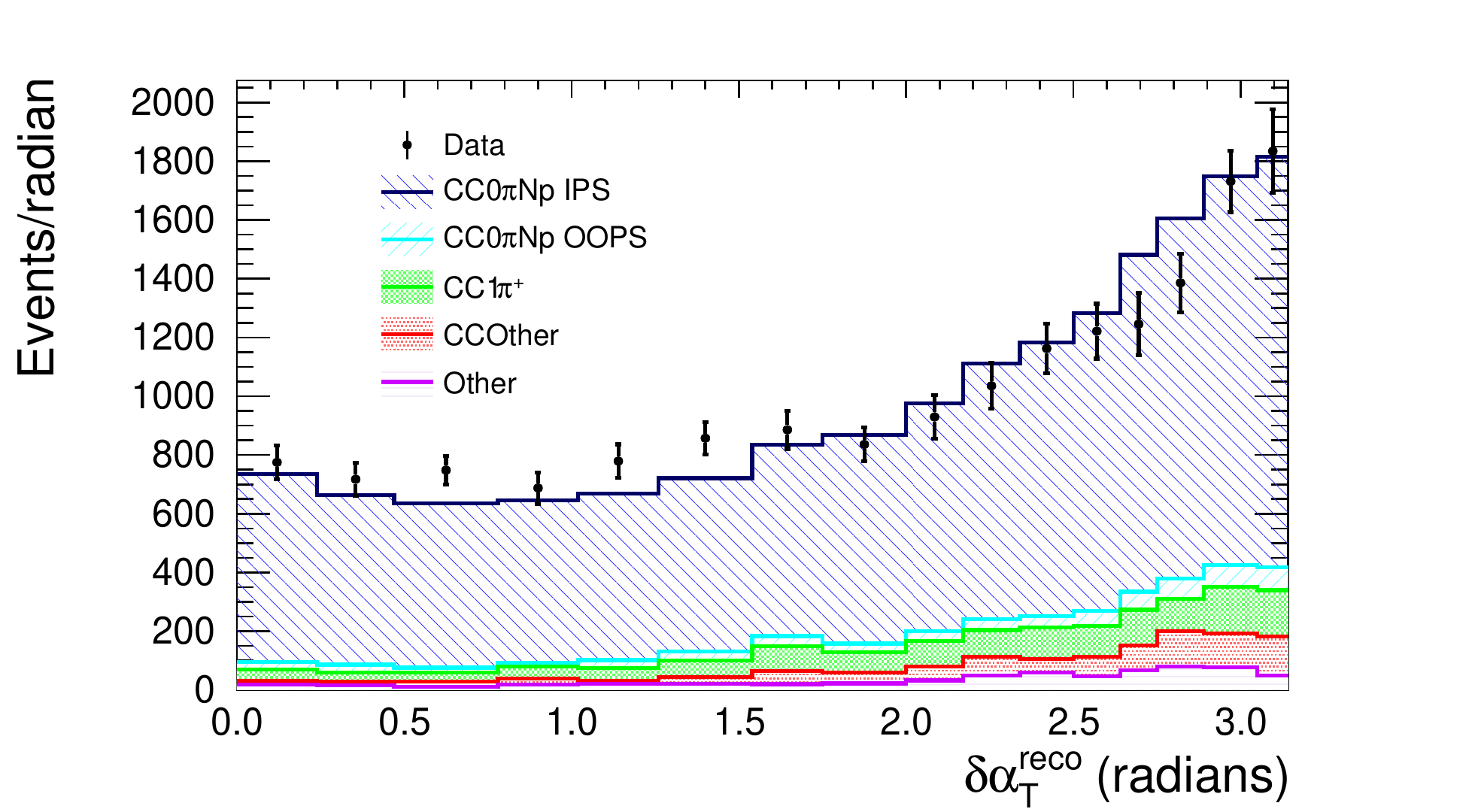}
\caption{The distribution of reconstructed observables used within the inferred kinematics and STV analyses following the event selection and phase-space restrictions on the reconstructed kinematics for both NEUT 5.3.2 and data. The plots are broken down by interaction topology and the CC0$\pi$ contribution is further split depending on whether the interaction falls in or out of the phase space constraints (IPS/OOPS) from Tab.~\ref{tab:phaseSpace} (the phase-space definitions for the STV or inferred kinematics plots follow the restrictions pertinent to their respective analyses). }
\label{fig:selikistv}
\end{figure*}

Following the event selection and the application of the phase-space restrictions in both true and reconstructed kinematics, the efficiency and purity of signal events for each analysis is shown in Tab.~\ref{tab:pssum}. The reconstructed muon and proton kinematics from the combined samples, broken down by topology, are shown in Fig.~\ref{fig:selmup}. The events are separated depending if, based in their true kinematics properties, they fall in or out of the phase-space restrictions (IPS/OOPS) listed in Tab.~\ref{tab:phaseSpace}. The distribution of the single-transverse and inferred kinematic observables are also shown in Fig.~\ref{fig:selikistv}. 

%Following the selection the overall efficiency and purity of selecting CC0$\pi$+Np interactions within the phase-space restrictions for each analysis is shown in table~\ref{tab:pssum}.

\begin{table}[h]
\begin{center}
\small
\centering
\begin{tabular}{|c|c|c|c|}
\hline
Analysis & Purity & Efficiency & Events \\
\hline
Multi-differential & 78.3\% & 20.5\% & 3674\\
Inferred kinematics & 79.4\% & 21.0\% & 3691\\
STV & 80.7\% & 24.1\% & 3073\\
Unconstrained & 81.2\% & 12.3\% & 4576\\
\hline
\end{tabular}
\caption[]{\label{tab:pssum}The purity, the efficiency (both from NEUT 5.3.2 and GENIE 2.8.0) and the number of selected events in data for each analysis in the restricted phase phase and before phase space restriction (unconstrained).}
\end{center}
\end{table}

%\begin{table}[h]
%\begin{center}
%\small
%\centering
%\begin{tabular}{|c|c|c|c|c|c|c|}
%\hline
%   & \multicolumn{6}{c|}{Topology} \\
%Sample & CC0$\pi$0p & CC0$\pi$1p & CC0$\pi$>1p & CC1$\pi^+$ & CCOther & Other \\
%\hline
%$\mu$TPC+pTPC  &0.01\%&46.1\%&36.1\%&9.55\%&4.23\%&3.98\%\\% &1677\\
%$\mu$TPC+pFGD  &0.00\%&50.1\%&36.2\%&8.01\%&2.35\%&2.40\%\\% &1222\\
%$\mu$FGD+pTPC  &0.00\%&60.3\%&23.8\%&5.36\%&2.60\%&7.92\%\\% &919\\
%$\mu$TPC+Np    &0.00\%&1.74\%&64.0\%&18.9\%&8.70\%&6.71\%\\% &154\\
%%$\mu$FGD+NpTPC &0.00\%&4.29\%&61.4\%&11.4\%&5.00\%&17.9\%\\
%%$\mu$FGD+NpFGD &0.00\%&12.9\%&51.4\%&24.3\%&4.29\%&7.14\%\\
%%CC1$\pi^+$ Control &0.00\%&3.28\%&1.53\%&72.5\%&13.7\%&9.10\%\\
%%CCOther Control  &0.00\%&0.31\%&0.37\%&11.3\%&70.9\%&17.2\%\\
%\hline
%\end{tabular}
%\caption[The proportion of each true signal topology in each reconstructed sample alongside the expected number of events in available data.]{\label{tab:topo}The proportion of each true signal topology in each reconstructed sample.}
%\end{center}
%\end{table}

%\subsubsection{Control regions}

Although the selection presented in this chapter identifies a high purity sample of CC0$\pi$+Np events, there are still non-negligible backgrounds. The majority of these come from CC1$\pi^+$ events, where the pion (and associated Michel electron) are missed, but there is also notable contribution from  other (multi-pion) CC events. These backgrounds are constrained through dedicated control samples which allow an improved background estimation and thereby smaller background modelling uncertainties. 

\begin{figure}[hbtp]
\begin{center}
\includegraphics[width=0.4\textwidth]{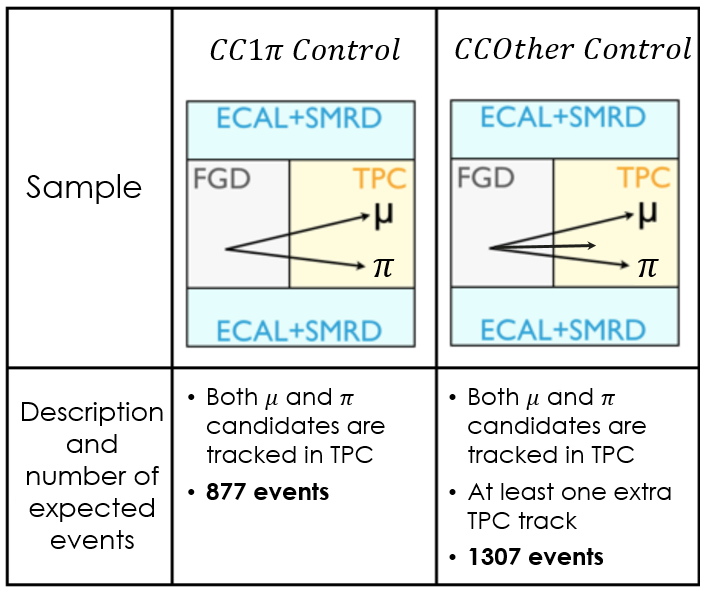}
\end{center}
\caption{A diagram summarising the different control samples used. The number of events selected in data for each sample is indicated.}
\label{fig:sbtopo}
\end{figure}

\begin{figure}[htbp]
\begin{center}
\includegraphics[width=0.49\textwidth]{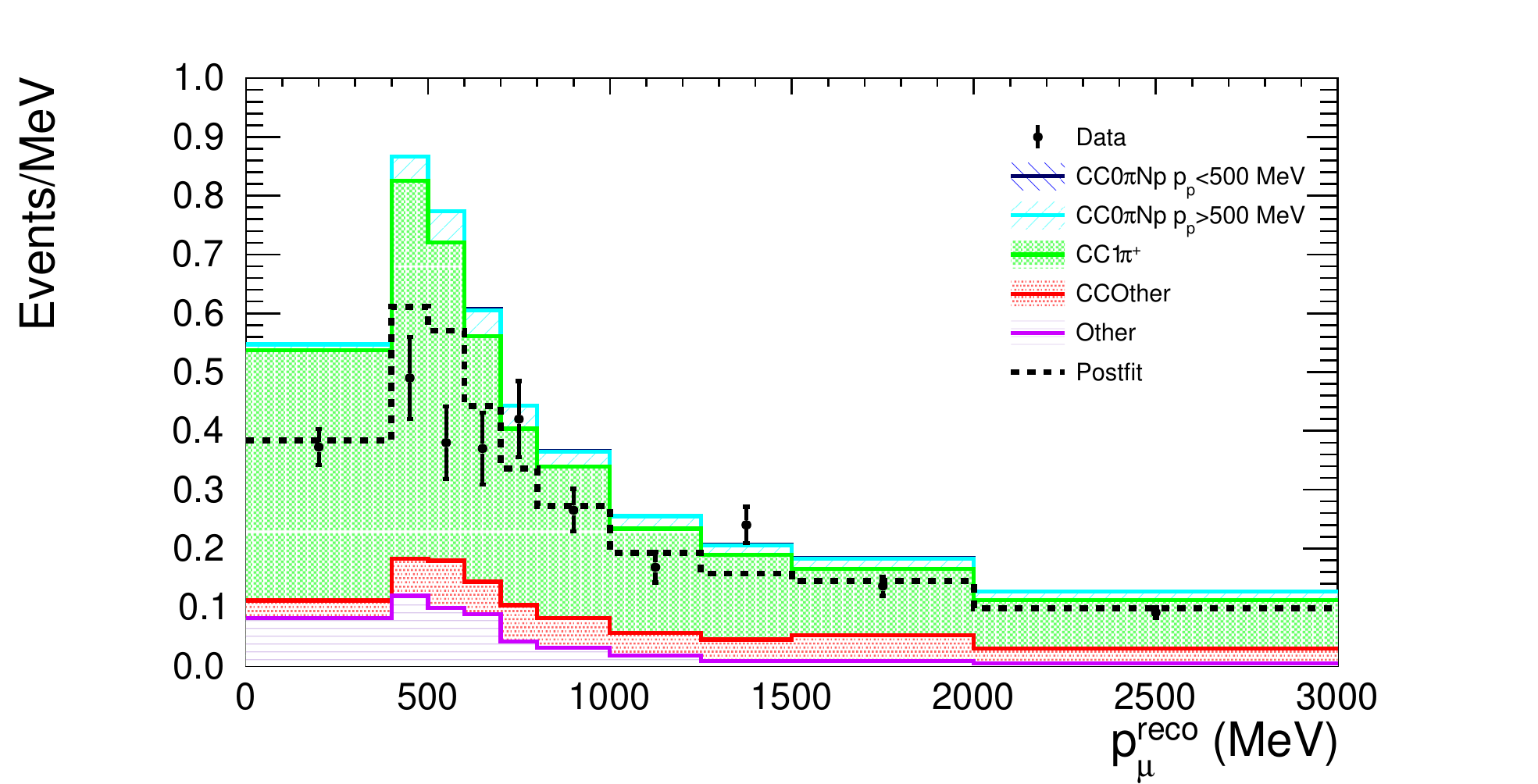}
\includegraphics[width=0.49\textwidth]{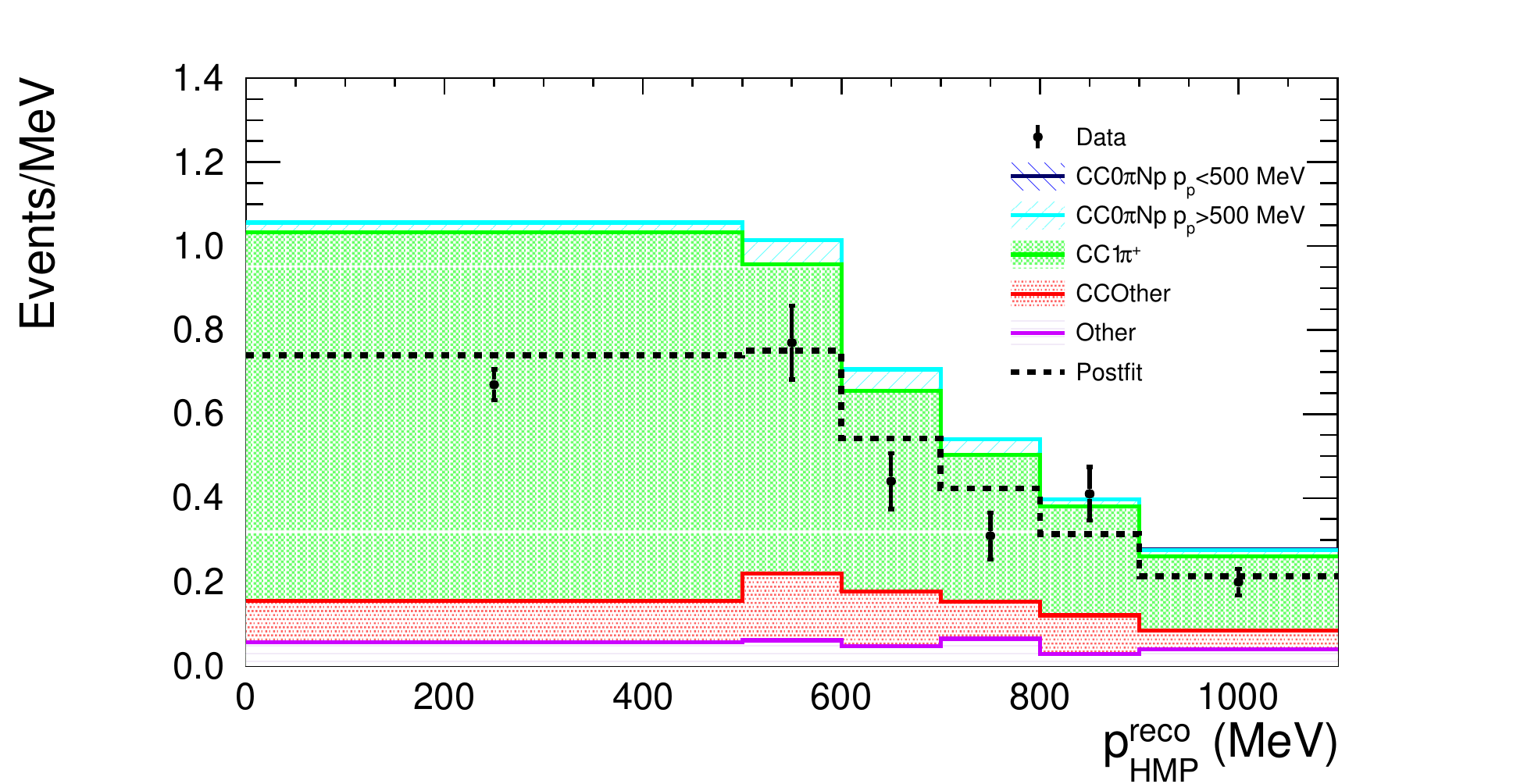}
\includegraphics[width=0.49\textwidth]{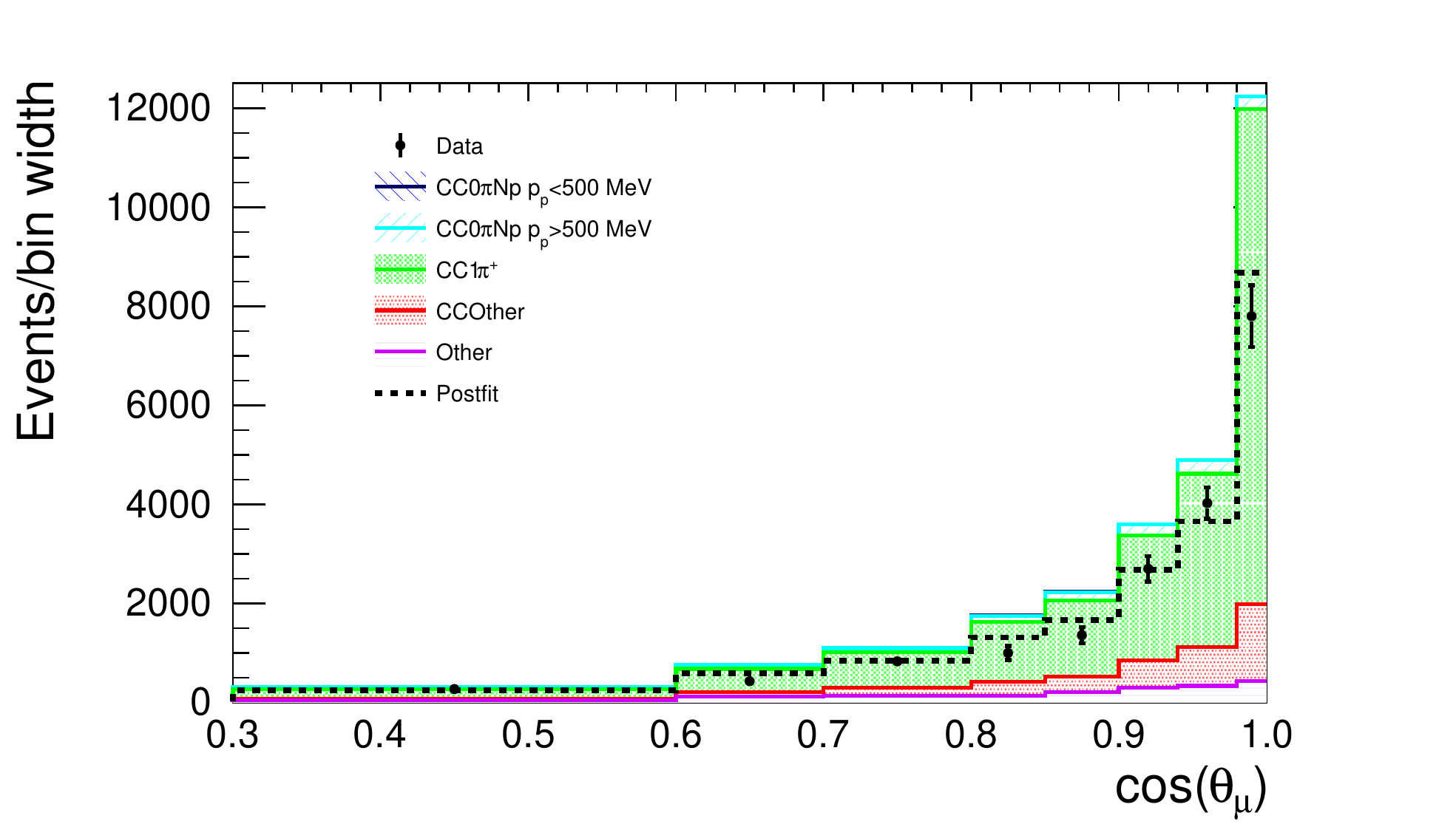}
\includegraphics[width=0.49\textwidth]{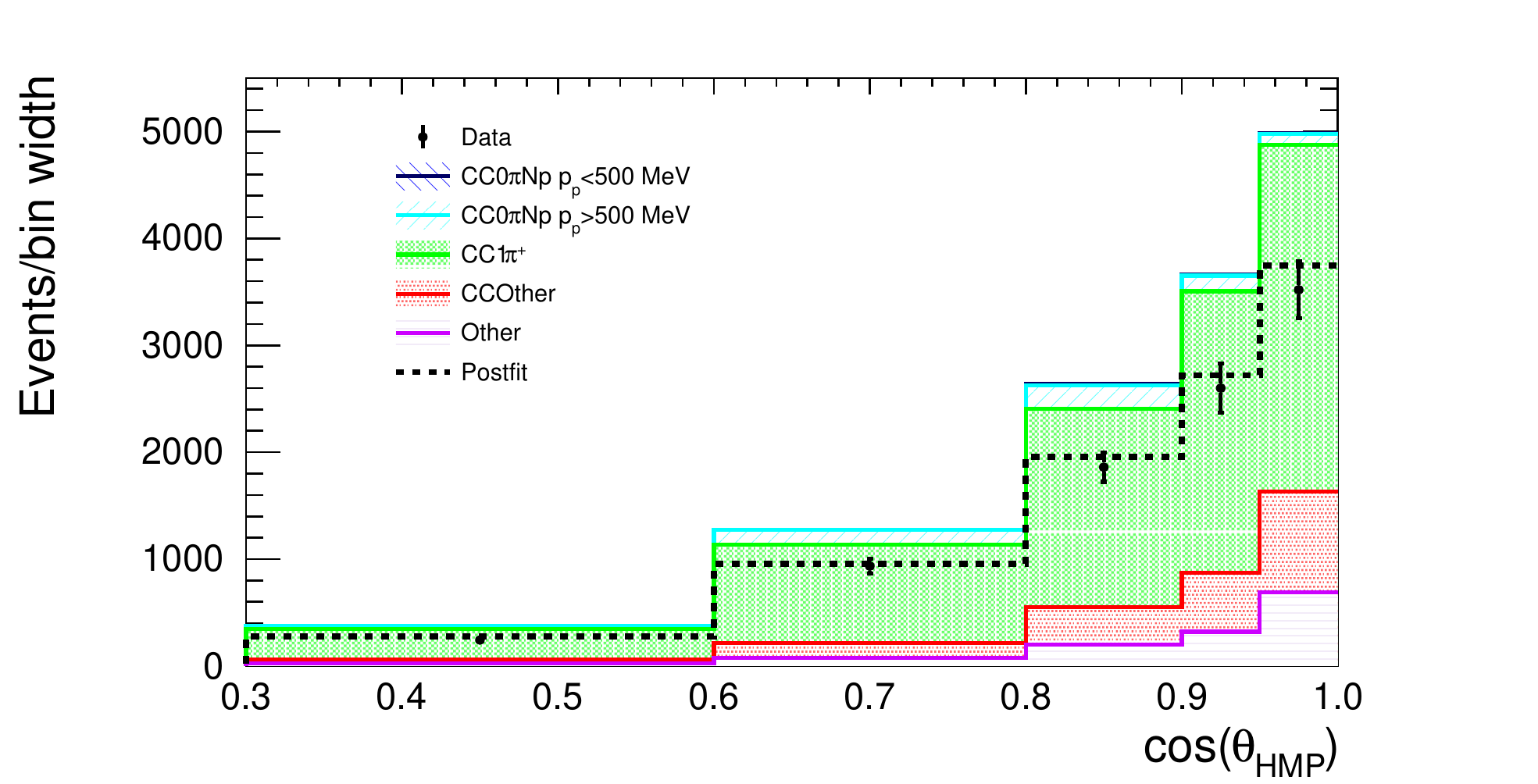}
\end{center}
\caption{The reconstructed kinematics of the muon and of the highest momentum positive (HMP) hadron for events selected within the CC1$\pi^+$ control region from both data and NEUT 5.3.2. The plots are broken down by interaction topology and the CC0$\pi$ contribution is further split depending on whether the interaction falls within the multi-differential analysis phase space constraints from Tab.~\ref{tab:phaseSpace}. The postfit NEUT prediction from the likelihood fit to extract a cross section as a function of $\dpt$ is also shown.}
\label{fig:sbkine}
\end{figure}

\begin{figure*}[!htbp]
\begin{center}
\includegraphics[width=0.49\textwidth]{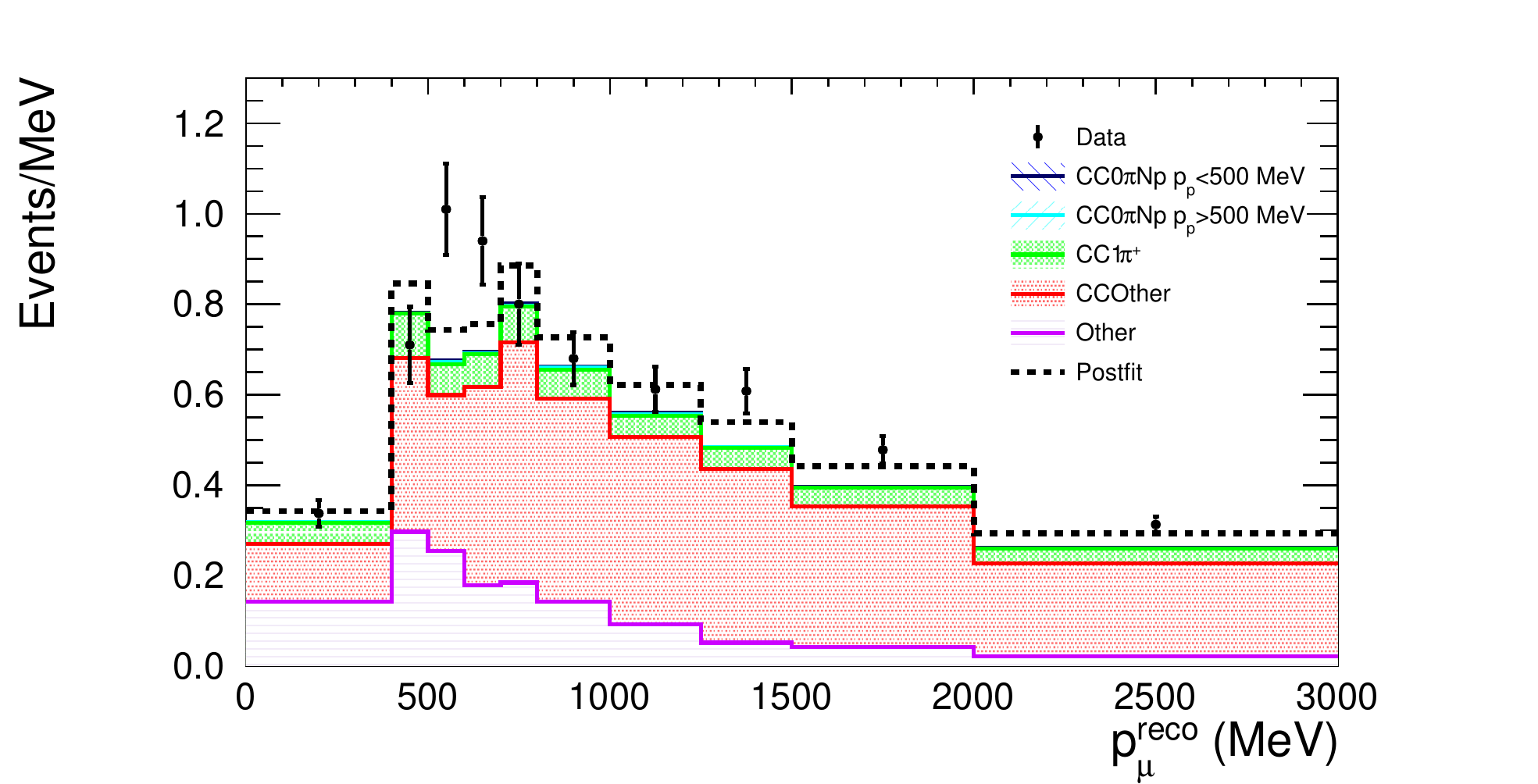}
\includegraphics[width=0.49\textwidth]{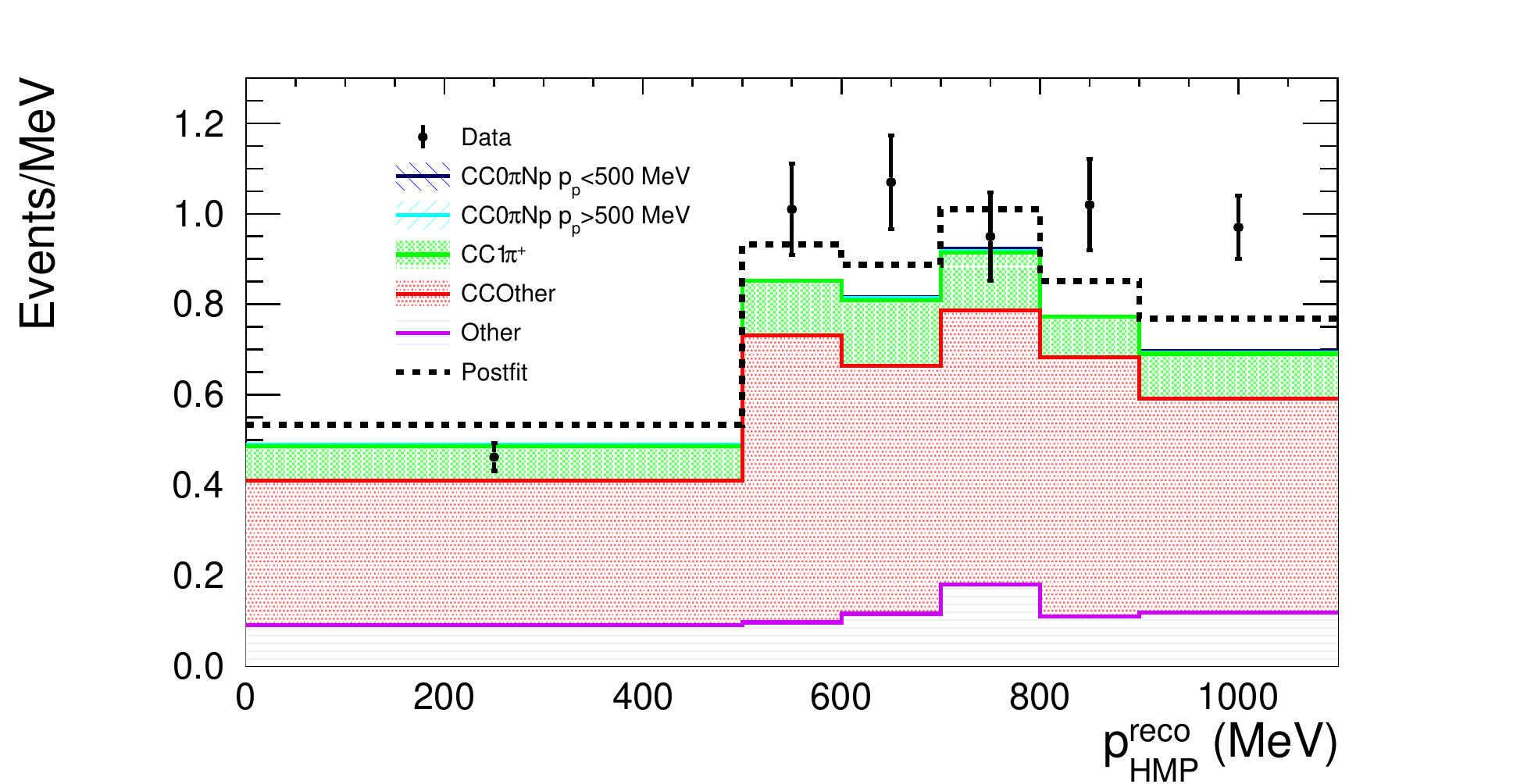}
\includegraphics[width=0.49\textwidth]{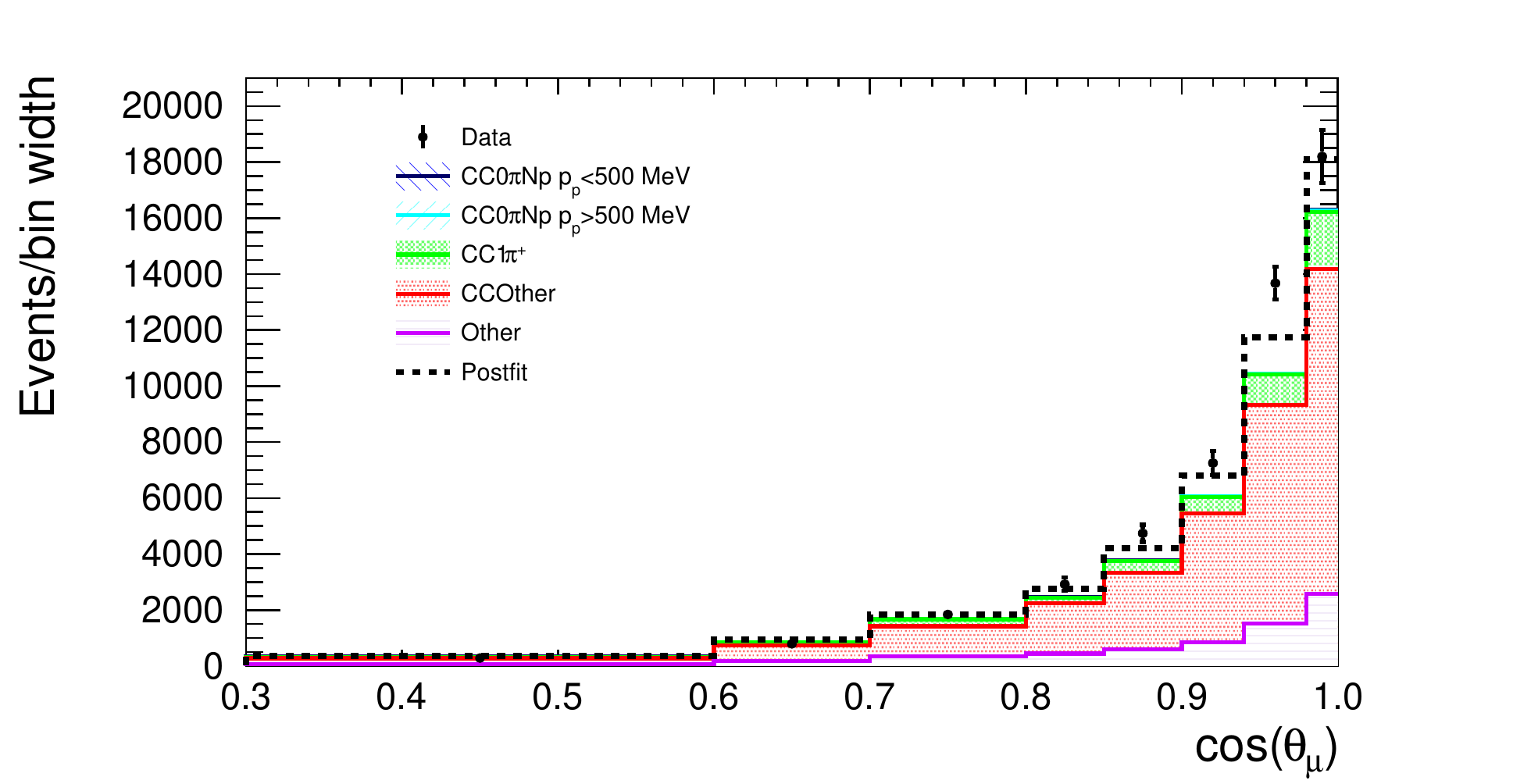}
\includegraphics[width=0.49\textwidth]{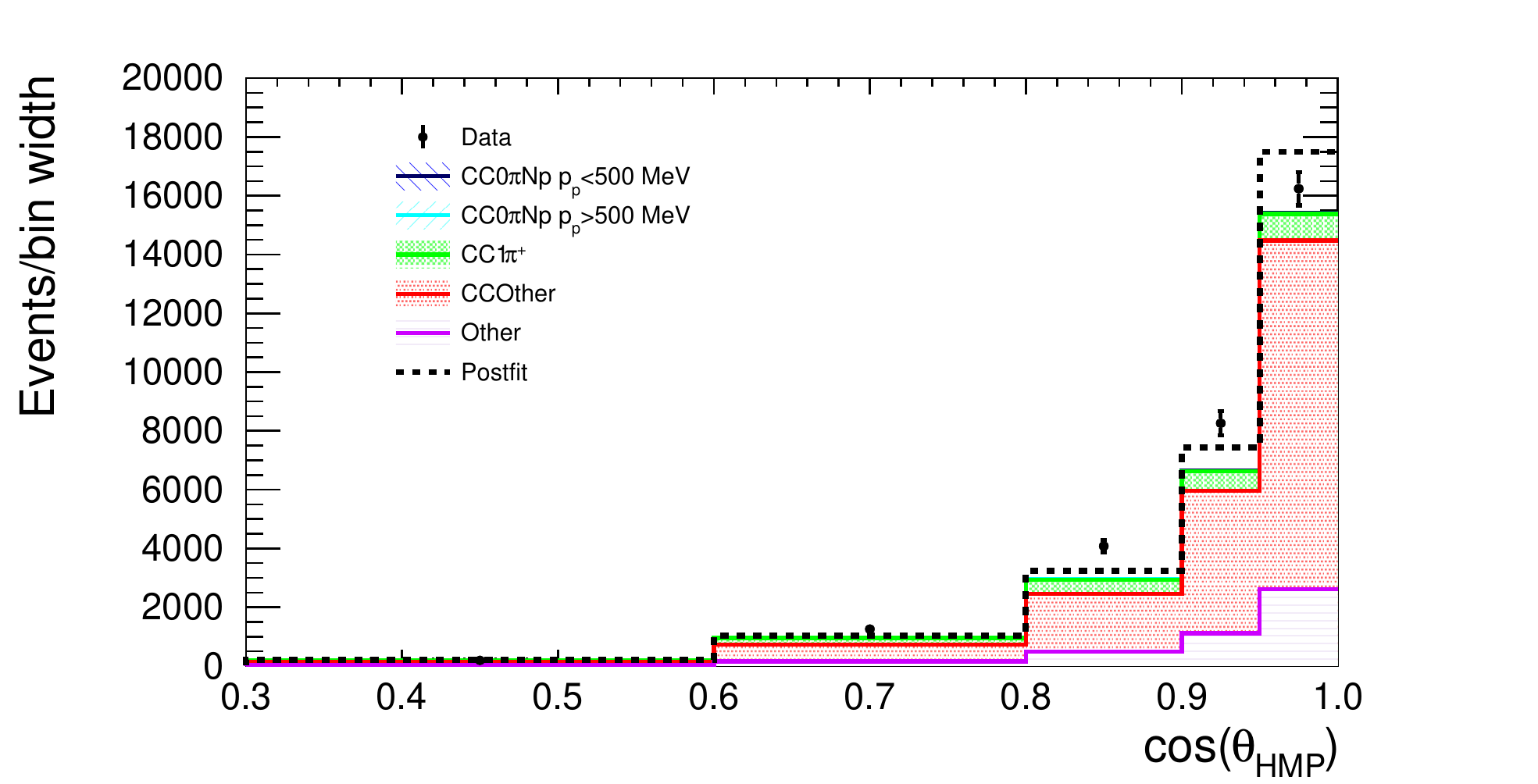}
\end{center}
\caption{The reconstructed kinematics of the muon and of the highest momentum positive (HMP) hadron for events selected within the CCOther control region from both data and NEUT 5.3.2. The plots are broken down by interaction topology and the CC0$\pi$ contribution is further split depending on whether the interaction falls within the multi-differential analysis phase space constraints from Tab.~\ref{tab:phaseSpace}. The postfit NEUT prediction from the likelihood fit to extract a cross section as a function of $\dpt$ is also shown.}
\label{fig:sbkine2}
\end{figure*}

In the multi-differential and STV analyses two control samples are employed for the background constraint. Both require the identification of a negatively charged muon-like track and a positively charged pion-like track in the TPC and are split depending on whether there are any extra tracks sharing a common vertex with the identified muon and pion candidates. The vertex must be contained in the FGD1. These control regions will be referred to as CC1$\pi^+$ and CCOther respectively. An illustration of the topologies these aim to identify is shown in Fig.~\ref{fig:sbtopo} while the distribution of the data and simulated events within each control sample are shown in Figs.~\ref{fig:sbkine} and~\ref{fig:sbkine2}. 

These figures highlight an initial large discrepancy between the NEUT prediction and the data, particularly in the CC1$\pi^+$ sample. This is understood to primarily come from an over-estimation of the contribution from neutrino induced coherent pion production, as demonstrated in Ref.~\cite{Higuera:2014}. However, the likelihood fit used in the multi-differential and the STV analyses allows to adapt the NEUT model to the data within the control regions. The postfit NEUT prediction from the likelihood fit performed in the $\dpt$ measurement is shown in the figures to be in much better agreement with the data (similar results are also obtained in the other STV and multi-differential analyses). 

\begin{figure*}[!htbp]
\begin{center}
\includegraphics[width=0.49\textwidth]{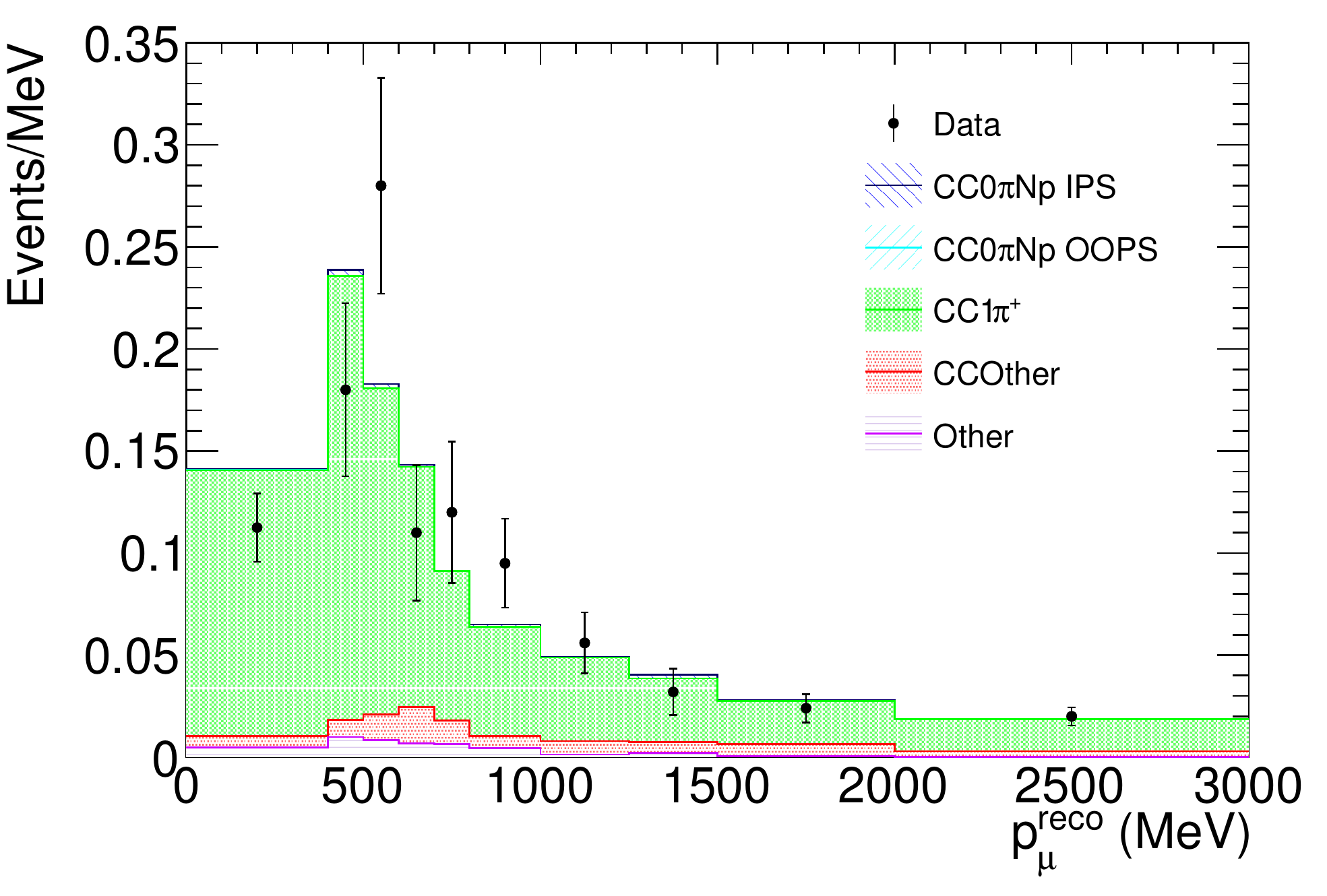}
\includegraphics[width=0.49\textwidth]{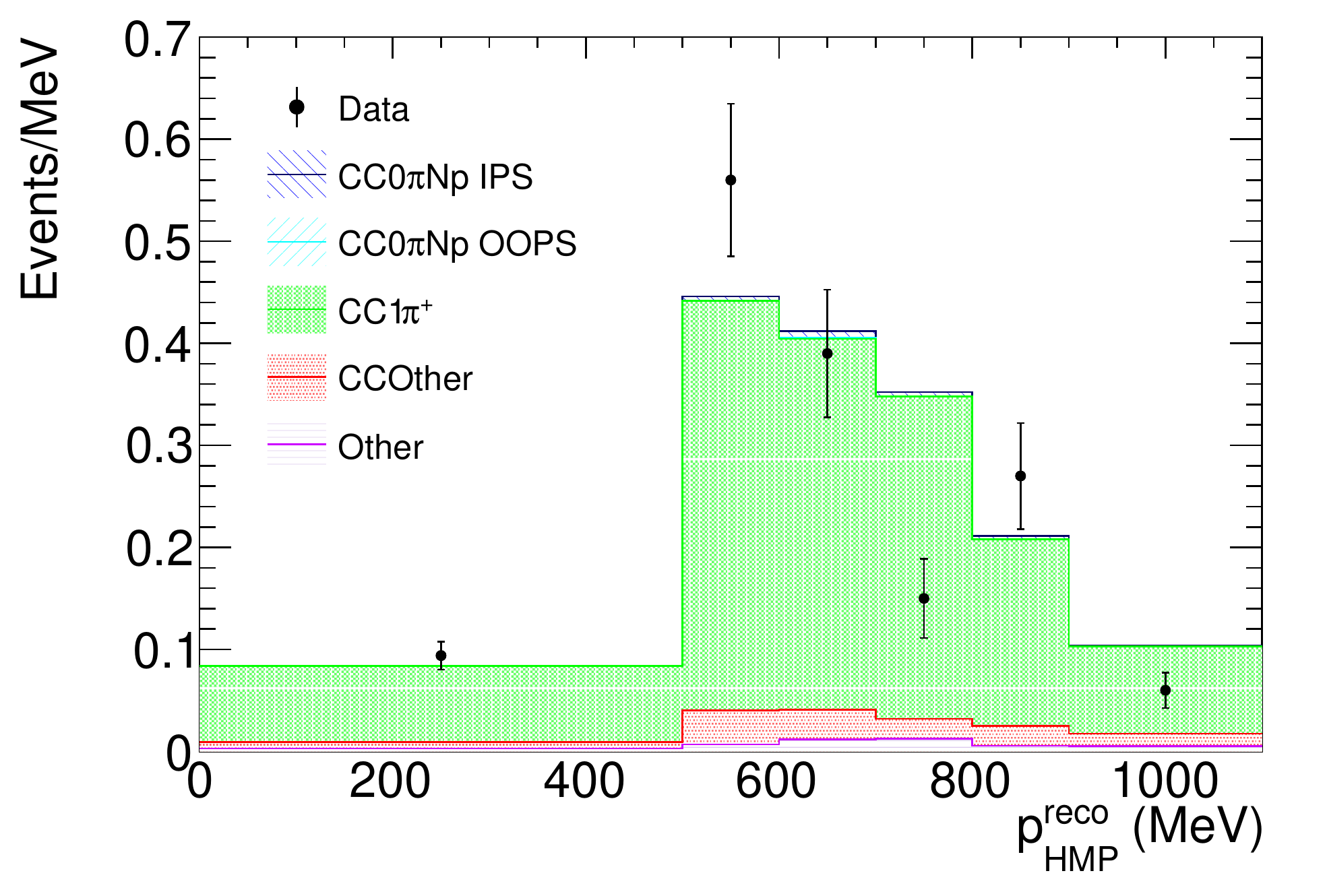}
\includegraphics[width=0.49\textwidth]{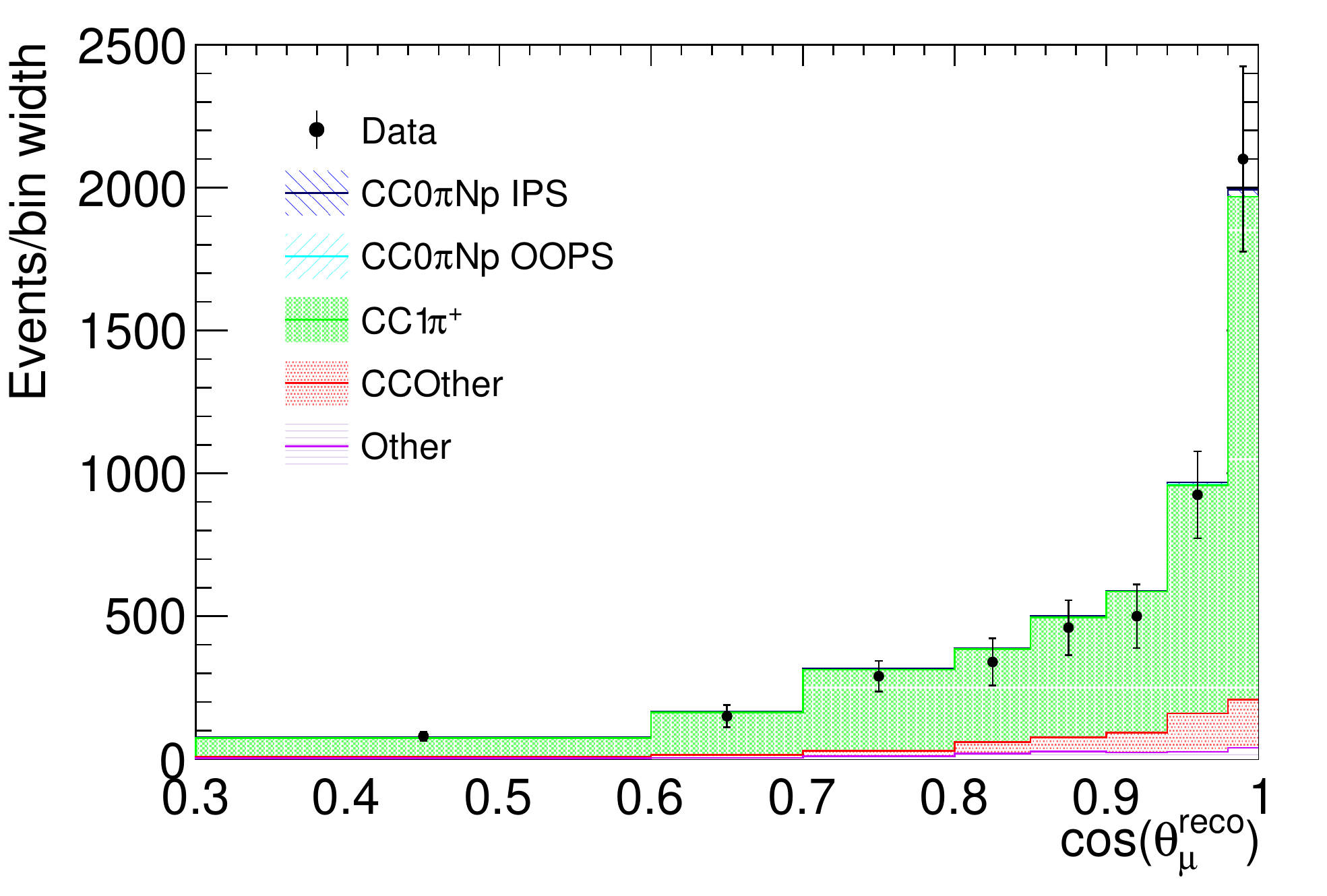}
\includegraphics[width=0.49\textwidth]{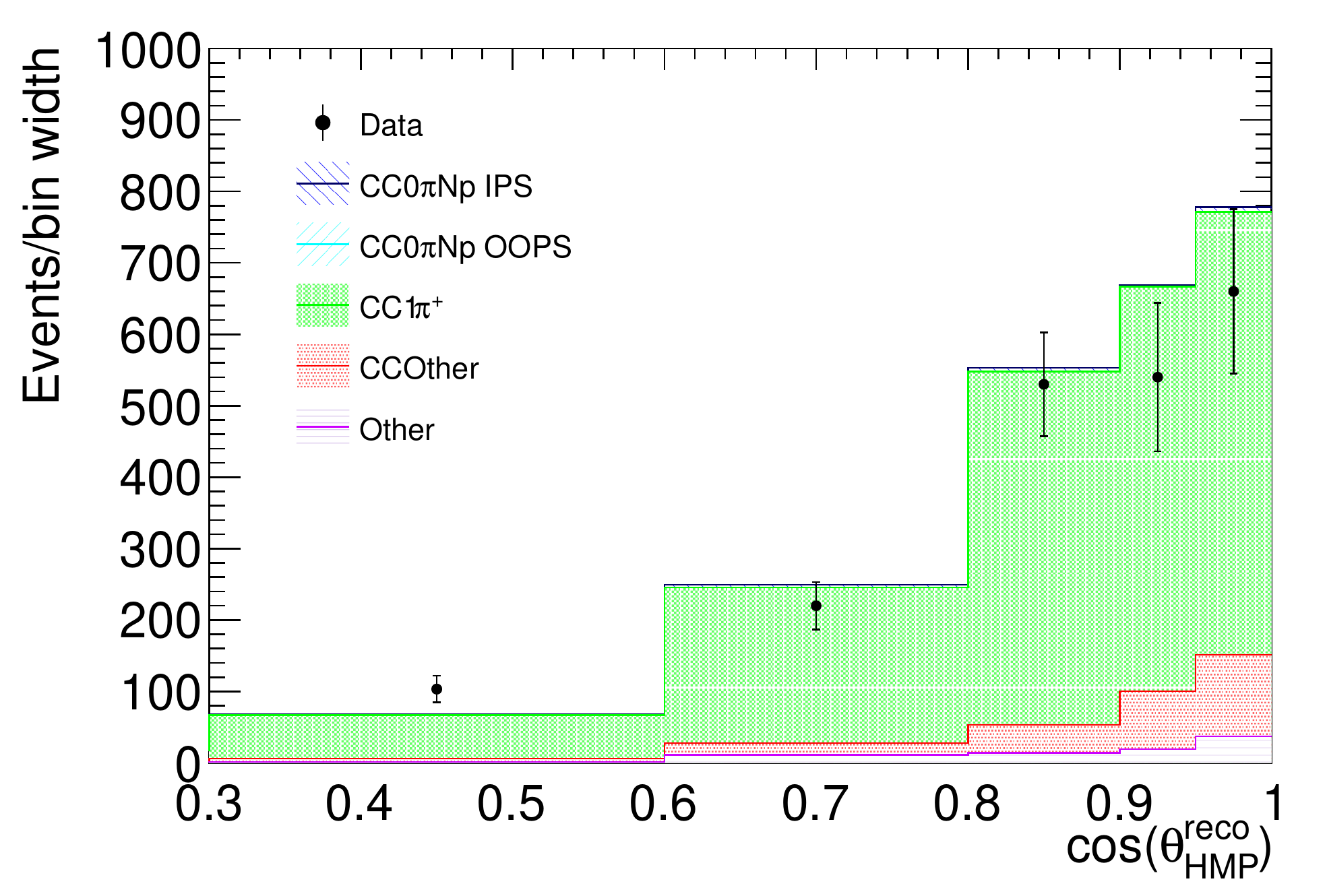}
\end{center}
\caption{The reconstructed kinematics of the muon and of the highest momentum positive (HMP) hadron for events selected within the Michel electron tagged control region from both data and NEUT 5.3.2. The plots are broken down by interaction topology and the CC0$\pi$ contribution is broken down by whether the true kinematics of the events adhere to the phase-space constraints (IPS) for the inferred kinematics analysis or fall outside of them (OOPS).}
\label{fig:sbinf}
\end{figure*}

In the inferred kinematics analysis a control sample is built by inverting the cut on Michel electrons in the signal samples. The resultant kinematic distributions of the selected data and MC events are shown in Fig.~\ref{fig:sbinf}. This control sample is then unfolded simultaneously with the signal regions to constrain the background.

%in the multi-differential and STV analyses while they are corrected on the basis of MC simulation in the measurement of inferred kinematics. The first approach allow to minimize the systematic uncertainties due to background modelling which are instead pretty large in the latter thus covering for possible biases in the background simulation. 

%In the fitting methodology used for analyses X and Y (detailed in sections X and Y), background model parameters are fit in the control regions simultaneously to the parameters describing the signal. Therefore the control regions are chosen to offer a high statistics model constraint of events that are approximately characteristic of the predominant backgrounds.

%\begin{table}[h]
%\begin{center}
%\small
%\centering
%\begin{tabular}{|c|c|c|c|c|c|c|c|}
%\hline
%   & \multicolumn{6}{c|}{Topology}\\
%Sample & CC0$\pi$0p & CC0$\pi$1p & CC0$\pi$Np & CC1$\pi^+$ & CCOther & Other \\
%\hline
%CC1$\pi^+$  &0.00\%&3.28\%&1.53\%&72.5\%&13.7\%&9.10\% \\ %&877\\
%CCOther  &0.00\%&0.31\%&0.37\%&11.3\%&70.9\%&17.2\%  \\ %&1307\\
%\hline
%\end{tabular}
%\caption[The proportion of each true signal topology in each control region.]{\label{tab:sbtopobd}The proportion of each true signal topology in each control region.}
%\end{center}
%\end{table}

%The background subtraction method of the inferred kinematics analysis relies on the kinematic phase space of the events in the control regions to almost exactly match those in the singal regions. For this reason a different sideband sample is used in which the Michel electron tag in the signal selection is reversed. 
 
%\paragraph{Iterative unfolding analysis}

%\clearpage
%\input{Systematics}
\subsection{Sources of systematic uncertainties \label{sec:systematics}}
The measurements presented in this paper account for the following systematic uncertainties:
\begin{itemize}
\item neutrino flux uncertainty. The flux simulation is tuned using external hadron-production measurements and INGRID monitoring, as discussed  in Sec.~\ref{sec:simulation}. The residual flux uncertainties affect the cross section measurements presented in this paper, mainly through an overall normalization uncertainty of approximately 8.5\%.
\item detector effects (efficiency and resolution) which are not perfectly reproduced in the simulation. To evaluate such uncertainties the simulation is compared to the data in dedicated and independent control samples, any observed bias is corrected and the statistical uncertainties in such data and simulated samples are used as residual uncertainties. 
\item modelling of the signal and background interactions, including nuclear effects. As previously discussed in Sec.~\ref{sec:modeldep}, a possible model-dependent bias may be introduced in the multi-differential and STV analyses through efficiency corrections, while the measurement of proton inferred kinematics is also affected through the unfolding procedure and the simulation-based background corrections. Such effects are covered by dedicated systematic uncertainties which are quantified by evaluating the variation of the measured cross section using modified simulation models; the theory parameters describing the signal and the background, including proton and pion FSI, are varied inside their prior uncertainty, based on theory expectations and comparisons to external data.
%Do we need to list the parameters? Maybe just citation to previous OA paper with same parameters? Different proton FSI in the analyses: GENIE vs NEUT  
\end{itemize}
Such uncertainties are implemented in the cross section extraction in different ways in each of the analyses presented in this paper, as will be described in the following sections. In general the systematic parameters considered and their variation is similar to that used for the near detector fit of T2K oscillation analyses described in Ref.~\cite{Abe:2017vif}: the most notable differences being the inclusion of proton FSI uncertainties and the usage of Gaussian priors for the parameters describing CCQE uncertainties.

\clearpage

\subsection{Method of cross section evaluation \label{sec:extraction}}

%Describe the likelihood fit (and highlight the differences between the multidifferential and the STV,
%eg: flux fitted or not, treatment of out of phase space background, other?)

%Describe the unfolding procedure

Each of the analyses take different approaches when extracting a cross section from the selected events detailed in Sec.~\ref{sec:selection}. All of these methods involve an effective background subtraction; an efficiency correction; and the deconvolution of detector effects either by a binned-likelihood fit for the multi-differential and STV analyses, or an iterative unfolding procedure for the analysis of the inferred kinematics.

\subsubsection{Binned likelihood fitting}
\label{subsec:mlfitting}
In order to produce a data spectrum that is de-convoluted from detector smearing, the input simulation is varied via a set of parameters, such that a best fit set can be extracted once the simulation best describes the observed data.
%These parameters should describe the signal to extract and any background processes which may affect either the best set of fit parameters or their uncertainty. 
%The signal is parametrised by weights on the nominal number of signal events in each \textit{true} analysis bin (with no prior constraint).  
The signal is parametrised using ``template signal weights'' ($c_i$) which alter the number of selected signal events in bins ($i$) of some truth-level observable(s) with no prior constraint. Parameters describing plausible systematic variations of the flux, detector response and background processes can also be fit simultaneously to the signal parameters. The effect of these parameter variations is then propagated through to the number of selected events in reconstructed bins of the same observable (using the expected smearing due to detector resolution and efficiency), such that the updated simulation prediction can be compared to the data. The best fit set of parameters are chosen by minimising the following negative log-likelihood:

\begin{equation}
\label{eqn:chi2tot}
-2\log(L)=-2\log(L_{stat})-2\log(L_{syst}).
\end{equation}
Where:
\begin{equation}
\label{eqn:chi2stat}
-2\log(L)_{stat}=\sum_{j}^{reco\:bins} 2(N_{j}^{sim} - N_{j}^{obs} + N_{j}^{obs} \log\frac{N_{j}^{obs}}{N_{j}^{sim}}),
\end{equation}
and
\begin{equation}
\label{eqn:chi2syst}
-2\log(L)_{syst} = (\vec{a}^{syst} - \vec{a}_{prior}^{syst})(V_{prior}^{syst})^{-1}(\vec{a}^{syst} - \vec{a}_{prior}^{syst}).
\end{equation}
The term in Eq.~\ref{eqn:chi2stat} is the Poisson likelihood, where $N_j^{sim}$ and $N_j^{obs}$ are the number of simulated and observed events in each reconstructed bin, $j$. The term in Eq.~\ref{eqn:chi2syst} characterises the prior-knowledge of the values of the systematic parameters ($\vec{a}^{syst}$) and their correlations, as a multi-variate Gaussian likelihood where $\vec{a}_{prior}^{syst}$ are the prior values of these parameters and $V_{prior}^{syst}$ is a covariance matrix describing the correlations between them.

As described above, $N_j^{sim}$ is described by alterations to the nominal input simulation based on the template signal weights and the systematic fit parameters:

\begin{equation}
\label{eqn:fit}
N_j^{sim}=\sum_{i}^{true\:bins} (c_{i} w_{i}^{sig} N_{i}^{sim\:sig} + w_{i}^{bkg} N_{i}^{sim\:bkg} ) U_{ij},
\end{equation}
where  $N_{i}^{sim\:sig}$ and $N_{i}^{sim\:bkg}$ are the number of signal and background events in true bin $i$ of the input simulation; $c_i$ are the signal template weights; $w_{i}^{signal}$ and $w_{i}^{bkg}$ describe the alterations to the input simulation from the aforementioned systematic parameters; and $U_{ij}$ is the smearing matrix describing the probability of finding an event in true bin $i$ in reconstructed bin $j$. This smearing matrix is also subject to change with the alteration of systematic parameters.

%Equation~\ref{eqn:chi2syst} thereby facilitates a constraint of the systematic parameters in the fit, which picks up a penalty for moving these parameters far from their prior values. 

The result of the fit is the $N^{CC0\pi}_{i}$ term from Eq.~\ref{eqn:xsec}: the number of selected signal events deconvoluted from detector smearing in each analysis bin. As shown in Eq.~\ref{eqn:xsec}, this must then account for the integrated T2K flux, the number of target nucleons and the bin width before being efficiency corrected to produce a differential cross section.

Such a method of deconvolution is entirely unregularised and is therefore equivalent to using D'Agostini iterative unfolding~\cite{dagostini1} with an infinite number of iterations or to simply inverting the detector response matrix providing this gives an entirely positive unsmeared spectrum. Provided that the analysis bins do not integrate over regions of phase space of rapidly changing efficiency, this method of unsmearing is completely unbiased but is susceptible to the so-called ``ill-posed problem'' of deconvolution - where relatively small statistical fluctuations in the reconstructed bins can cause large variations in the fitted contents of true kinematic bins~\cite{Kuusela:2015mzh}. %~\cite{kussela}
%As such, this method can produce results which, whilst potentially unbiased, oscillate wildly with corresponding exteamely large anticorrelations. 
These results are fully correct and perfectly suitable for further use, for example in fits to constrain parameters in model predictions or to compare the suitability of different models, but they cannot easily be interpreted ``by-eye'', since they often contain large anticorrelation between adjacent bins which causes the result to strongly 'oscillate' between such bins.  Moreover, within the pertinent observables in these analyses, neutrino-interaction cross sections are not expected to follow such an 'oscillating' behaviour. 
These large variations between neighbouring bins can be suppressed by regularising the results, i.e. imposing smoothness of the fitted parameters $c_i$, thus inducing a small overall reduction of the uncertainties and some dependence of the results on the input signal simulation model. As such, the STV analysis provides both regularised and unregularised results.
To achieve this, a regularisation term is optionally added to the likelihood in Eq.~\ref{eqn:chi2tot}: 
\begin{equation}
\label{eq:regpen}
-2\log(L)_{reg} = p_{reg} \sum_{i}^{true\:bins-1} (c_{i}-c_{i+1})^2.
\end{equation}
Here $c_i$ is the signal weight for the $i^{th}$ true bin and $p_{reg}$ controls the regularisation strength. It is clear that this implementation of regularisation adds a constraint which can bias the fit toward the shape of the signal model in the input simulation. 
%One may think that a way to avoid this model bias would be to regularise the difference between the absolute number of events in adjacent bins, rather than the relative change from nominal (equivalent to using a flat input simulation) but this would bias the result toward a flat result, which is less realistic than any simulation. It is also potentially possible to reduce bias by using a different term within summation, for example using the numerical approximation of the second derivative rather than the first may lead to less impactful input model. However, in this case the first derivative is used since there are not many bins in the STV analysis and the second derivative (which requires each bins next to nearest neighbour to calculate) would have too many edge effects. 
%The difficulty with implementing regularisation often comes with the selection of an appropriate regularisation strength. 
%If the strength is too low then the regularisation will not smooth the unstable fit result but if the strength is too high then the extracted unsmeared distribution will be biased toward the shape of the input simulation. 
However, the impact of the bias can be mitigated by the careful selection of an appropriate regularisation strength. A simple method of choosing $p_{reg}$ in such a regularisation scheme is the `L-curve' technique presented in Ref.~\cite{lcurve}. In this approach a compromise is found between the impact of the regularisation (defined by the normalised regularisation penalty: $-2\log(L)_{reg}/p_{reg}$) and the goodness of fit (decreased $\log(L)_{reg}$). One of the significant advantages of this method, over those typically used to choose the regularisation strength (like tuning the number of iterations) in iterative unfolding methods, is that it is ``data-driven'': the regularisation strength is determined from assessing the properties of real data and is not solely reliant on simulation studies. 

It is important to emphasise that the application of regularisation produces a result that is easier to interpret without statistical methods but is at least slightly biased. A regularised result is therefore particularly well suited for result-theory comparison plots but the unregularised result is likely more suitable for forming quantitative conclusions. For this reason unregularised results will be provided in both the multi-differential and STV analyses.

%Overall, the fit to the data gives the number of selected signal events in true bins. To get to a flux-integrated differential cross section result, it is still necessary to correct for the efficiency, the integrated flux and the number of targets in the fiducial volume of the target detector as discussed in section~\ref{sec:anaStrategy} and quantified in equation~\ref{eqn:xsec}.

\subsubsection{Iterative D'Agostini unfolding}
\label{subsec:unfolding}
Unfolding accounts for smearing between the true spectrum and reconstructed spectrum due to the detector efficiency and resolution. 
The relation between true and measured spectrum can be written as :
\begin{equation}
E_j = \sum_{i=1}^{N_t} S_{ji} C_i
\end{equation}
where $C_i$ is a number of events in true bin $i$, $E_j$ is a number of events in measured bin $j$, $S_{ji}$ is a smearing matrix, and 
$N_t$ is the number of true bins.

The smearing matrix is constructed from MC predictions which gives the information of event migrations. The Iterative unfolding, proposed by D'Agostini~\cite{dagostini1, dagostini2}, uses Bayes' theorem to obtain an unsmearing matrix from the smearing matrix as :

\begin{equation}
U_{ij} = \frac{\mbox{P}_{eff}(E_j|C_i)\mbox{P}_0(C_i)}{\sum_{i=1}^{N_t}\mbox{P}(E_j|C_i)\mbox{P}_0(C_i)},
\end{equation}

where $\mbox{P}(E_j|C_i)$ is a probability of the true events in bin $i$ measured in bin $j$ written as :

\begin{equation}
\mbox{P}(E_j|C_i) = \frac{N_{ji}}{C_i},
\end{equation}
where $N_{ji}$ is the number of true events in bin $i$ measured in bin $j$. $\mbox{P}_{eff}(E_j|C_i)$ is defined as :

\begin{equation}
\mbox{P}_{eff}(E_j|C_i) = \frac{\frac{N_{ji}}{C_i}}{\sum_{j=1}^{N_m}\frac{N_{ji}}{C_i}},
\end{equation}
where $N_m$ is number of measured bins.

$\mbox{P}_0(C_i)$ is a prior probability representing the predicted number of events in bin $i$, written as :
\begin{equation}
\mbox{P}_0(C_i) = \frac{C_i}{\sum_{i=1}^{N_t}C_i}.
\end{equation}

Therefore, the unfolded spectrum is :

\begin{equation}
C^{'}_i = \sum_{j=1}^{N_m} U_{ij} E^{data}_j,
\end{equation}
where $N_m$ is the number of bins of measured spectrum. After each iteration, $\mbox{P}_0(C_i)$ is updated with the posterior of the previous iteration. 

This method is regularised by choosing the number of iterations, inducing a bias toward the input simulation used. Such bias is tested through multiple mock datasets with alternative simulation models. The number of iterations was chosen by requiring the $\chi^2$ values obtained between the unfolded result and the truth of these mock datasets to reach a stable value: 2-iterations for $\Delta p_{p}$, 6-iterations for $\Delta \theta_{p}$ and 4-iterations for $|\Delta \overrightarrow{p}_{p}|$.  The bias in the results was shown to always be well within the uncertainties.

Overall this produces an efficiency corrected and unfolded distribution of signal events which must then account for the flux normalisation, the number of target nucleons and the bin width to form a differential cross section, as described by Eq.~\ref{eqn:xsec}.

\subsection{Multi-differential muon and proton kinematics \label{sec:multidiff}}

This analysis measures the multi-differential cross section of CC0$\pi$ events as a function of the muon and proton kinematics and the proton multiplicity. As previously described, a multi-dimensional efficiency correction is applied, the cross section is evaluated with a binned likelihood fit and the background is constrained by using dedicated control regions.
The binning, reported in Tab.~\ref{tab:multiDiffBinning}, is chosen to keep the systematic uncertainty smaller than the statistical uncertainty and to cope with the track reconstruction capabilities of the detector.  Due to the small available statistics, the events with two or more protons are all collected in a single bin.

\begin{figure*}[hp!]
\begin{center}
 \includegraphics[width=\textwidth]{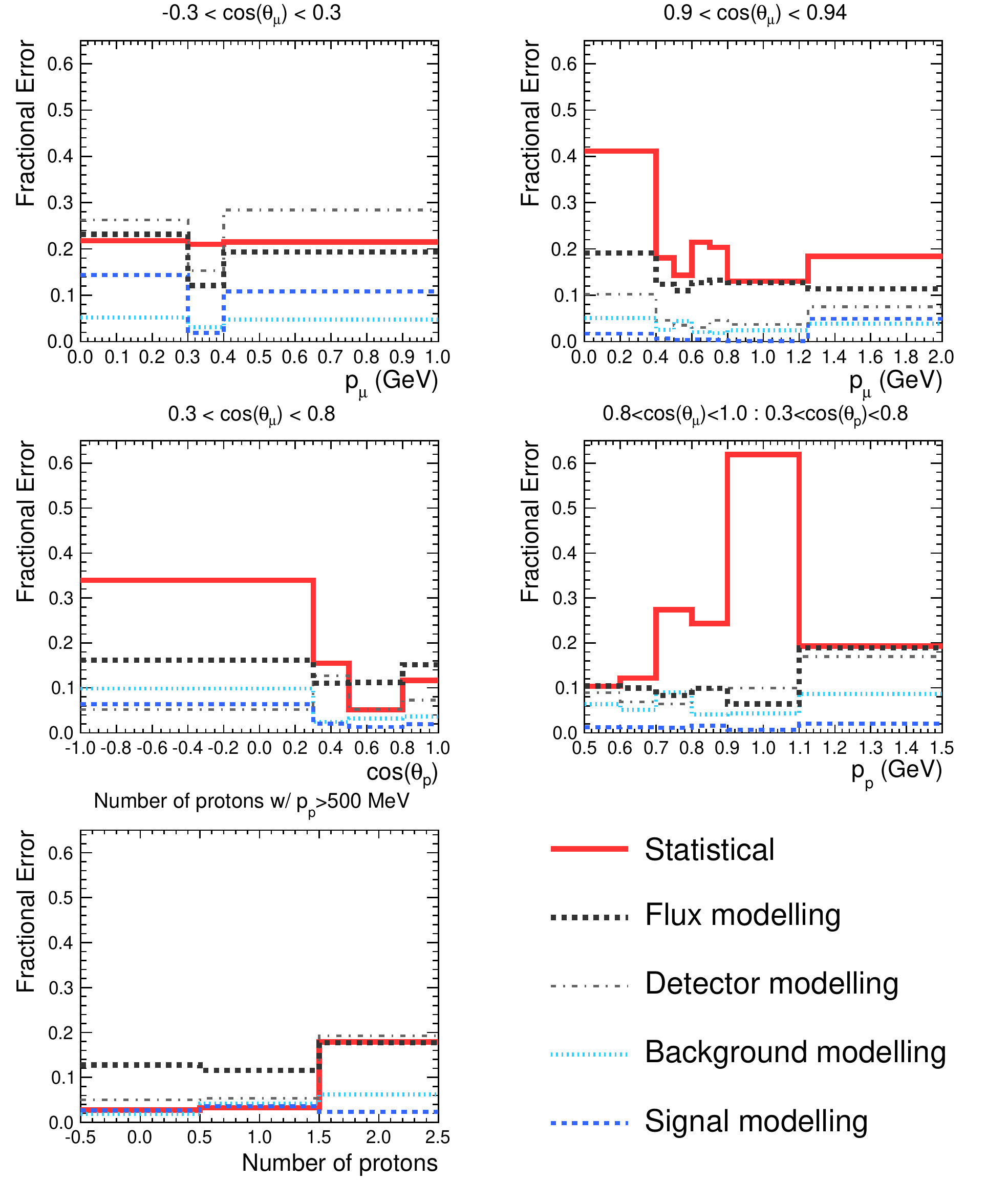}
  \end{center}
\caption{Statistical and systematic uncertainty (separated by source) within some representative bins of the multi-differential analysis.}
\label{fig:systStatMultiDiff}
\end{figure*}

The statistical uncertainties are evaluated by fluctuating the total number of observed event in each bin with a Poisson probability and running the fit multiple times. The systematic uncertainties are evaluated by running the analysis on many toy datasets produced by varying the parameters describing the systematics effects detailed in Sec.~\ref{sec:systematics}. The uncertainties are then found by computing the covariance of the resultant cross sections between every pair of analysis bins.
% The uncertainties on the measurement are evaluated using the RMS of the distribution of the extracted cross section over all the toys sampling the statistical and systematic uncertainties.
The fractional uncertainties are shown in Fig.~\ref{fig:systStatMultiDiff} for some representative bins. The different sources of systematic uncertainties are shown separately and the total systematic uncertainty is evaluated by simultaneously varying all the nuisance parameters corresponding to the different source of uncertainties.
The flux uncertainty is the largest, followed by detector effects. The background modelling uncertainty is sizeable, i.e. of the same order of detector effects, only in the regions with high momentum forward going muons where the background is larger. Finally the signal modelling uncertainty is only non-negligible in the region of backward, low momentum muons where the detector reconstruction capabilities are limited and, due to low available statistics, the angular bin is large, averaging over
angles with different reconstruction efficiencies. This is also the region where the backgrounds coming from outside the FGD1 fiducial volume are larger. All these effects tend to increase the dependence of the results on the signal modelling in this particular region of phase space.
The statistical uncertainty dominates in most of the bins, except in the regions where the width of the bins is driven by the detector performances. For instance the bin of low proton momentum cannot be further subdivided due to the limited resolution for short tracks. Analogously, in the regions where the muon or the proton have an angle almost perpendicular to the neutrino direction, in order to match the reconstruction capabilities of the detector in absence of a TPC track, the measurement is reported in large bins, and thus systematic uncertainties are larger than statistical uncertainty.
Fig.~\ref{fig:systStatMultiDiff} shows the uncertainties on the measurement of proton multiplicity.  In this case, integrating over all the bins of muon and proton kinematics, the statistical uncertainty is always smaller than the systematic ones. The dominant uncertainty is still due to the flux, followed by the detector effects, which become very important in the events with two or more protons where the outgoing nucleons have very low momentum and are thus difficult to reconstruct. The uncertainty due to background is completely negligible, while the effect of nucleon FSI becomes important at increasing proton multiplicity due to migration effects between different multiplicity bins. The uncertainty due to signal modelling is very small and more or less constant across all the multiplicities.

\begin{table}[h]
\footnotesize
\begin{subtable}{}
\begin{tabular}{|l|l|}
\hline
cos$\theta_\mu$ & $p_\mu$ (GeV) \\
\hline
-1.0, -0.3 & \\
-0.3, 0.3 & 0.0, 0.3, 0.4, 30 \\
0.3, 0.6 & 0.0, 0.3, 0.4, 0.5, 0.6, 30 \\
0.6, 0.7 & 0.0, 0.3, 0.4, 0.5, 0.6, 30 \\
0.7, 0.8 & 0.0, 0.3, 0.4, 0.5, 0.6, 0.7, 0.8, 30 \\
0.8, 0.85 & 0.0, 0.4, 0.5, 0.6, 0.7, 0.8, 30 \\
0.85, 0.9 & 0.0, 0.3, 0.4, 0.5, 0.6, 0.7, 0.8, 1.0, 30 \\
0.9, 0.94 & 0.0, 0.4, 0.5, 0.6, 0.7, 0.8, 1.25, 30 \\
0.94, 0.98 & 0.0, 0.4, 0.5, 0.6, 0.7, 0.8, 1.0, 1.25, 1.5, 2.0, 30 \\
0.98, 1.0 & 0.0, 0.5, 0.65, 0.8, 1.25, 2.0, 3.0, 5.0, 30 \\
\hline
\end{tabular}
\end{subtable}
\vspace{10mm}
\begin{subtable}{}
\begin{tabular}{|l|l|l|}
\hline
cos$\theta_\mu$ & cos$\theta_p$ & $p_p$ (GeV) \\
\hline
-1.0, -0.3 & -1.0, 0.87, 0.94, 0.97, 1.0 & \\
-0.3, 0.3 & -1.0, 0.75, 0.85 & \\
          & 0.85, 0.94 & 0.5, 0.68, 0.78, 0.9, 30\\
          & 0.94, 1.0 & \\
0.3, 0.8 & -1.0, 0.3, 0.5 & \\
          & 0.5, 0.8 & 0.5, 0.6, 0.7, 0.8, 0.9, 30\\
          & 0.8, 1.0 & 0.5, 0.6, 0.7, 0.8, 1.0, 30\\
0.8, 1.0 & -1.0, 0.0, 0.3 & \\
         & 0.3, 0.8 & 0.5, 0.6, 0.7, 0.8, 0.9, 1.1, 30\\
         & 0.8, 1.0 & \\
\hline
\end{tabular}
\end{subtable}
\caption{\label{tab:multiDiffBinning}Bins in muon and proton kinematics in which the multi-differential cross section is measured for the 0 proton sample (top) and the 1 proton sample (bottom).}
\end{table}

\begin{figure*}[hp!]
\begin{center}
\includegraphics[width=1.0\textwidth]{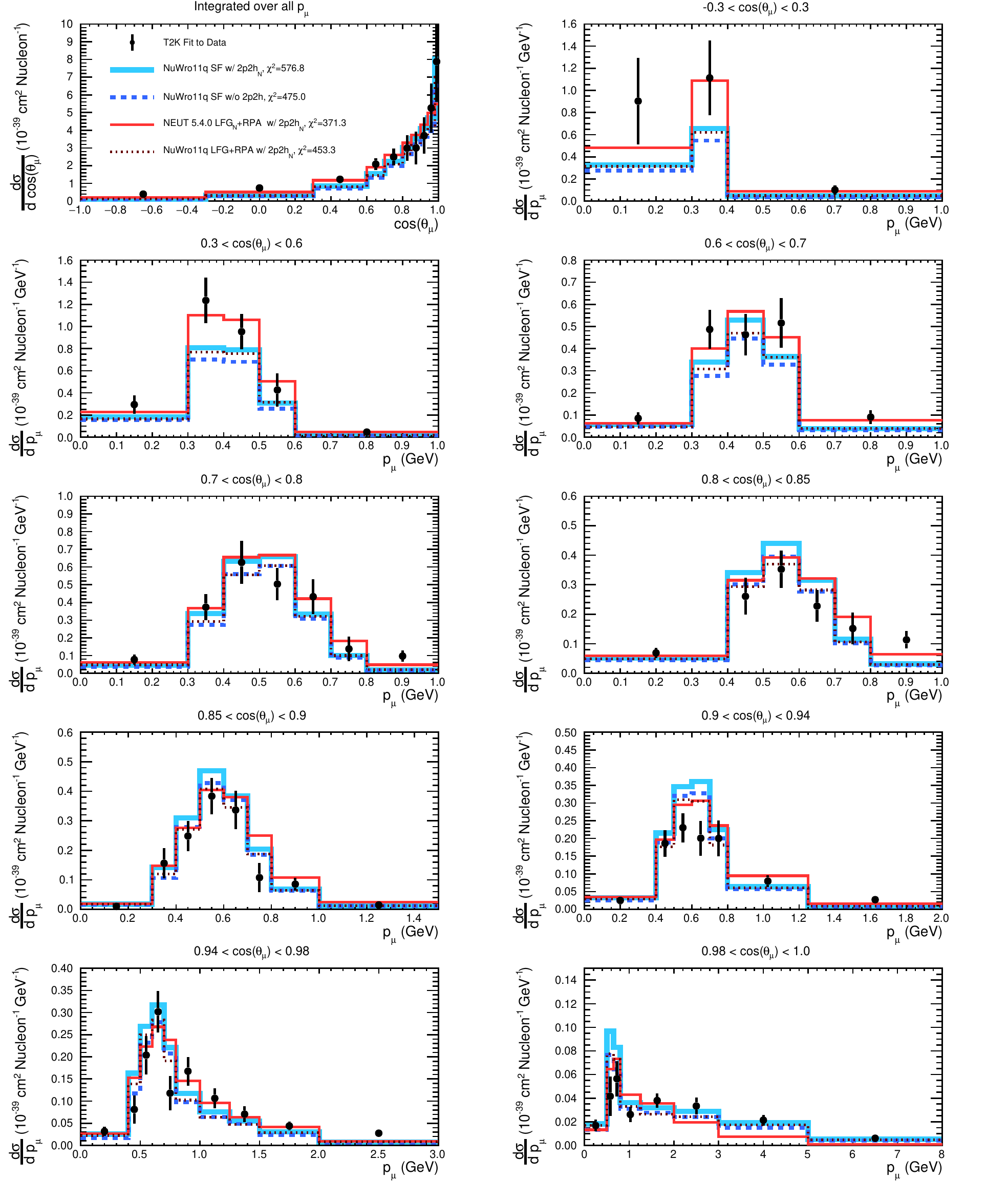} 
\end{center}
\caption{Measurement of the cross section as a function of the muon kinematics when there are no protons (with momenta above 500 MeV). The data are compared to: NuWro 11q with the SF nuclear model, both with and without additional 2p2h contribution; NEUT 5.4.0, which uses an LFG+RPA model that includes 2p2h predictions; and NuWro 11q with an LFG+RPA nuclear model and a separate 2p2h prediction. 2p2h$_N$ indicates the 2p2h model is an implementation of the Nieves et. al. model of Ref.~\cite{Nieves:2012}. The `N' subscript after LFG indicates that the model is using both a 1p1h and 2p2h prediction from the aforementioned model of Nieves et. al. More details of these models can be found in Sec.~\ref{sec:simulation}. Note that the last momentum bin in each plot is shortened for readability.}
\label{fig:MultiDiff_0p}
\end{figure*}

\begin{figure*}[hp!]
\begin{center}
 \includegraphics[width=\textwidth]{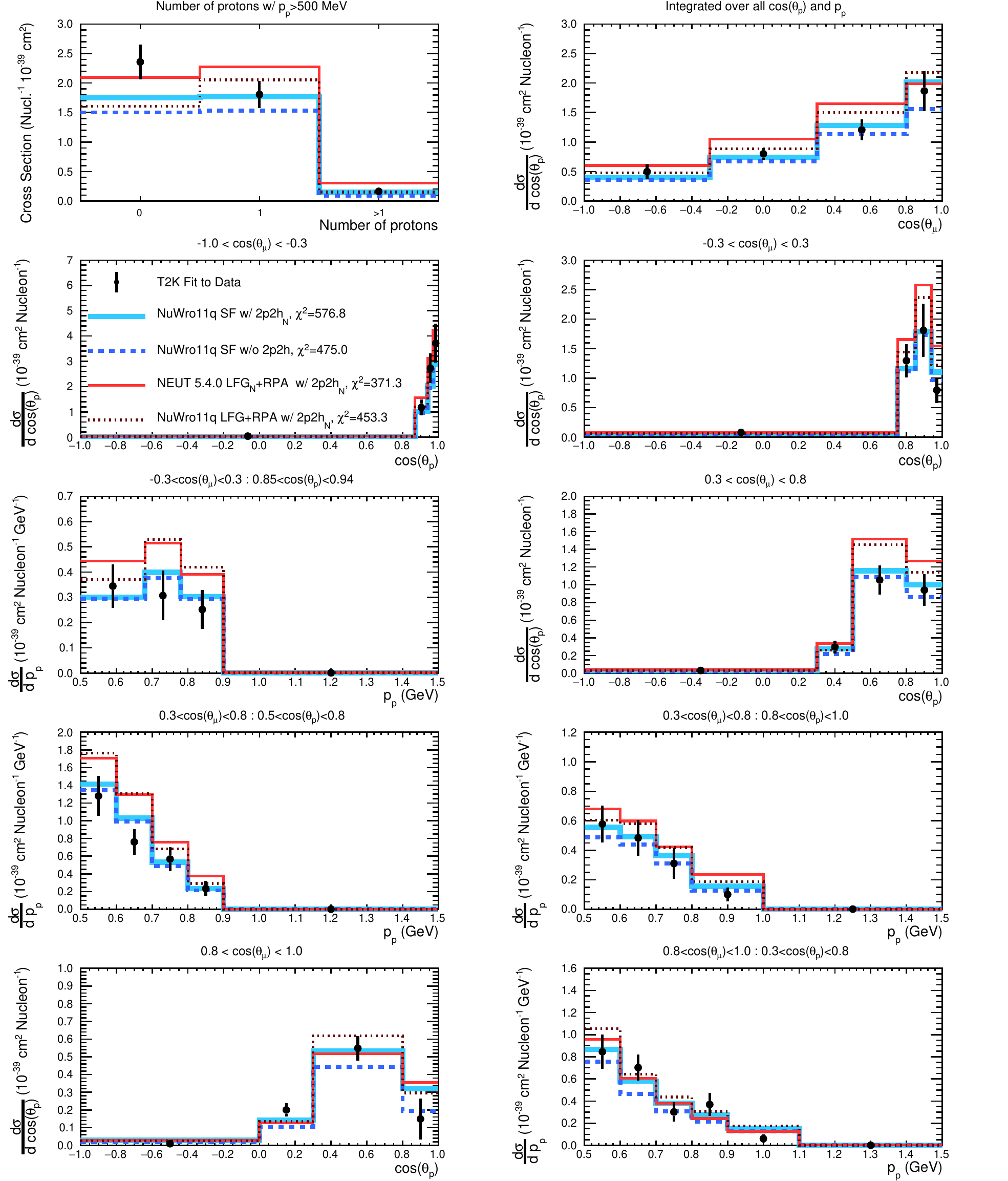} 
  \end{center}
\caption{Measurement of the cross section as a function of the proton multiplicity (top left) and as a function of proton and muon kinematics where there is exactly one proton (with momentum above 500 MeV). The data are compared to: NuWro 11q with the SF nuclear model, both with and without an additional 2p2h contribution; NEUT 5.4.0, which uses an LFG+RPA model that includes 2p2h predictions; and NuWro 11q with an LFG+RPA nuclear model and a separate 2p2h prediction. The same models are also compared to the cross section as a function of proton multiplicity. 2p2h$_N$ indicates the 2p2h model is an implementation of the Nieves et. al. model of Ref.~\cite{Nieves:2012}. The `N' subscript after LFG indicates that the model is using both a 1p1h and 2p2h prediction from the aforementioned model of Nieves et. al. More details of these models can be found in Sec.~\ref{sec:simulation}. Note that the last momentum bin in each plot is shortened for readability.}
\label{fig:MultiDiff_Np}
\end{figure*}

The extracted CC0$\pi$ multi-differential cross section results are shown in Figs.~\ref{fig:MultiDiff_0p}~and~\ref{fig:MultiDiff_Np}, compared to a variety of different model predictions. Value of the comparisons' $\chi^2$ are reported in the figures. Additional model comparisons to assess the suitability of RFG+RPA nuclear models (Figs.~\ref{fig:MultiDiff_0p_rfgComp}-\ref{fig:MultiDiff_Np_rfgComp}) and the impact of FSI (Figs.~\ref{fig:MultiDiff_0p_FSIComp}-\ref{fig:MultiDiff_Np_FSIComp}) are also shown in appendix~\ref{app:multidim}. The last bin of each of the momenta bins is shortened to improve the plot's readability but these bins are normalised by their total width (as specified in Tab.~\ref{tab:multiDiffBinning}) in accordance with Eq.~\ref{eqn:xsec}. These results are discussed in details in Sec.~\ref{sec:discussion}.

The total CC0$\pi$ cross section extracted is given in Tab.~\ref{tab:mdtotxsec} alongside a prediction from the default NuWro 11q simulation (which will be used as a standard reference for the comparison of the integrated measured cross section in all the analyses). 

\begin{table}[h]
\begin{center}
\small
\centering
\begin{tabular}{|c|c|c|}
\hline
Cross section & NuWro prediction\\
\hline
$4.329 \pm 0.502$ & $3.669$ \\
\hline
\end{tabular}
\caption[The total CC0$\pi$ cross section extracted in units of $10^{-39}$ cm$^2$ within the multi-differential analysis alongside the prediction from NuWro 11q using an SF nuclear model and with the 2p2h model of Nieves et. al.~\cite{Nieves:2012}.]{\label{tab:mdtotxsec}The total CC0$\pi$ cross section extracted in units of $10^{-39}$ cm$^2$ within the multi-differential analysis alongside the prediction from NuWro 11q using an SF nuclear model and with the 2p2h model of Nieves et. al.~\cite{Nieves:2012}. }
\end{center}
\end{table}

\subsection{Single Transverse Variables  \label{sec:STV}}

%general introduction aboout what we measure in this analysis and what is different with respect to the other analyses

%\textbf{SB: this section was definitively too long (I commented it out for now, the commented text can be reused by copying and paste in the Discussion section). In particular: STV already defined in 'Analysis strategy'. Sensitivity of the various variables to nuclear effects should be discussed in the 'Discussion' section in view of the actual results. This section should be really short and just show binning, plots of systematics and results}

%\textbf{Should this sort of thing be in the introduction rather than here?} 

In this analysis a novel set of observables is used to directly probe nuclear effects in measurements of CC0$\pi$+Np interactions. As previously discussed, these observables exploit the kinematic imbalance between the outgoing lepton and highest momentum proton in the plane transverse to the incoming neutrino, which can act as a powerful probe of nuclear effects both in the initial state and in FSI. 

%\subsubsection{Cross section extraction}

To extract the CC0$\pi$+Np differential cross section in the STV, a binned likelihood fit to the number of selected events in reconstructed STV bins is used, as described in Sec.~\ref{sec:extraction}. The uncertainties are evaluated from the postfit covariance matrix, thereby assuming they are Gaussian distributed. The binning for each of the STV are shown in Tab.~\ref{tab:stvbins} but it should be noted that these STV bins are also restricted to the reduced muon and proton kinematic phase space discussed in Sec.~\ref{sec:anaStrategy}. The binning is chosen such that the statistical error is comparable to the systematic error and that the bin widths are comparable to the detector resolution.

\begin{table}[h]
\begin{center}
\small
\centering
\begin{tabular}{|c|c|c|}
\hline
$\dpt$ (GeV) & $\dphit$ (radians) & $\dalphat$ (radians) \\
\hline
0.0-0.08 & 0.0-0.067 & 0.0-0.47\\
0.08-0.12 & 0.067-0.14 & 0.47-1.02\\
0.12-0.155 & 0.14-0.225 & 1.02-1.54\\
0.155-0.2 & 0.225-0.34 & 1.54-1.98\\
0.2-0.26 & 0.34-0.52 & 1.98-2.34\\
0.26-0.36 & 0.52-0.85 & 2.34-2.64\\
0.36-0.51 & 0.85-1.50 & 2.64-2.89\\
0.51-1.1 & 1.50-$\pi$ & 2.89-$\pi$\\
\hline
\end{tabular}
\caption[The STV analysis binning.]{\label{tab:stvbins}The chosen binning for STV cross section extraction.}
\end{center}
\end{table}

As described in Sec.~\ref{sec:extraction}, the fit must include effects from plausible variation of detector, flux, and neutrino interaction models. This is achieved by fitting the systematic parameters described in Sec.~\ref{sec:systematics} alongside the signal weights in both the signal region and control samples simultaneously (as described in Eq.~\ref{eqn:fit}). 
The systematic uncertainty due to the impact of these parameters is assessed alongside the statistical uncertainty from a post-fit covariance matrix of the parameters, which itself is constructed from the shape of the likelihood surface close to the best-fit point. All parameters are then marginalised in order to project the uncertainty onto the true number of selected signal events in bins of the STV. For a more conservative error estimation, the prefit uncertainty on the detector and flux systematic parameters is considered when evaluating the uncertainty on the subsequent flux normalisation and efficiency correction.
Parameters describing the signal which are overly degenerate with the signal weights (such as $M_{A}^{QE}$) are not fit. In such cases their uncertainty is taken into account, without any constraints from the data, by computing the effect of their variation on the efficiency corrections. Such effect is found to be relatively small (less than 2\% in all but the last bin of $\dpt$ and $\dphit$ where it is about 4\%). Further details regarding the uncertainty calculation and the handling of systematic uncertainties can be found in Ref.~\cite{dolanThesis}.
%The fit is free to move all of these parameters in accordance to a covariance matrix defined alongside them which parametrises the prior knowledge, picking up a penalty in the log-likelihood as it does. The implementation of these systematic parameters is very similar to that used for the near detector fit of T2K oscillation analyses, describe in Ref.~\cite{t2k2017longoa}.
%The systematic uncertainty due to the impact of these parameters and of the statistics is assessed through their influence on the local curvature of the log-likelihood surface around the best-fit point. This curvature is parametrised by the inverse of the matrix of second derivatives (i.e. the inverse Hessian) of the negative log-likelihood, with respect to the fit parameters. 
%
% Additional uncertainties are also included to take into account signal modelling effects on the efficiency correction and the number of targets and these are both found to be relatively small (less than 2\% in all but the last bin of $\dpt$ and $\dphit$ where it is about 4\%). 
%In case of flux normalization and detector efficiency corrections the prefit uncertainty is considered in order to be more conservative on these parameters, which are strongly correlated with the signal cross section. 
%The truth-level distribution of the STV is then used to calculate a cross section as in equation~\ref{eqn:xsec}. 

The extracted cross sections from the regularised fit in each of the STV are compared to the latest predictions from the current state-of-the-art models from the NEUT 5.3.2.2, NEUT 5.4.0, GENIE 2.12.4, NuWro~11q and GiBUU 2016 neutrino interaction simulations in a variety of configurations. The $\chi^2$ of each comparison is reported within the figures. As discussed in Sec.~\ref{subsec:mlfitting}, it can be more useful to instead consider the $\chi^2$ formed from the unregularised result but in this case the conservative data-driven regularisation strength means that the difference in the regularised and unregularised $\chi^2$ is marginal. This is demonstrated and discussed in appendix~\ref{app:unregchi2} (which provides the $\chi^2$ from the comparisons with the unregularised result).

\begin{figure*}[!hp]
\begin{center}
\includegraphics[width=0.49\textwidth]{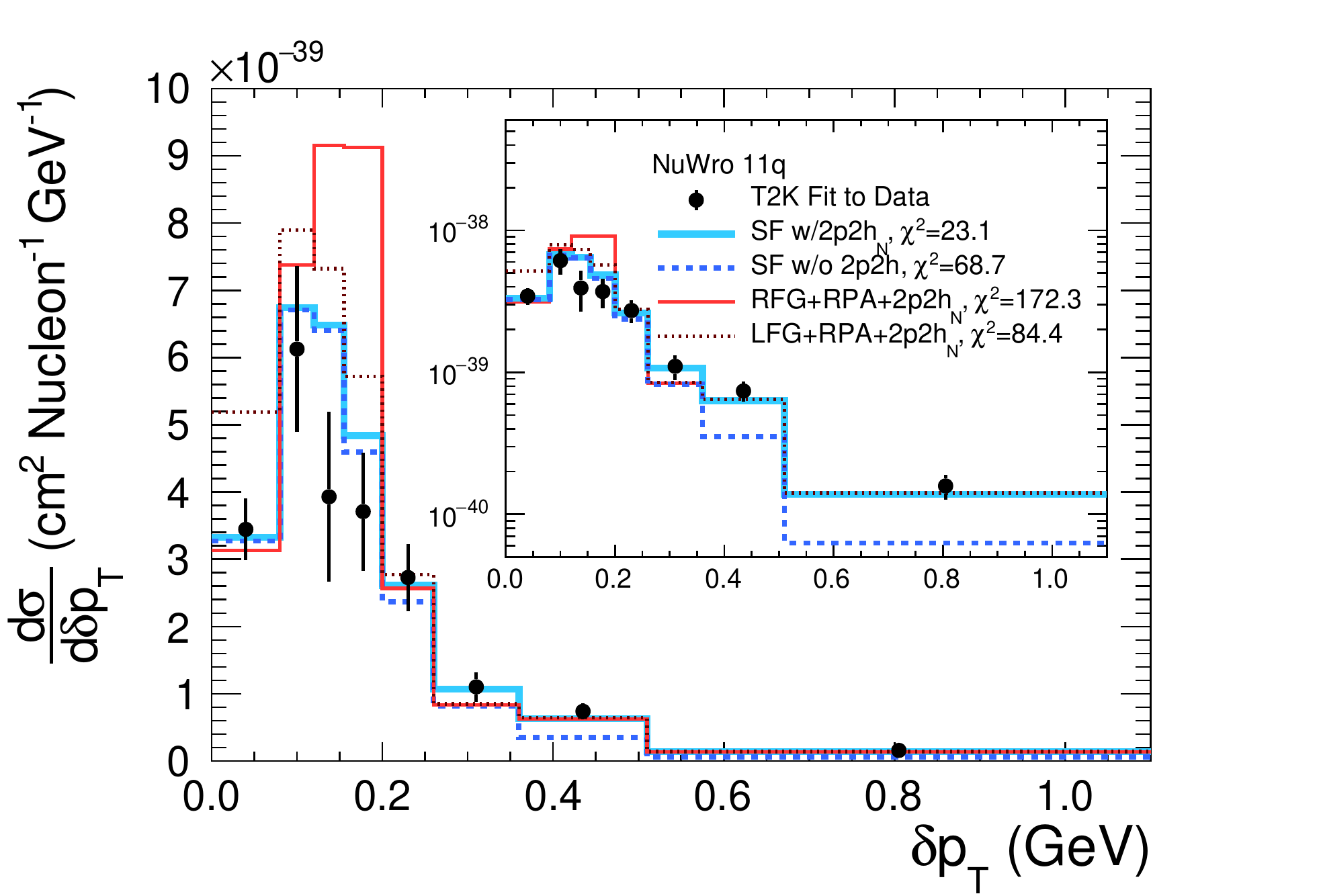}
\includegraphics[width=0.49\textwidth]{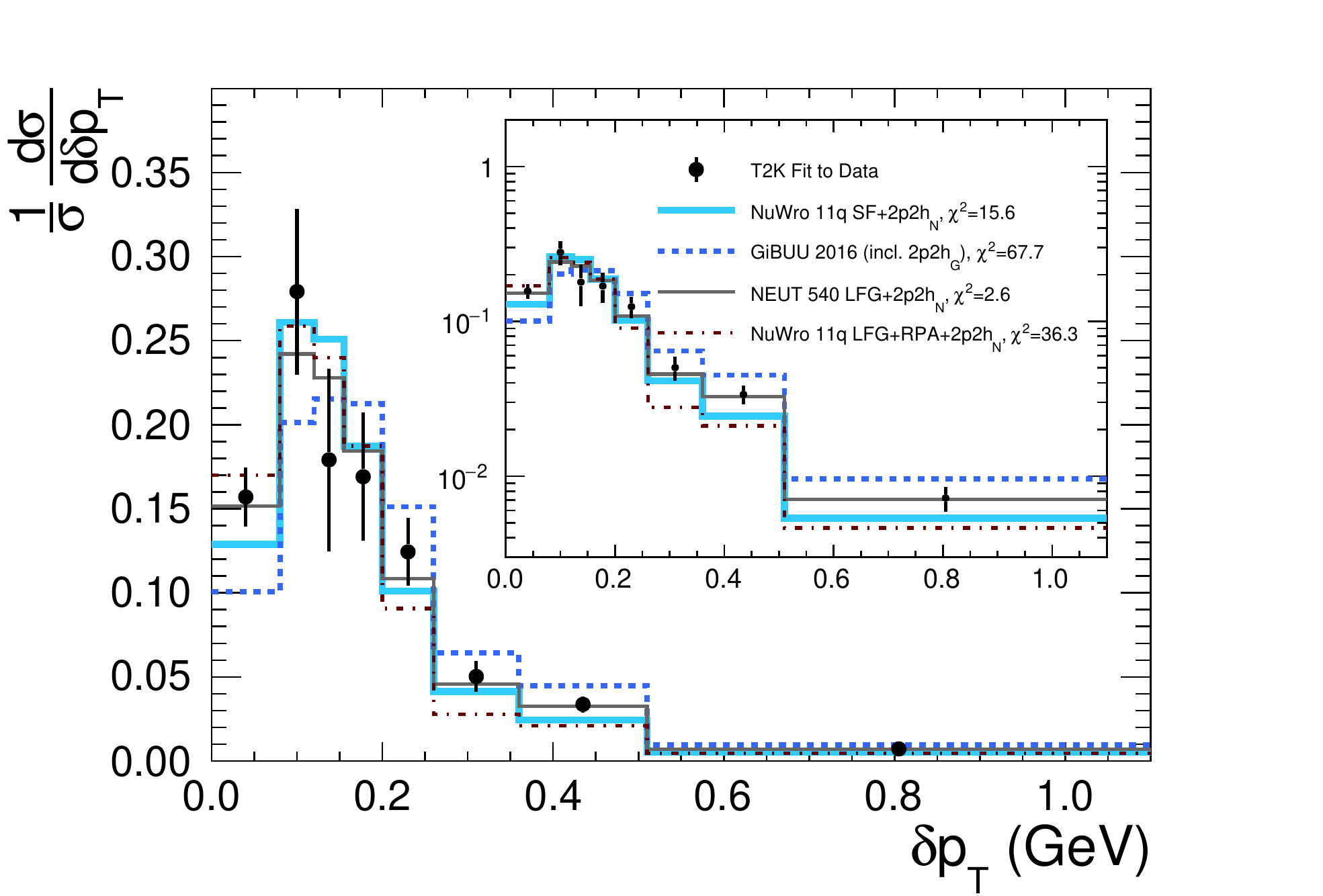}

\includegraphics[width=0.49\textwidth]{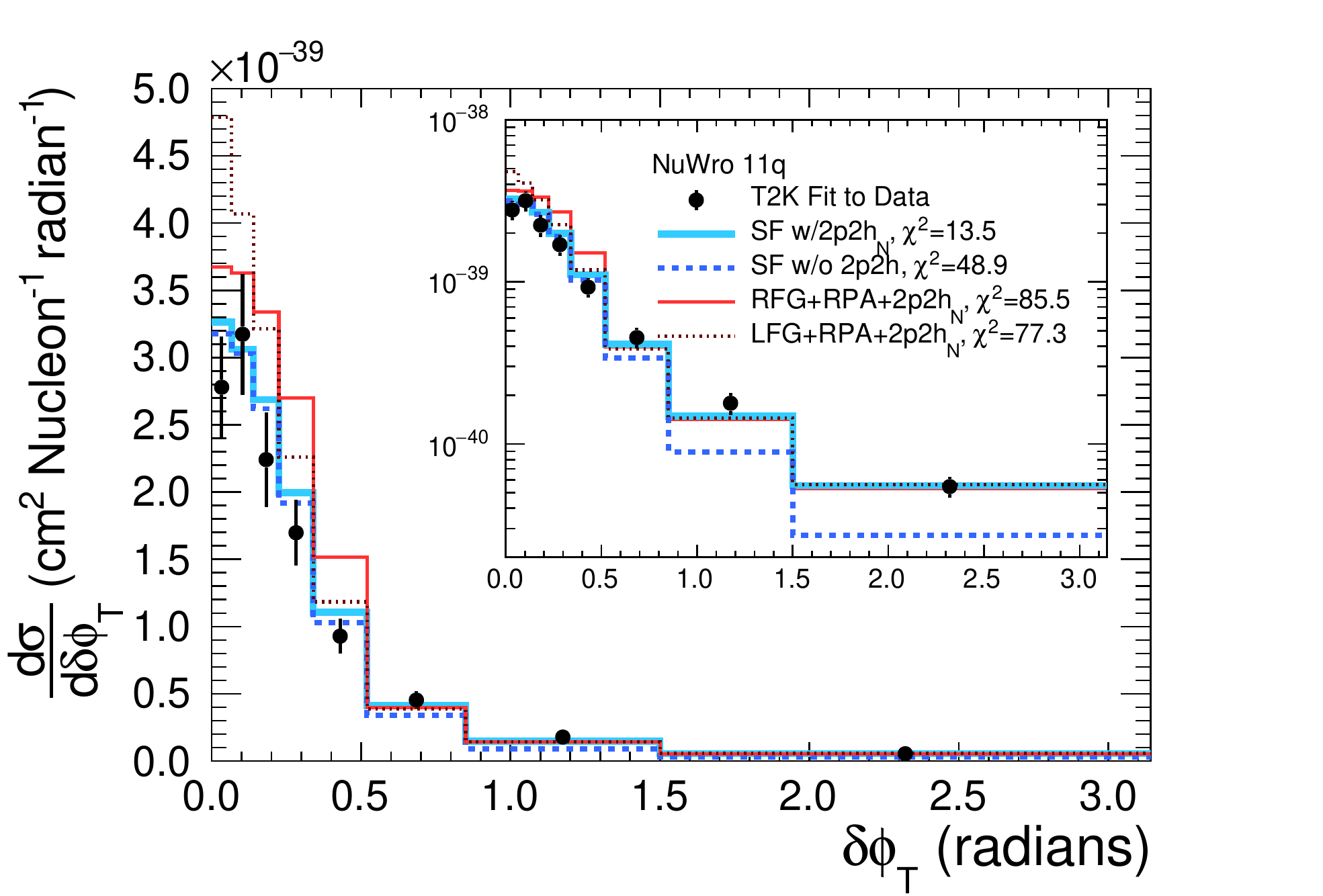}
\includegraphics[width=0.49\textwidth]{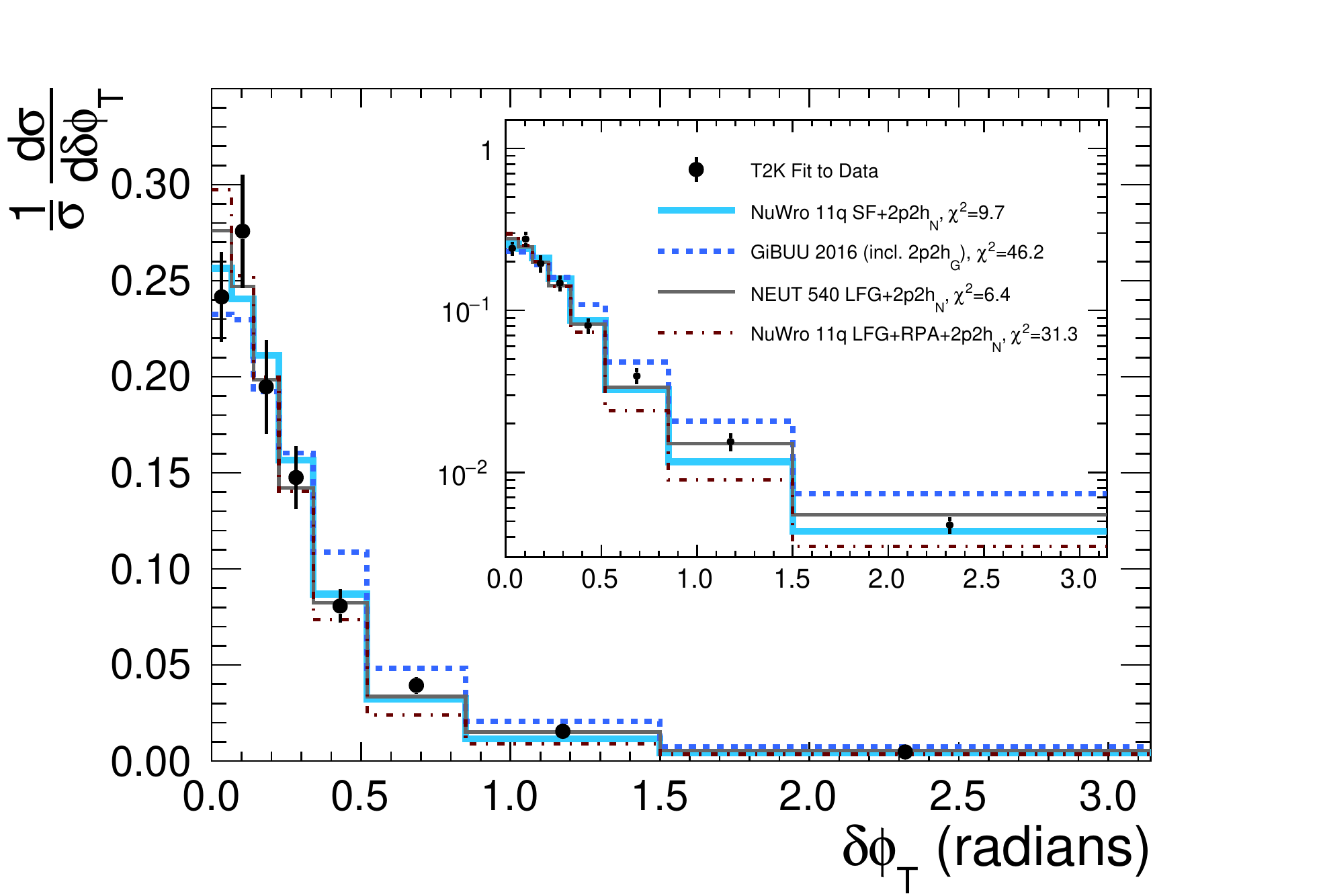}

\includegraphics[width=0.49\textwidth]{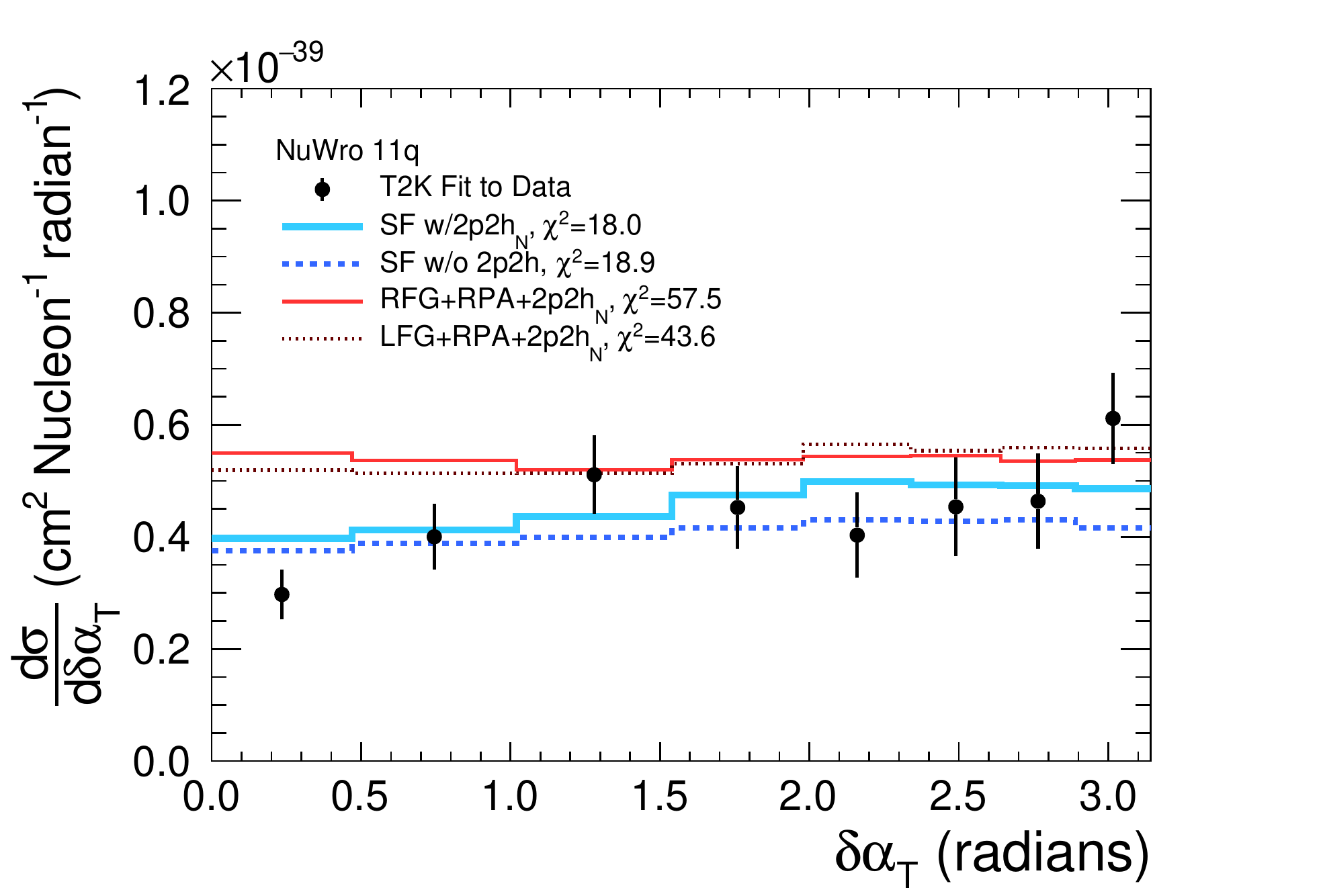}
\includegraphics[width=0.49\textwidth]{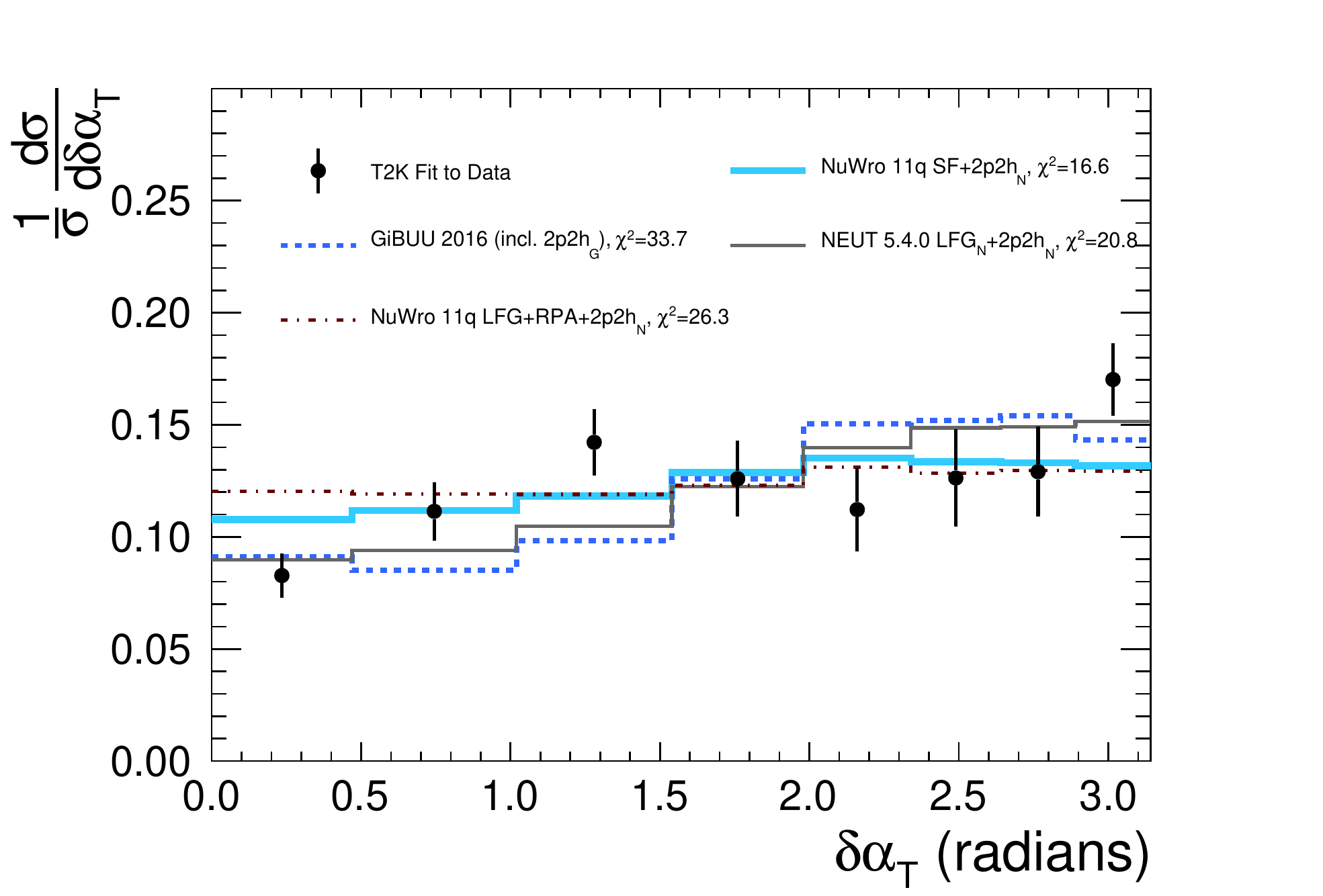}

\end{center}
\caption{The extracted differential cross section as a function of the single transverse variables compared to: different initial state models in the NuWro 11q simulation (\textbf{left}); shape only predictions from NuWro 11q and GiBUU 2016 (\textbf{right}). Although it is not shown the NEUT 5.3.2.2 SF prediction has an almost identical shape to the NuWro 11q SF prediction. The NuWro 11q RFG+RPA prediction shown is similar to the NEUT model used as a starting point for T2K's oscillation analyses. 
2p2h$_N$ indicates the Nieves et. al. model of Ref.~\cite{Nieves:2012} as implemented in NEUT or NuWro, while 2p2h$_G$ indicates an extrapolation from electron-scattering data implemented in the GiBUU 2016 simulation~\cite{Buss:2011mx}. More details of these models can be found in Sec.~\ref{sec:simulation}. The inlays show the same comparisons on a logarithmic scale. }
\label{fig:STV_NuWro}
\end{figure*}

\begin{figure*}[!hp]
\begin{center}
\includegraphics[width=0.49\textwidth]{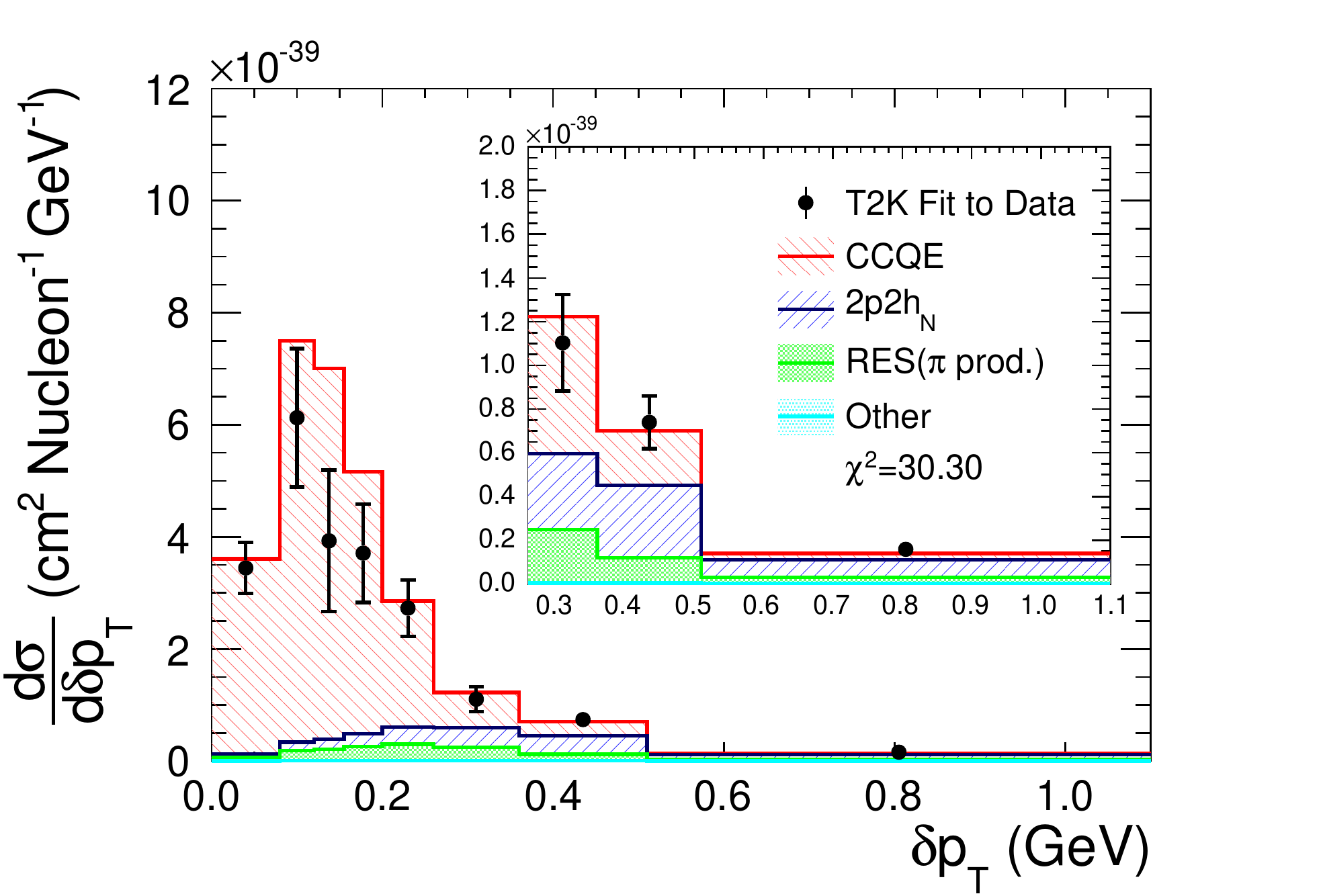}
\includegraphics[width=0.49\textwidth]{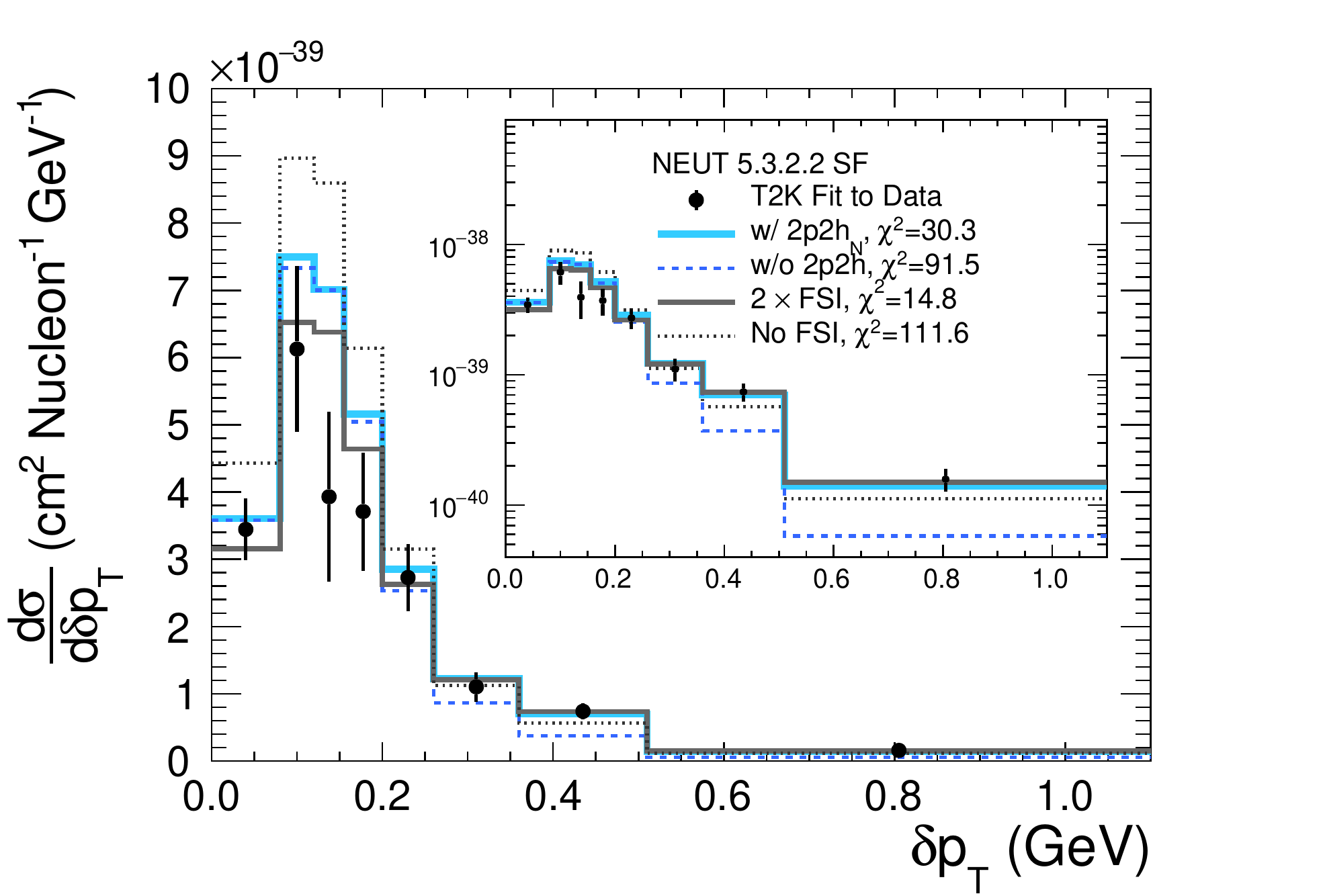}

\includegraphics[width=0.49\textwidth]{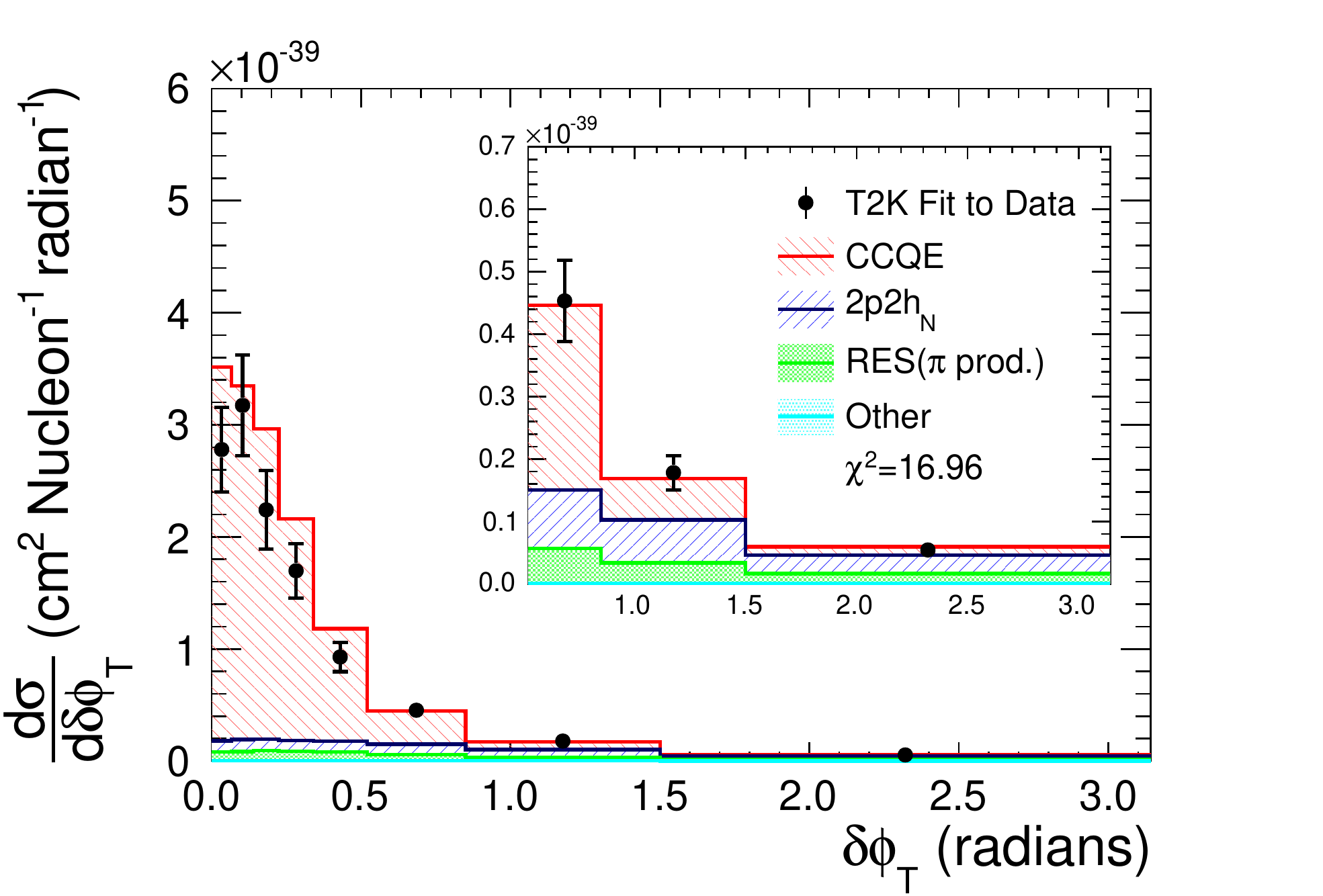}
\includegraphics[width=0.49\textwidth]{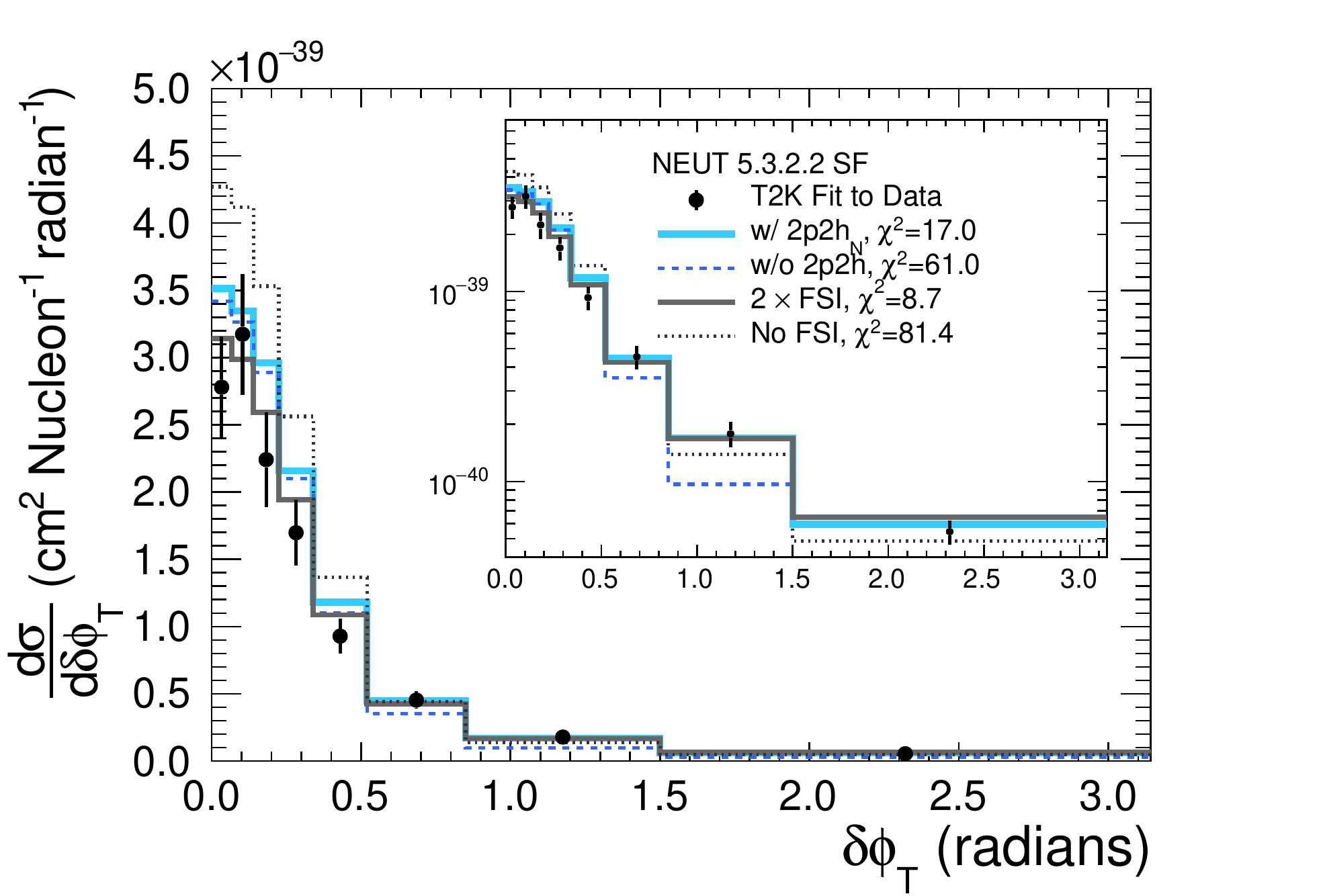}

\includegraphics[width=0.49\textwidth]{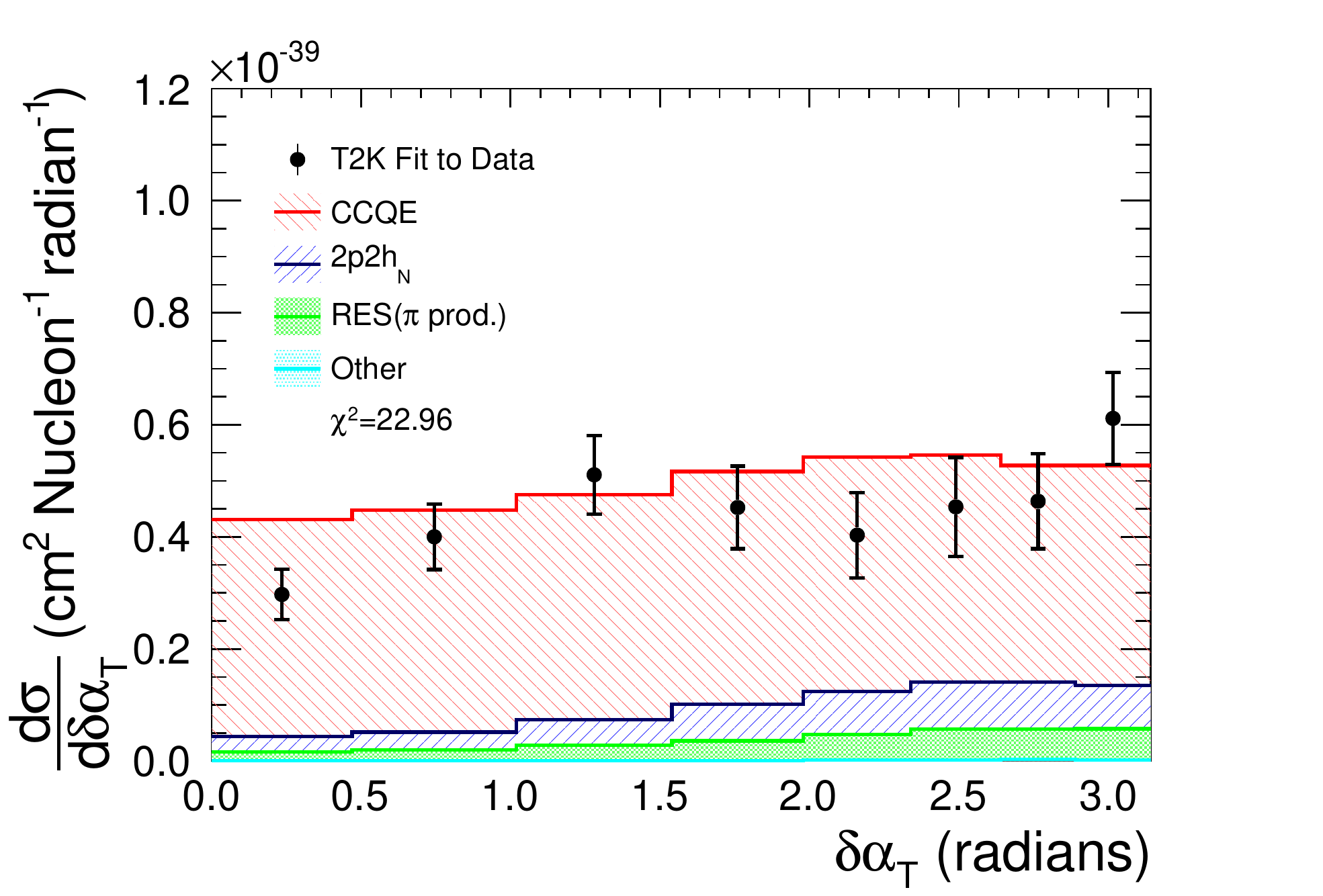}
\includegraphics[width=0.49\textwidth]{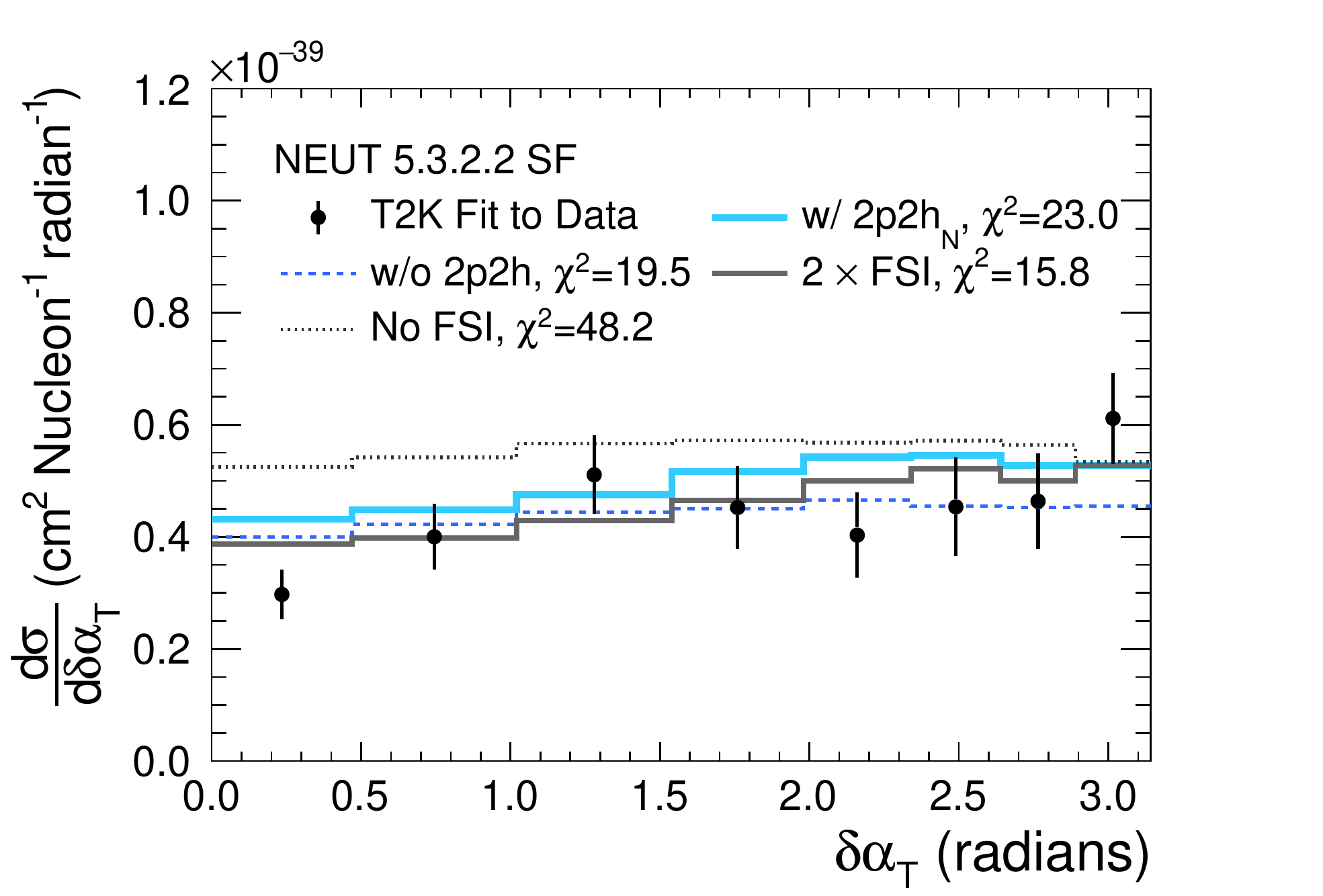}
\end{center}
\caption{The extracted differential cross section as a function of the single transverse variables compared to: the NEUT 5.3.2.2 simulation with the SF initial state model and an ad hoc 2p2h model (\textbf{left}); the same NEUT simulation with various scalings of the mean free path of nucleons undergoing FSI processes to simulate different FSI strengths (\textbf{right}). 2p2h$_N$ indicates the Nieves et. al. model of Ref.~\cite{Nieves:2012} implemented in NEUT. A comparison of the NEUT prediction without a 2p2h contribution is also shown.  More details of these models can be found in Sec.~\ref{sec:simulation}. The `N' subscript after LFG indicates that the model is using both a 1p1h and 2p2h prediction from the aforementioned model of Nieves et. al. The inlays on the left plots show a close-up of the tail regions of $\dpt$ and $\dphit$ whilst those on the right show the same comparisons on a logarithmic scale.}
\label{fig:STV_NEUT6D}
\end{figure*}

\begin{figure*}[!hp]
\begin{center}
\includegraphics[width=0.49\textwidth]{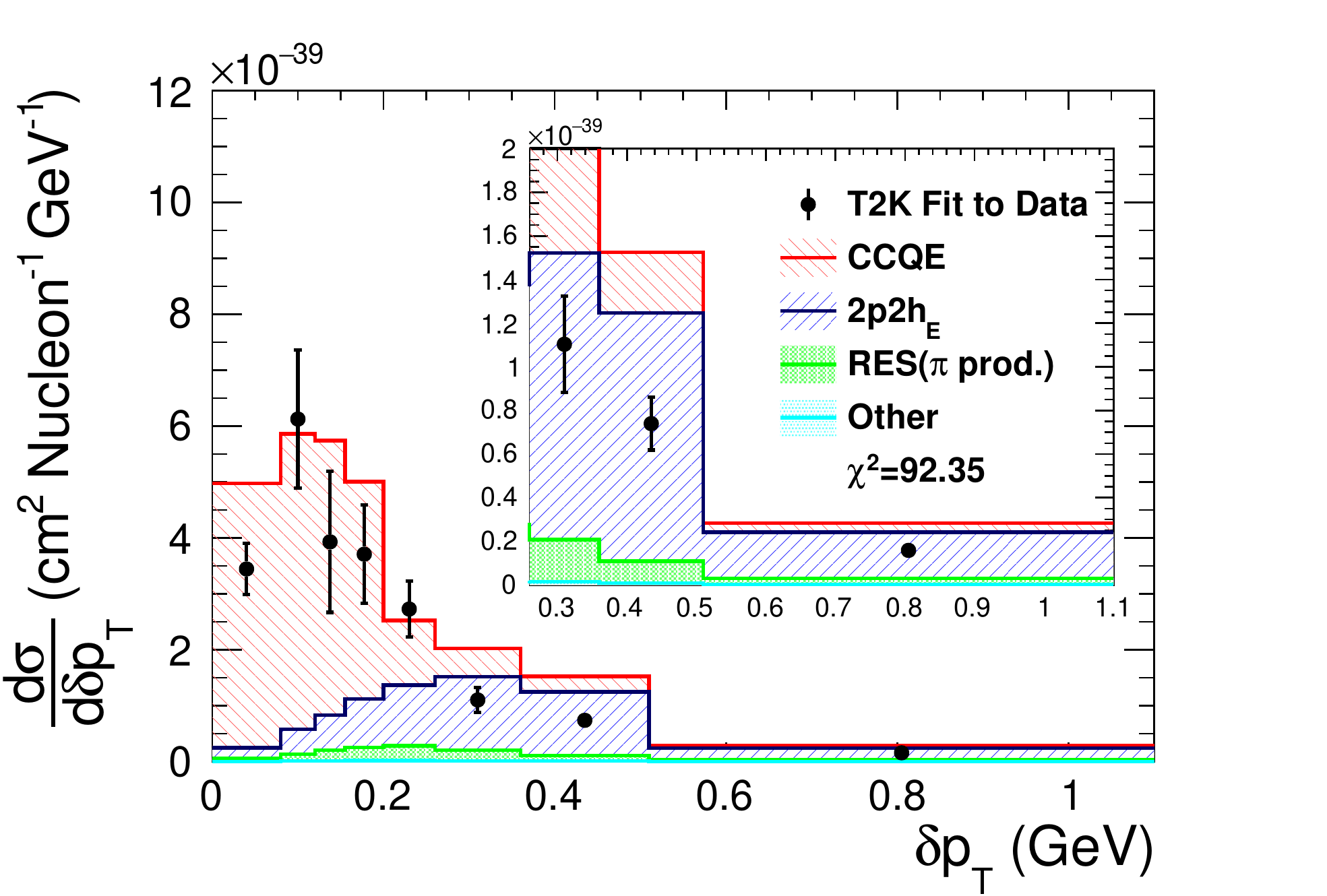}
\includegraphics[width=0.49\textwidth]{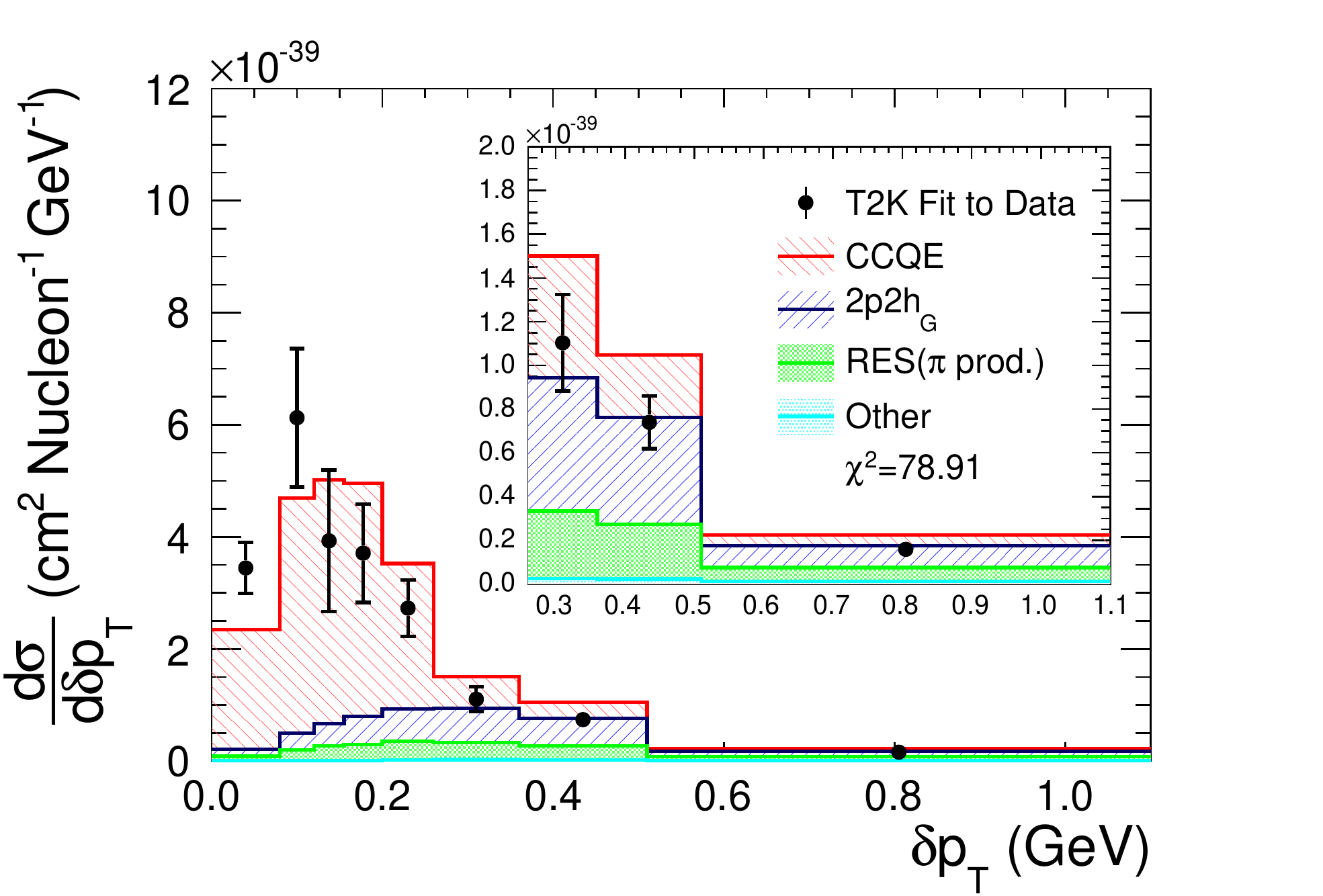}

\includegraphics[width=0.49\textwidth]{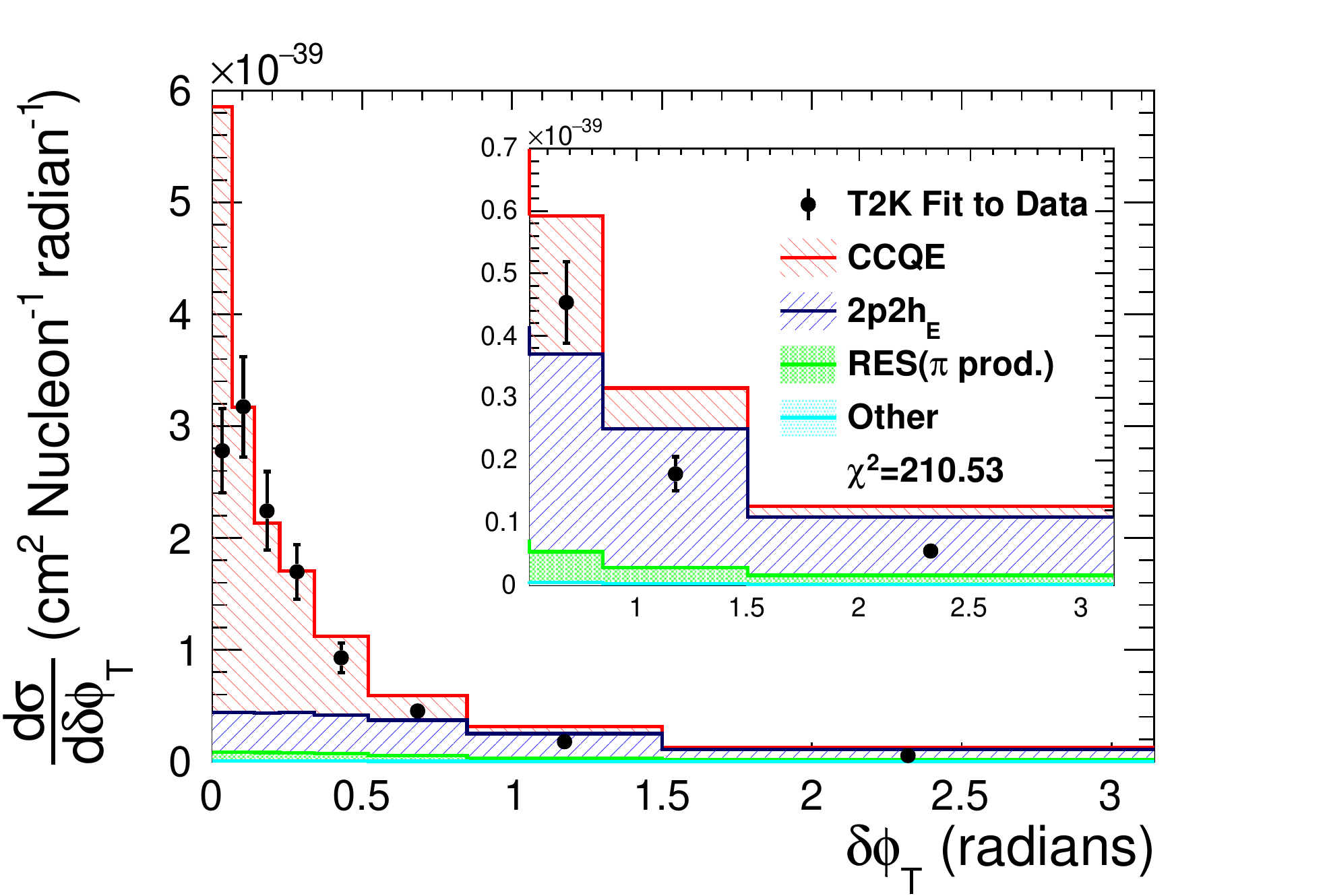}
\includegraphics[width=0.49\textwidth]{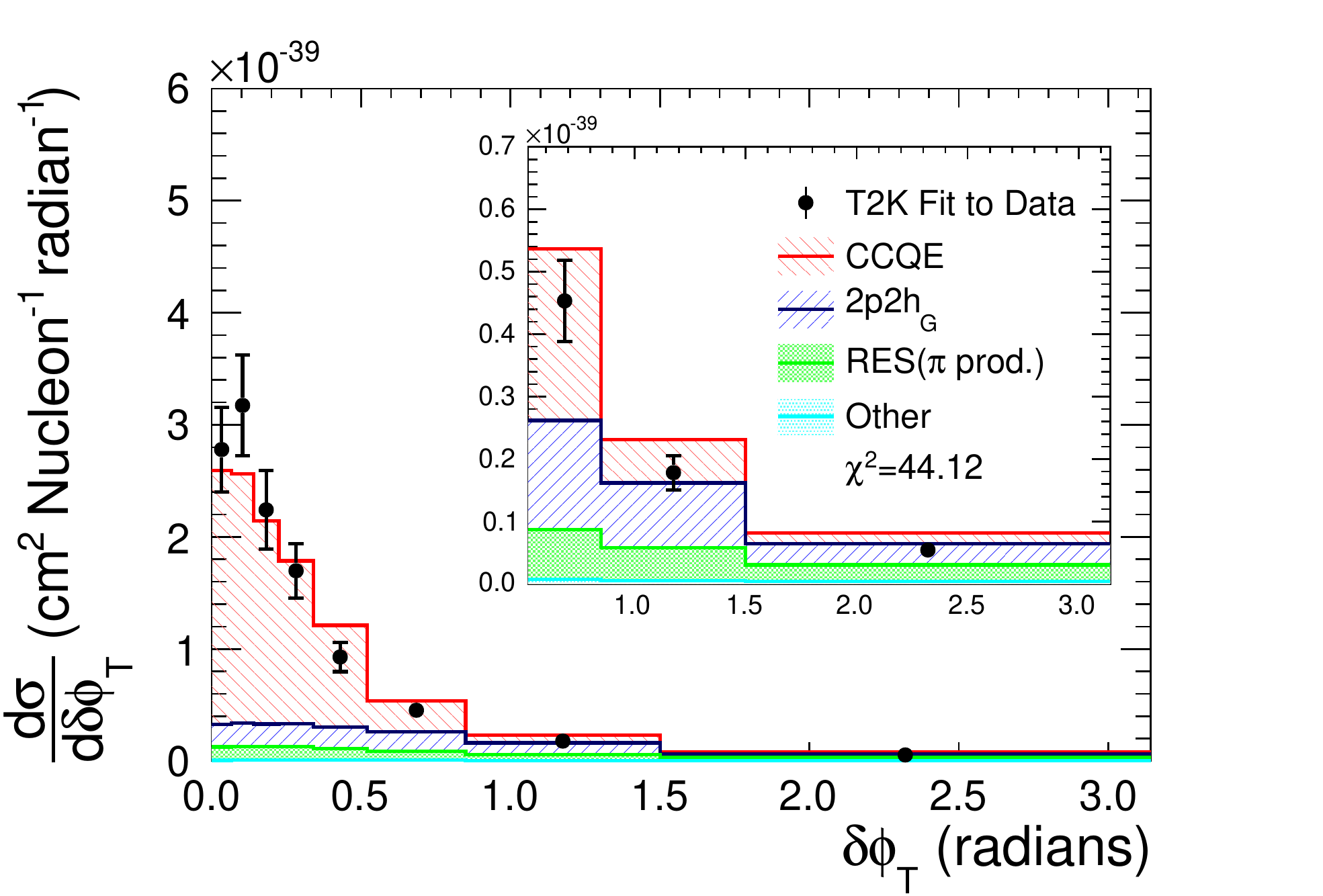}

\includegraphics[width=0.49\textwidth]{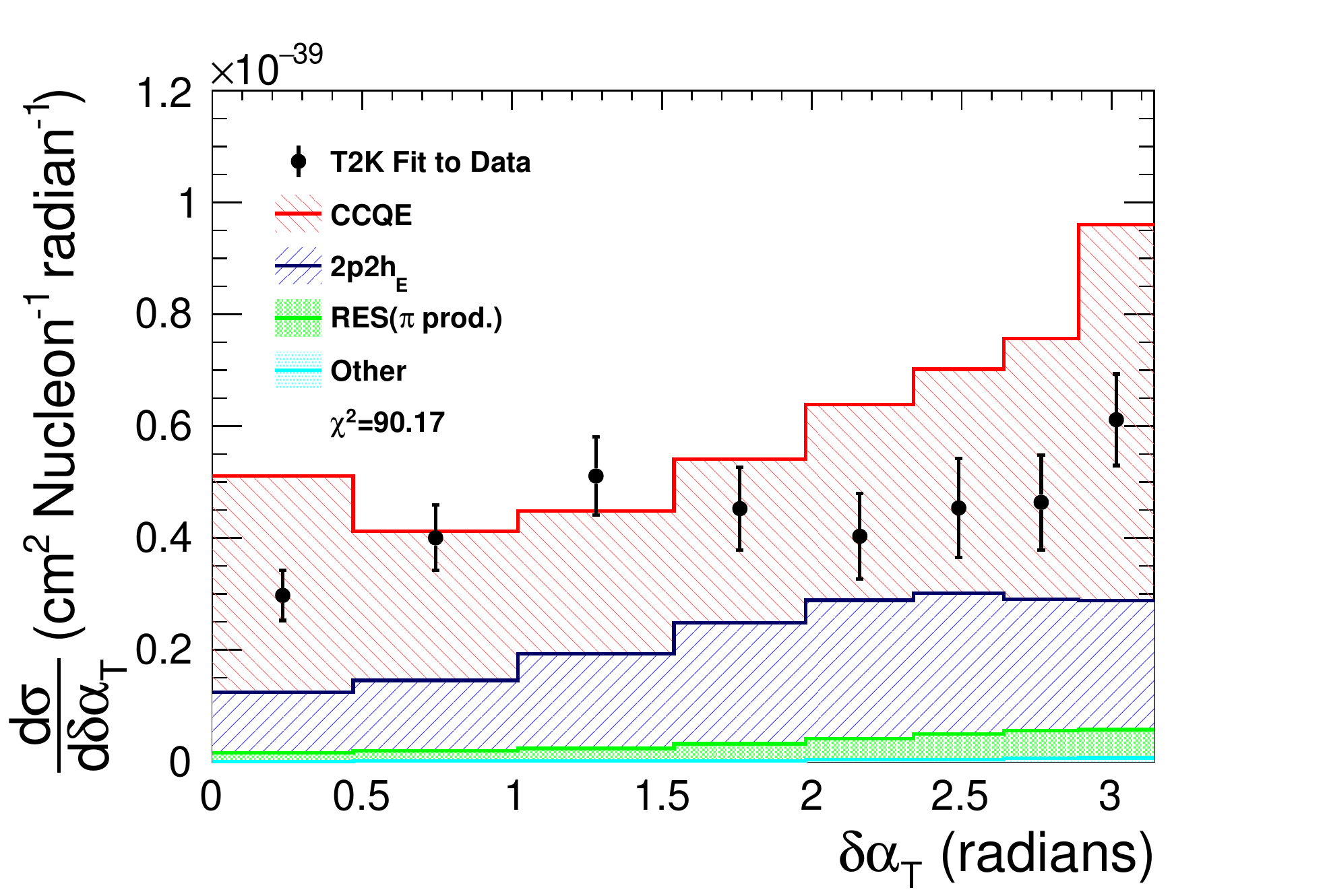}
\includegraphics[width=0.49\textwidth]{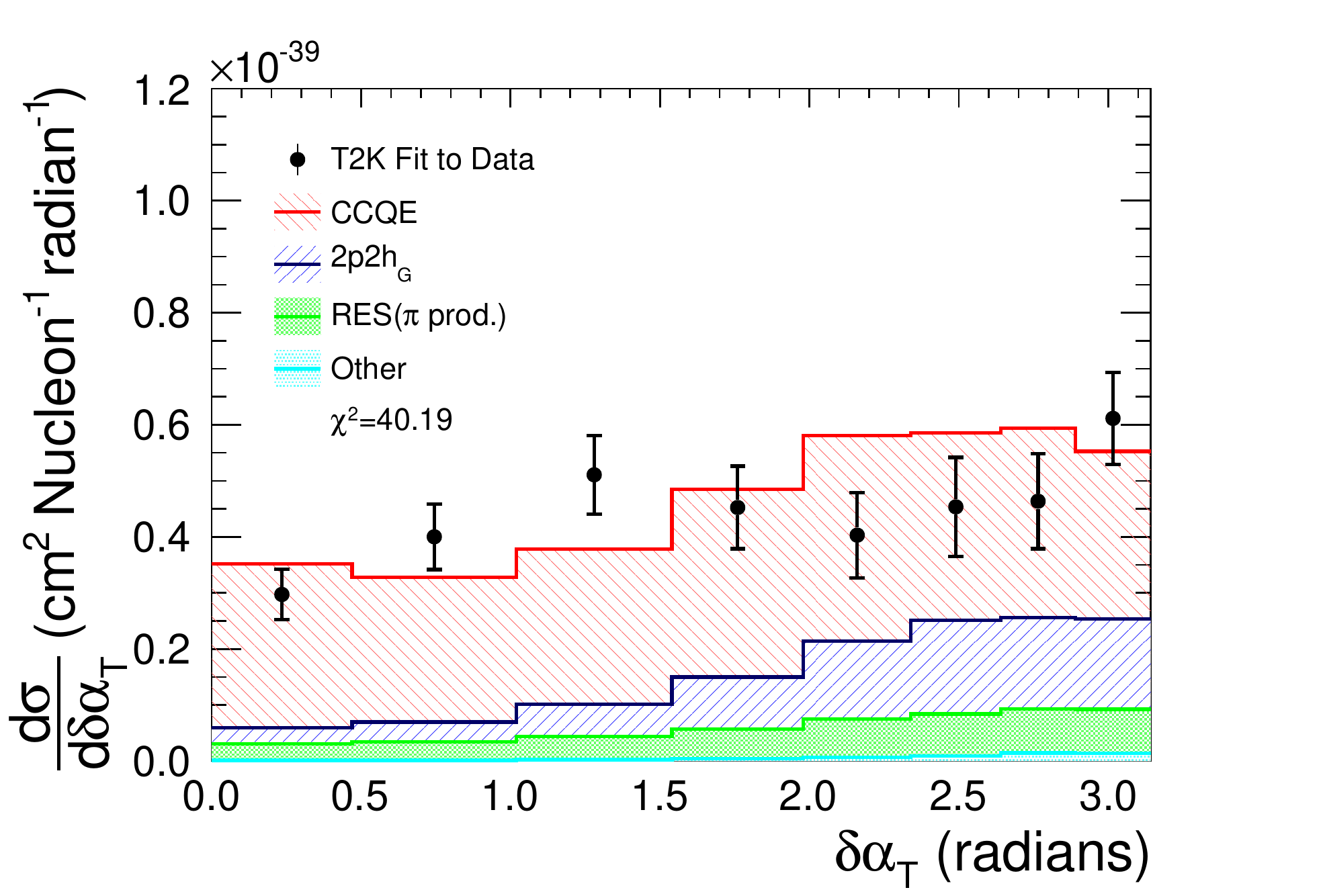}

\end{center}
\caption{The extracted differential cross section as a function of the single transverse variables compared to: the GENIE 2.12.4 simulation (\textbf{left}) and the GiBUU 2016 simulation (\textbf{right}). GENIE uses the Bodek and Richie RFG initial state model and this prediction also includes GENIE's empirical 2p2h prediction (2p2h$_E$). This GENIE prediction is similar that used as a starting point for the NO$\nu$A experiment's oscillation analyses. More details of these models can be found in Sec.~\ref{sec:simulation}. The inlays on the plots show a close-up of the tail regions of $\dpt$ and $\dphit$.}
\label{fig:STV_GENIE}
\end{figure*}

The left plots of Fig.~\ref{fig:STV_NuWro} compares the results to a variety of different initial state models whilst a shape-only comparison to a subset of these shown on the right alongside the GiBUU 2016 prediction. The contribution from each interaction mode predicted from NEUT 5.3.2.2 is shown in Fig.~\ref{fig:STV_NEUT6D}, alongside the impact of altering the simulations with and without a 2p2h contribution and modifying the FSI strength by varying the mean free path of nucleons within the NEUT FSI cascade model.  Finally a similar breakdown by interaction mode is then made for GiBUU and GENIE in Fig.~\ref{fig:STV_GENIE}. In GENIE the empirical 2p2h contribution, used for instance in the neutrino-interaction model of the NO$\nu$A experiment~\cite{novaOA2017}, is enabled. 
Figures to evaluate the impact of RPA and the role of regularisation of the cross section extraction are shown in appendix~\ref{app:stv}.   These results are discussed in details in Sec.~\ref{sec:discussion}.
%Within all these plots the subscript after 2p2h indicates whether it is an implementation of the Nieves et. al. model of Ref.~\cite{Nieves:2012} in NEUT or NuWro (N); an extrapolation from electron-scattering data in the GiBUU 2016 simulation~\cite{Buss:2011mx} (G); or from the empirical MEC model in GENIE (E). The `N' subscript after LFG indicates that the model is using both a 1p1h and 2p2h prediction from the aforementioned model of Nieves et. al.

The total CC0$\pi$+Np cross section extracted (within the phase space constraints listed in Tab.~\ref{tab:phaseSpace}) for each of the STV is given in Tab.~\ref{tab:stvtotxsec} alongside a prediction from NuWro 11q. These total cross sections are not identical since the best-fit parameters are altered slightly depending on which projection of the event selection is used as an input. 

\begin{table}[h]
\begin{center}
\small
\centering
\begin{tabular}{|c|c|c|}
\hline
Observable & Cross section & NuWro prediction \\
\hline
$\dpt$     & $1.303 \pm 0.127$&  $1.422$ \\
$\dphit$   & $1.326 \pm 0.124$&  $1.422$ \\
$\dalphat$ & $1.375 \pm 0.130$&  $1.422$ \\
\hline
\end{tabular}
\caption[The total CC0$\pi$+Np cross section in units of $10^{-39}$ cm$^2$ extracted (within the phase space constraints listed in Tab.~\ref{tab:phaseSpace}) for each of the STV shown alongside the prediction from NuWro 11q using an SF nuclear model and with the 2p2h model of Nieves et. al.~\cite{Nieves:2012}.]{\label{tab:stvtotxsec}The total CC0$\pi$+Np cross section in units of $10^{-39}$ cm$^2$ extracted (within the phase space constraints listed in Tab.~\ref{tab:phaseSpace}) for each of the STV shown alongside the prediction from NuWro 11q using an SF nuclear model and with the 2p2h model of Nieves et. al.~\cite{Nieves:2012}.}
\end{center}
\end{table}

%% \begin{figure}
%% \begin{center}
%% \includegraphics[width=0.45\textwidth]{figures/STV/neut5322_SF_out_dpt_reacStackXsec.png}
%% \includegraphics[width=0.45\textwidth]{figures/STV/neut5322_SF_out_dphit_reacStackXsec.png}
%% \includegraphics[width=0.45\textwidth]{figures/STV/neut5322_SF_out_dat_reacStackXsec.png}
%% \end{center}
%% \caption{A comparison of analysis results to predictions from NEUT 5.3.2.2 using $M_A^{QE}=1.03$ and an SF nuclear model, split by interaction mode. \textbf{PLACEHOLDER, value of MAQE is not consistent with NEUT 6D - should be 1.21.}}
%% \label{fig:defgenneut}
%% \end{figure}

%\begin{figure*}
%\begin{center}
%\end{center}
%\caption{The extracted area-normalised differential cross section as a function of the single transverse variables compared to predictions from NuWro 11q and GiBUU 2016. Although it is not shown the NEUT 5.3.2.2 SF prediction is almost identical to the NuWro prediction.}
%\label{fig:STV_allMC}
%\end{figure*}

%\begin{figure*}
%\begin{center}
%\includegraphics[width=0.45\textwidth]{figures/STV/results/gibuu_dpt.png}
%\includegraphics[width=0.45\textwidth]{figures/STV/results/gibuu_dphit.png}
%\includegraphics[width=0.45\textwidth]{figures/STV/results/gibuu_dat.png}
%\end{center}
%\caption{Measurements of cross section as a function of STV compared to the GiBUU 2016 simulation.}
%\label{fig:STV_Gibuu}
%\end{figure*}

%\begin{figure*}
%\begin{center}
%\end{center}
%\caption{The extracted differential cross section as a function of the single transverse variables compared to the NEUT 5.3.2.2 simulation shown in figure~\ref{fig:STV_NEUT6D} various scalings of the mean free path of nucleons undergoing FSI processes to simulate different FSI strengths. A comparison of the NEUT prediction without a 2p2h contribution is also shown.}
%\label{fig:STV_FSI}
%\end{figure*}

%\clearpage
%\input{Inferred}
%
\subsection{Proton inferred kinematics \label{sec:inferred}}

%In section~\ref{subsec:binning}, the motivation and the definition of the muon phase-space binning as a part of the signal definition will be discussed in details. The selection efficiencies for each muon phase-space bin will be shown at the end of the section. 
%In section~\ref{subsec:unfolding}, the iterative Bayesian unfolding is explained, followed by the description on the propagation of uncertainties in section~\ref{subsec:uncertainties}. 
%The unfolded data using the validated machinery will be shown in section~\ref{subsec:results}.

%\subsection{$\mu$ Phase-Space Binning}
%\label{subsec:binning}

%Instead of restricting the phase space, this analysis includes the
%effect by defining bins in the muon phase-space variables, and performing the analysis in them separately.
%There are various signal regions where the selection is corrected to, as following :

As outlined in Sec.~\ref{sec:anaStrategy}, this analysis uses the inferred kinematic imbalance between measured proton kinematics and what would be inferred from the measured muon kinematics under a QE approximation, which can act as a metric for the extent to which the QE approximation is reliable for events that are approximately characteristic of the dominant sample used in T2K neutrino oscillation analyses.   

In this analysis the muon phase-space is divided into multiple bins in order to correct for the different selection efficiencies and CC-non-QE contributions across the phase-space:
\begin{itemize}
    \item Bin 0 : $\cos \theta_\mu < -0.6$.
    \item Bin 1 : $-0.6 < \cos \theta_\mu < 0.0 \mbox{ \& } p_{\mu} < 250 \mbox{ MeV}$.
    \item Bin 2 : $-0.6 < \cos \theta_\mu < 0.0 \mbox{ \& } p_{\mu} > 250 \mbox{ MeV}$.
    \item Bin 3 : $\cos \theta_\mu > 0.0 \mbox{ \& } p_{\mu} < 250 \mbox{ MeV}$.
    \item Bin 4 : $0.0 < \cos \theta_\mu < 0.8 \mbox{ \& } p_{\mu} > 250 \mbox{ MeV}$.
    \item Bin 5 : $0.8 < \cos \theta_\mu \mbox{ \& } 250 < p_{\mu} < 750 \mbox{ MeV}$.
    \item Bin 6 : $0.8 < \cos \theta_\mu \mbox{ \& } p_{\mu} > 750 \mbox{ MeV}$.
\end{itemize}
Within each muon angular bin, the same binning in the inferred kinematic variables is used, which is shown in Tab.~\ref{tab:infkbins}. The binning is chosen to ensure that the efficiency is suitably flat within each bin and that the bin width is not less than the detector resolution.

\begin{table}[h]
\begin{center}
\small
\centering
\begin{tabular}{|c|c|c|}
\hline
$\Delta p$ (GeV) & $| \Delta \textit{\textbf{p}}|$ (GeV) & $\Delta \theta$ (degrees) \\
\hline
-5.0, -0.3 & 0.0, 0.3 & -360, -5 \\
-0.3, 0.0  & 0.3, 0.4 & -5, 5   \\
0.0, 0.1   & 0.4, 0.5 & 5, 10   \\
0.1, 0.2   & 0.5, 0.6 & 10, 20  \\
0.2, 0.3   & 0.6, 0.7 & 20, 360  \\
0.3, 0.5   & 0.7, 0.9 &         \\
0.5, 5.0   & 0.9, 5.0 &         \\
\hline
\end{tabular}
\caption{\label{tab:infkbins}The bins of proton inferred variables in which the cross section is measured.}
\end{center}
\end{table}

To estimate the systematic uncertainties the entire unfolding procedure is repeated for a comprehensive set of plausible variations of the T2K reference model according to the systematic uncertainty sources discussed in Sec.~\ref{sec:systematics}. The covariance of the ensemble of results from the different pseudo-experiments is then taken to characterise the uncertainty:
\begin{equation}
V_{ij} = \frac{1}{N} \sum_{i = 1}^{N} (\sigma^{variation}_{i} - \sigma^{nominal}_{i})(\sigma^{variation}_{j} - \sigma^{nominal}_{j})
\end{equation}
where $\sigma^{variation}_{i}$ is an extracted cross section in bin $i$ for a particular variation of the input simulation and $\sigma^{nominal}_{i}$ is a nominal cross section in bin $i$.

%\subsubsection{Results}
%\label{subsec:results}

\begin{figure*}[!hp]
\begin{center}
 \includegraphics[width=\textwidth]{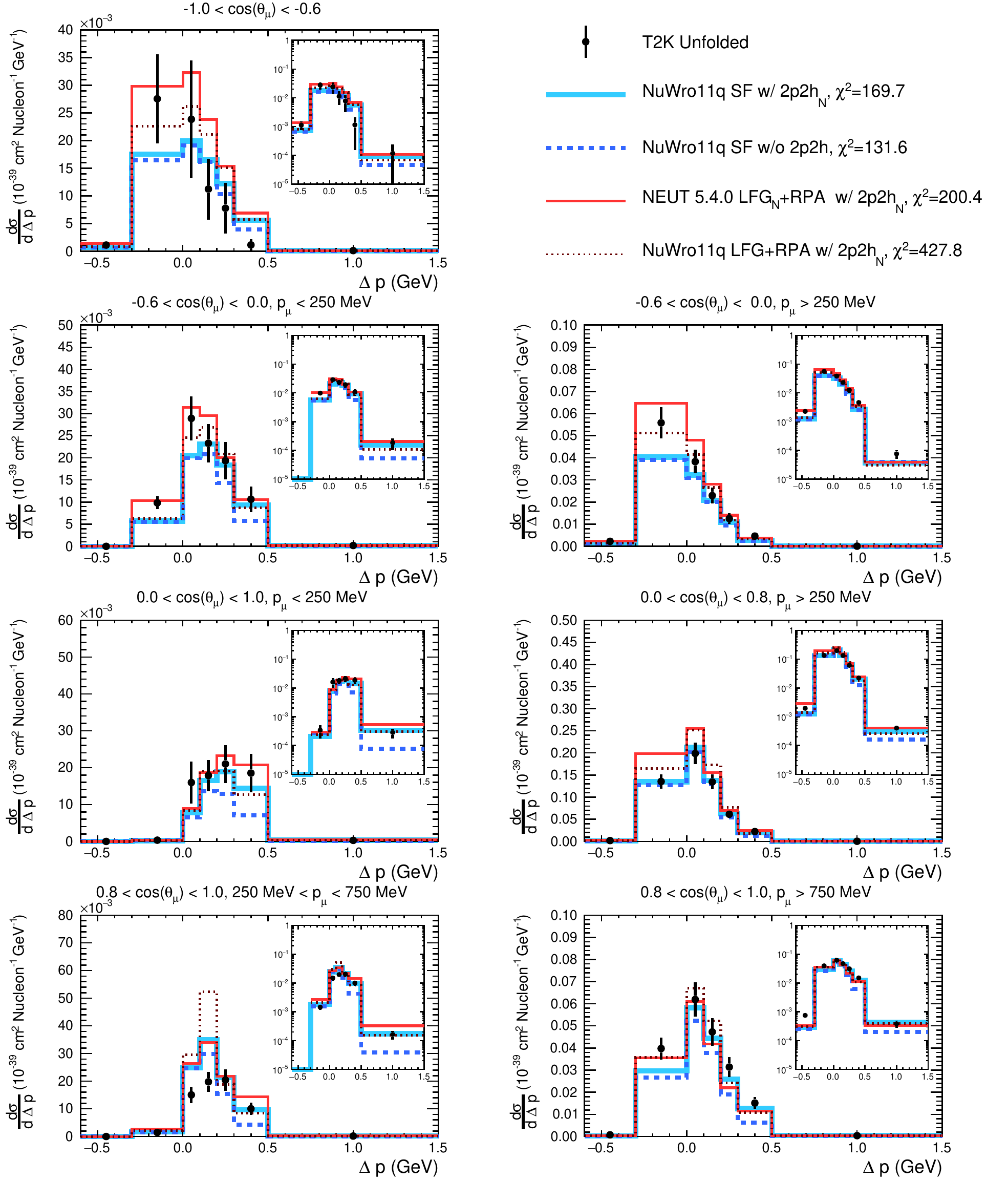} 
  \end{center}
\caption{The extracted differential cross section as a function of the inferred and true proton momentum difference in different muon kinematic bins, within a restricted proton kinematic phase space, compared to a variety of model predictions: NuWro 11q with the SF nuclear model both with and without an additional ad hoc 2p2h contribution; NEUT 5.4.0, which uses an LFG+RPA model that includes 2p2h predictions; and NuWro 11q with an LFG+RPA nuclear model and a separate 2p2h prediction. 2p2h$_N$ indicates the 2p2h model is an implementation of the Nieves et. al. model of Ref.~\cite{Nieves:2012}. The `N' subscript after LFG indicates that the model is using both a 1p1h and 2p2h prediction from the aforementioned model of Nieves et. al. More details of these models can be found in Sec.~\ref{sec:simulation}. Note that the first and last bin in each plot is shortened for improved readability. The inlays show the same comparisons on a logarithmic scale. }
\label{fig:inf_model_mom}
\end{figure*}

\begin{figure*}[!hp]
\begin{center}
 \includegraphics[width=\textwidth]{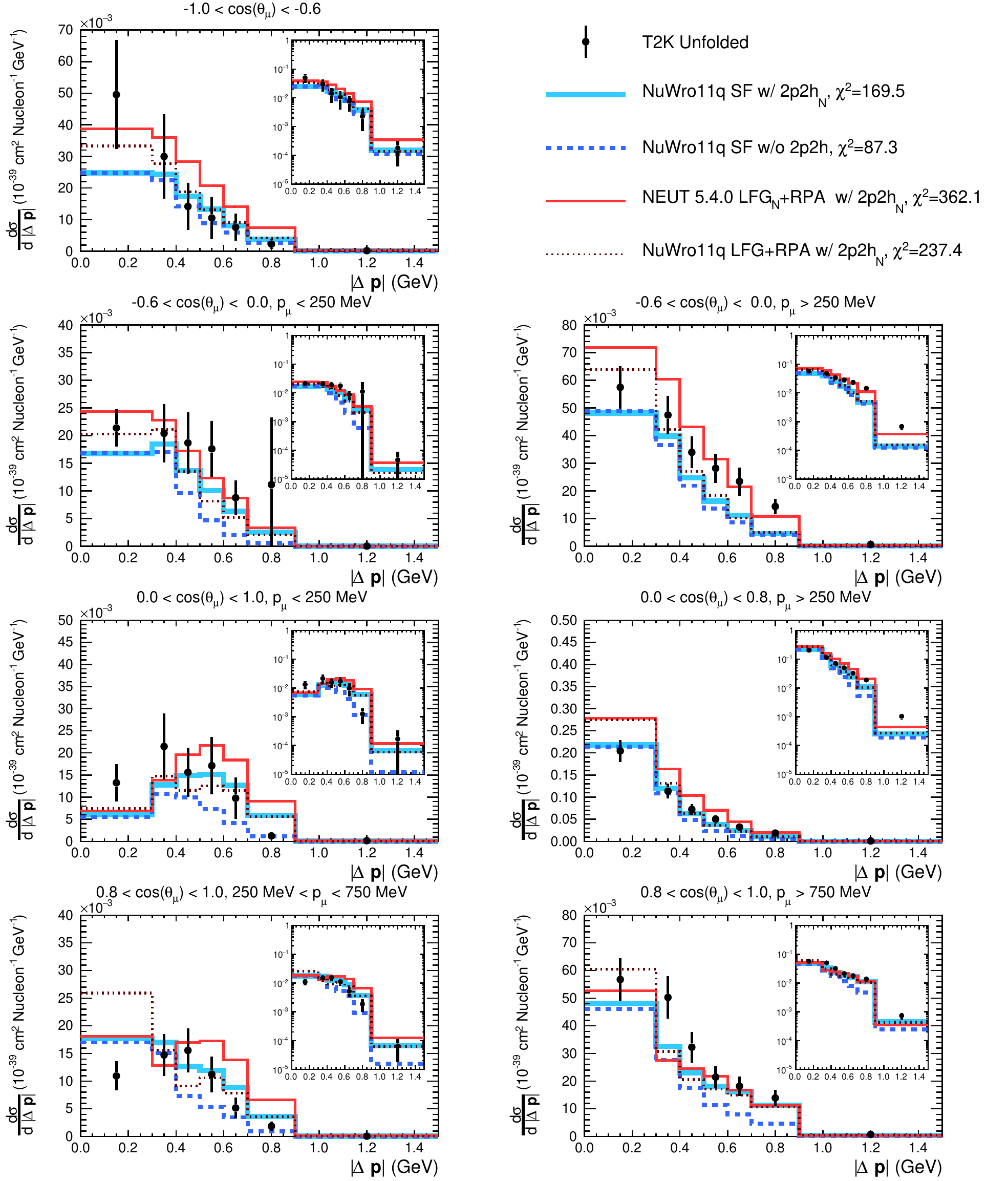} 
  \end{center}
\caption{The extracted differential cross section as a function of the modulus of the inferred and true proton three-momentum difference in different muon kinematic bins, within a restricted proton kinematic phase space, compared to a variety of model predictions: NuWro 11q with the SF nuclear model both with and without an additional ad hoc 2p2h contribution; NEUT 5.4.0, which uses an LFG+RPA model that includes 2p2h predictions; and NuWro 11q with an LFG+RPA nuclear model and a separate 2p2h prediction. 2p2h$_N$ indicates the 2p2h model is an implementation of the Nieves et. al. model of Ref.~\cite{Nieves:2012}. The `N' subscript after LFG indicates that the model is using both a 1p1h and 2p2h prediction from the aforementioned model of Nieves et. al. More details of these models can be found in Sec.~\ref{sec:simulation}. Note that the last bin in each plot is shortened for improved readability. The inlays show the same comparisons on a logarithmic scale. }
\label{fig:inf_model_tmom}
\end{figure*}

\begin{figure*}[!hp]
\begin{center}
 \includegraphics[width=\textwidth]{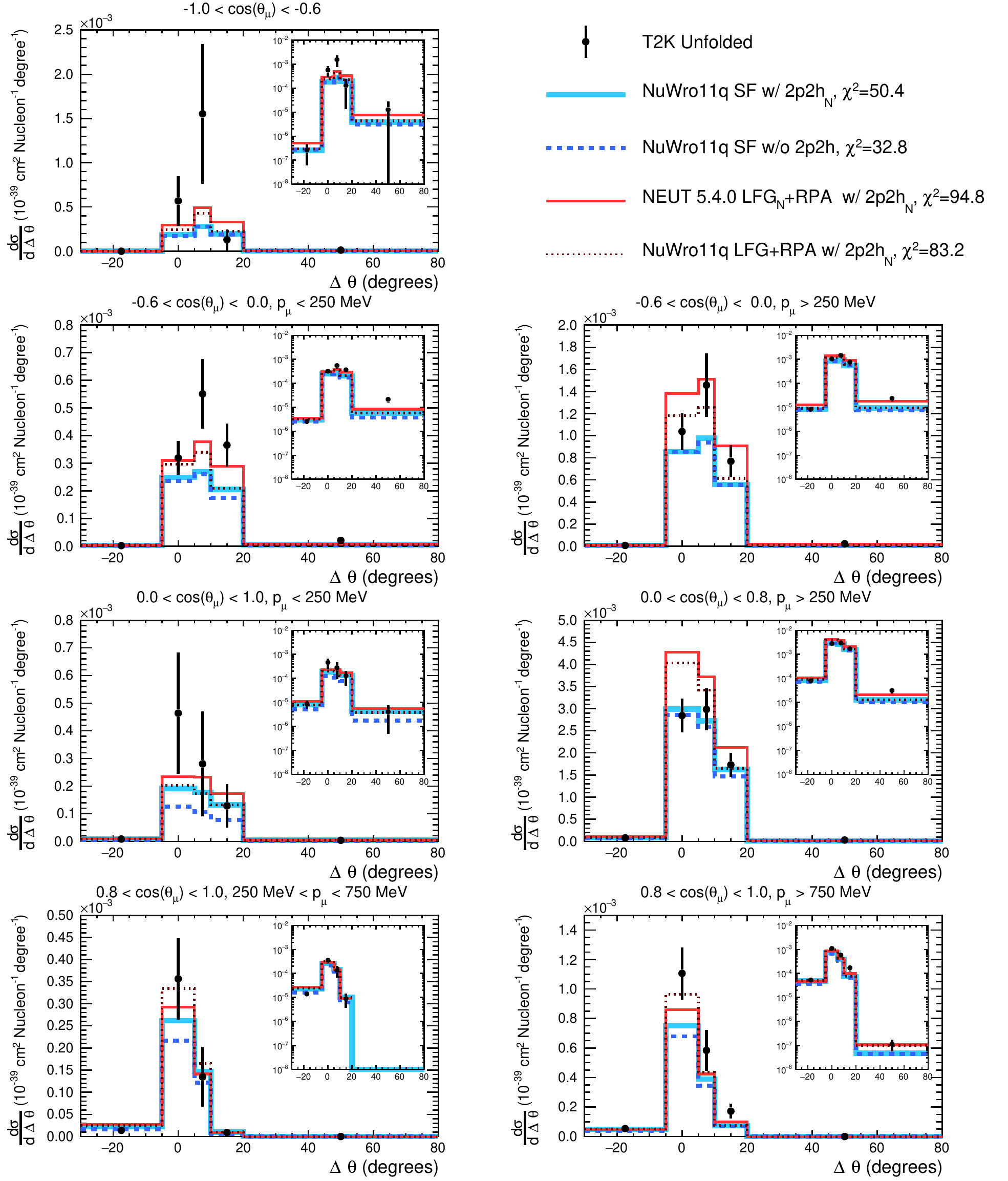} 
  \end{center}
\caption{The extracted differential cross section as a function of the inferred and true proton outgoing-angle difference in different muon kinematic bins, within a restricted proton kinematic phase space, compared to a variety of model predictions: NuWro 11q with the SF nuclear model both with and without an additional ad hoc 2p2h contribution; NEUT 5.4.0, which uses an LFG+RPA model that includes 2p2h predictions; and NuWro 11q with an LFG+RPA nuclear model and a separate 2p2h prediction. 2p2h$_N$ indicates the 2p2h model is an implementation of the Nieves et. al. model of Ref.~\cite{Nieves:2012}. The `N' subscript after LFG indicates that the model is using both a 1p1h and 2p2h prediction from the aforementioned model of Nieves et. al. More details of these models can be found in Sec.~\ref{sec:simulation}. Note that the first and last bin in each plot is shortened for improved readability. The inlays show the same comparisons on a logarithmic scale. }
\label{fig:inf_model_ang}
\end{figure*}

\begin{figure*}[!hp]
\begin{center}
 \includegraphics[width=\textwidth]{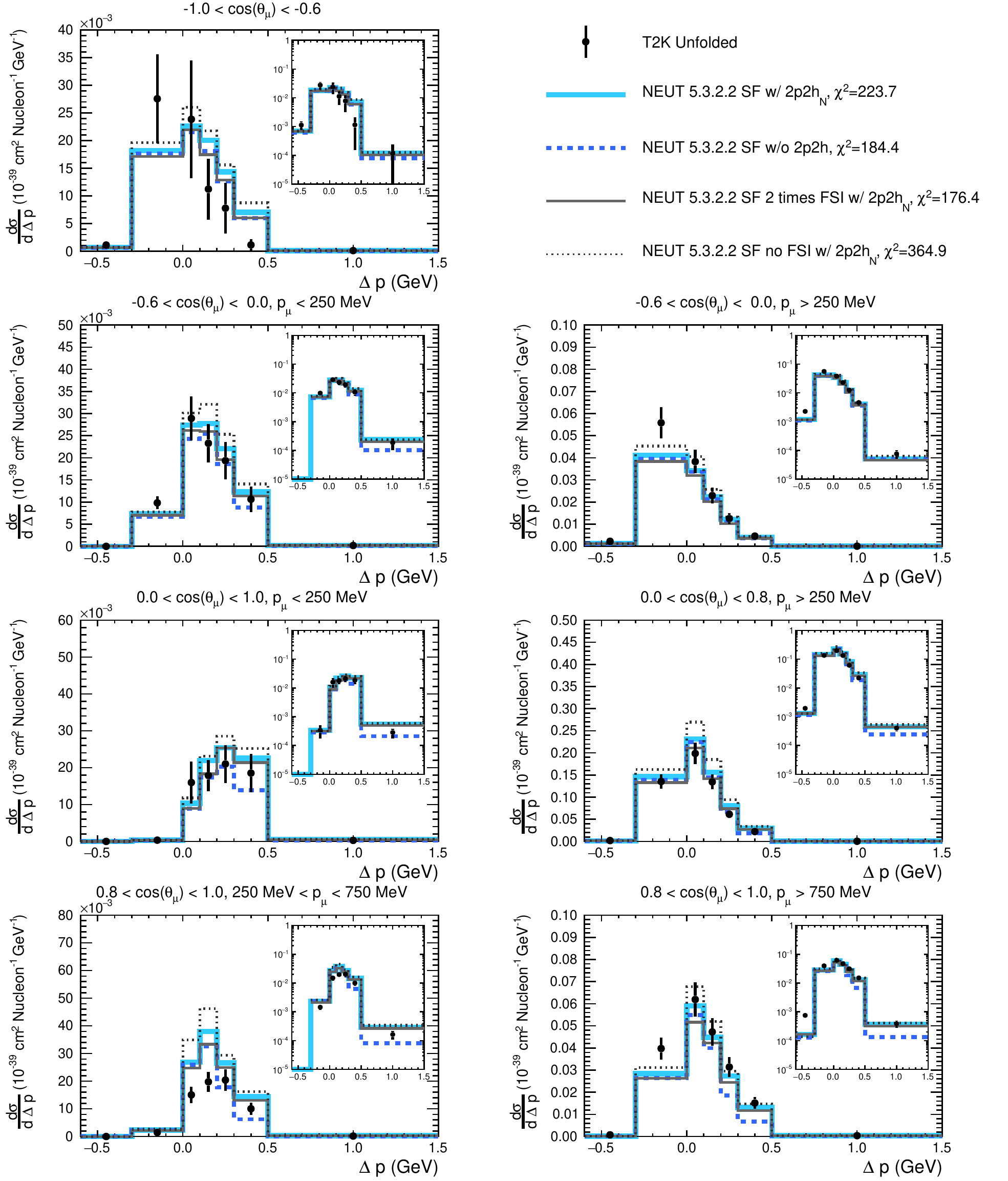} 
  \end{center}
\caption{The extracted differential cross section as a function of the inferred and true proton momentum difference in different muon kinematic bins, within a restricted proton kinematic phase space, compared to the NEUT 5.3.2.2 simulation with various scalings of the mean free path of nucleons undergoing FSI processes to simulate different FSI strengths. A comparison of the NEUT prediction without a 2p2h contribution is also shown. 2p2h$_N$ indicates the 2p2h model is an implementation of the Nieves et. al. model of Ref.~\cite{Nieves:2012}. More details of these models can be found in Sec.~\ref{sec:simulation}. Note that the first and last bin in each plot is shortened for improved readability. The inlays show the same comparisons on a logarithmic scale. }
\label{fig:inf_fsi_mom}
\end{figure*}

\begin{figure*}[!hp]
\begin{center}
 \includegraphics[width=\textwidth]{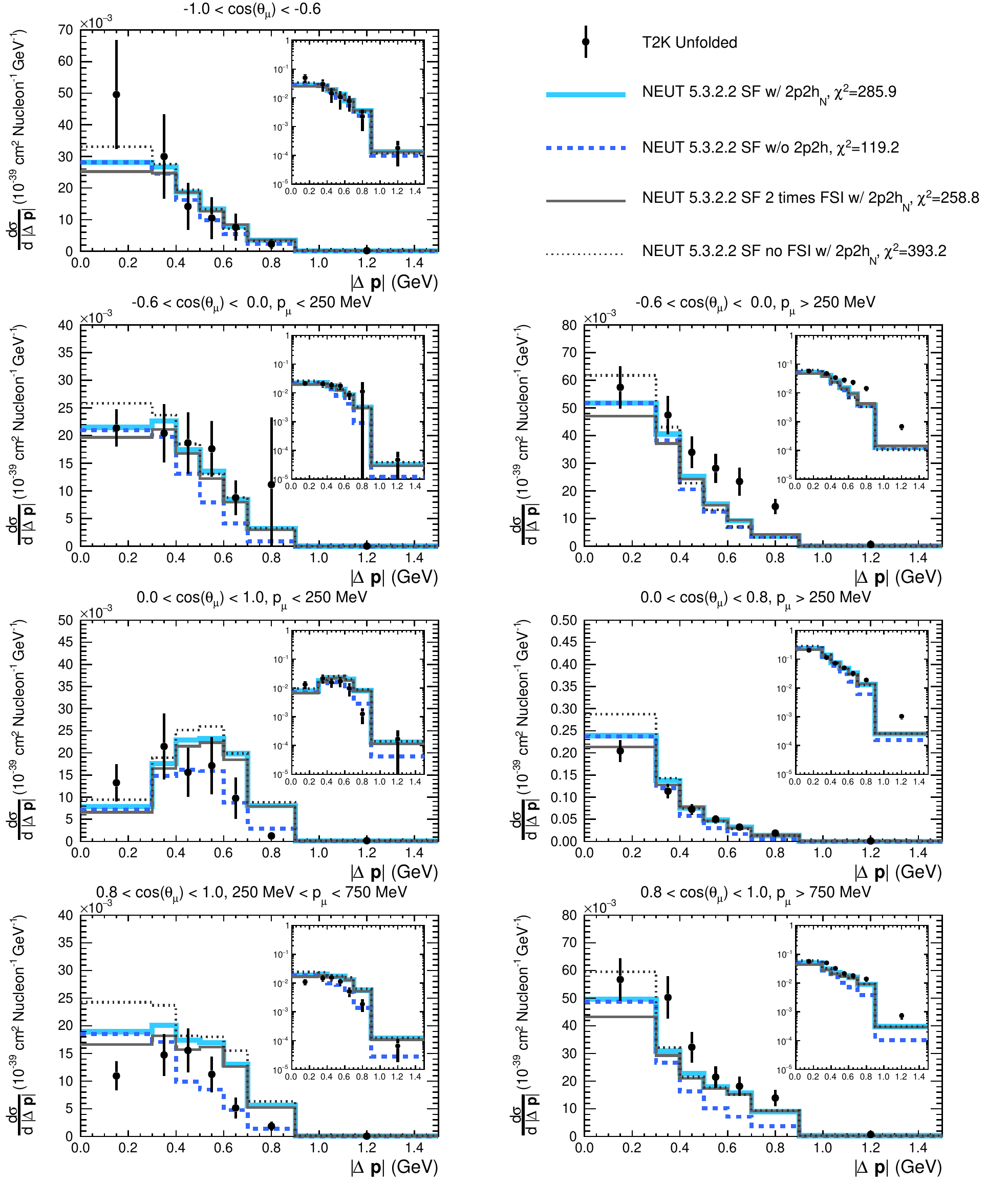} 
  \end{center}
\caption{The extracted differential cross section as a function of the inferred and true proton three-momentum difference in different muon kinematic bins, within a restricted proton kinematic phase space, compared to the NEUT 5.3.2.2 simulation with various scalings of the mean free path of nucleons undergoing FSI processes to simulate different FSI strengths. A comparison of the NEUT prediction without a 2p2h contribution is also shown. 2p2h$_N$ indicates the 2p2h model is an implementation of the Nieves et. al. model of Ref.~\cite{Nieves:2012}. More details of these models can be found in Sec.~\ref{sec:simulation}. Note that the last bin in each plot is shortened for improved readability. The inlays show the same comparisons on a logarithmic scale.}
\label{fig:inf_fsi_tmom}
\end{figure*}

\begin{figure*}[!hp]
\begin{center}
 \includegraphics[width=\textwidth]{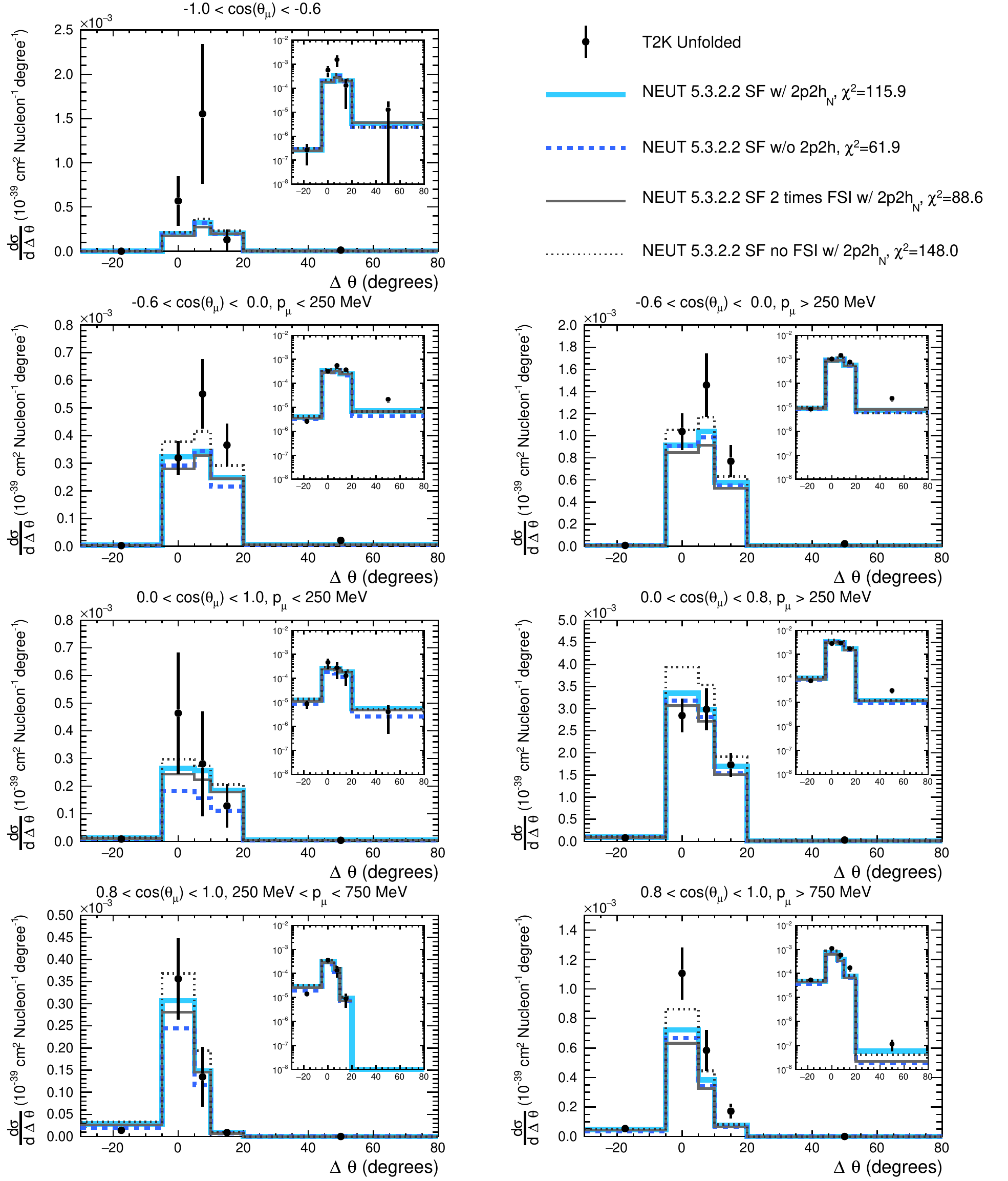} 
  \end{center}
\caption{The extracted differential cross section as a function of the inferred and true proton outgoing-angle difference in different muon kinematic bins, within a restricted proton kinematic phase space, compared to the NEUT 5.3.2.2 simulation with various scalings of the mean free path of nucleons undergoing FSI processes to simulate different FSI strengths. A comparison of the NEUT prediction without a 2p2h contribution is also shown. 2p2h$_N$ indicates the 2p2h model is an implementation of the Nieves et. al. model of Ref.~\cite{Nieves:2012}. More details of these models can be found in Sec.~\ref{sec:simulation}. Note that the first and last bin in each plot is shortened for improved readability. The inlays show the same comparisons on a logarithmic scale.}
\label{fig:inf_fsi_ang}
\end{figure*}

A set of model comparisons, similarly to the one in Sec.~\ref{sec:STV} and \ref{sec:multidiff}, is shown here for the proton inferred kinematic observables. Figures~\ref{fig:inf_model_mom}-~\ref{fig:inf_model_ang} show the results compared to LFG and SF models with and without a 2p2h contribution from the NEUT 5.4.0 and NuWro 11q simulations. Figures~\ref{fig:inf_fsi_mom}-~\ref{fig:inf_fsi_ang} show the impact of altering FSI strength and removing a (Nieves-like) 2p2h contribution within the NEUT 5.3.2.2 simulation.  Value of the comparisons $\chi^2$ are reported in the figures. Comparisons of the results to RFG nuclear models are shown in appendix~\ref{app:infk}. These results are discussed in details in Sec.~\ref{sec:discussion}.

The total CC0$\pi$+Np cross section extracted (within the phase space constraints listed in Tab.~\ref{tab:phaseSpace}) for each of the inferred kinematic observables is given in Tab.~\ref{tab:inftotxsec} alongside a prediction from NuWro 11q. The total cross sections and uncertainties are not identical since the unfolding method couples the systematic parameters between the signal and control regions.  

\begin{table}[h]
\begin{center}
\small
\centering
\begin{tabular}{|c|c|c|}
\hline
Observable & Cross section & NuWro prediction\\
\hline
$\Delta p$                        & $2.169 \pm 0.235$ & $1.916$ \\
$| \Delta \textit{\textbf{p}}|$   & $2.220 \pm 0.243$ & $1.916$ \\
$\Delta \theta$                   & $2.247 \pm 0.244$ & $1.916$ \\
\hline
\end{tabular}
\caption[The total CC0$\pi$+Np cross section in units of $10^{-39}$ cm$^2$ extracted (within the phase space constraints listed in Tab.~\ref{tab:phaseSpace}) for each of the inferred kinematic observables alongside the prediction from NuWro 11q using an SF nuclear model and with the 2p2h model of Nieves et. al.~\cite{Nieves:2012}.]{\label{tab:inftotxsec}The total CC0$\pi$+Np cross section in units of $10^{-39}$ cm$^2$ extracted (within the phase space constraints listed in Tab.~\ref{tab:phaseSpace}) for each of the inferred kinematic observables alongside the prediction from NuWro 11q using an SF nuclear model and with the 2p2h model of Nieves et. al.~\cite{Nieves:2012}. }
\end{center}
\end{table}

%\clearpage
%\input{Discussion}
\section{Discussion of the results}
\label{sec:discussion}

In all of the results shown in Sec.~\ref{sec:anaDescription} $\chi^2$ statistics are quoted to indicate the agreement between each model and the result, calculated using the full covariance matrix from the cross-section extraction. However, since no model describes the results well over the entire phase space, these $\chi^2$ statistics should be treated carefully: 
such quantitative estimation of the global data-model agreement is reliable only when a model is capable of describing the whole phase-space of the measurement and ideally when it is fit to the result using a well-motivated parametrisation of the signal predictions. Moreover, since multiplicative normalisation uncertainties make a significant contribution to the covariance of the results, such $\chi^2$ statistics can also suffer from Peelle's Pertinent Puzzle~\cite{ppp} and may therefore not accurately characterise the agreement (this does not effect shape-only $\chi^2$). For this reason these statistics should only be taken as an approximate metric and this section will mainly focus on discussing and interpreting discrepancies between the model and the simulations in specific regions of kinematic phase-space, rather than on the overall concurrence indicated by the $\chi^2$. However, when doing this it is important to be aware of the significant correlations between each bin in the extracted results. The significant detector smearing in the STV leads to adjacent bins being fairly anti-correlated (up to $\sim$35\%, with $\delta p_T$ the most affected) whilst the large flux normalisation uncertainty correlates all other bins (by $\sim$10\% to 35\%). Due to the larger regularisation strength in the inferred kinematics analysis and the coarser binning in the multi-differential analysis, the anti-correlations are generally much less prominent and most bins are positively correlated by the flux. The full covariance matrices are available in the data release for these results. A summary of the full and shape only $\chi^2$ for the regularised and unregularised STV results are provided and discussed in appendix~\ref{app:unregchi2}.

The measurement of proton multiplicity and proton and muon kinematics from the multidifferential analysis (Figs.~\ref{fig:MultiDiff_0p},~\ref{fig:MultiDiff_Np}) 
shows the phase space regions where the present models fail to describe the data. From this measurement it can be seen that, when there is no proton above threshold in the final state, while the SF prediction gives a reasonable agreement with the extracted result in the region ($0.6<$ cos $\theta<0.8$) , it clearly underestimates the cross section in the region of backward muon angle and overestimates it in the region with forward muons for intermediate muon momentum (0.5-0.8 GeV). Conversely, the SF model describes very well the rate of events with one proton above the momentum threshold in the final state, where a slight preference for the presence of 2p2h is observed in the region with forward muons (cos $\theta > 0.8$).

The LFG predictions from NEUT~5.4.0 and NuWro~11q differ when describing events without protons above threshold in the final state in the region with muons at high angle. Here the NEUT implementation describes the results well, while NuWro underestimates the cross section, in a similar manner to its SF prediction. In the phase space with intermediate and low muon angle, both LFG implementations describe the result reasonably well. However, both also overestimate the cross section with one above-threshold proton in the final state. Indeed, the two LFG models predict a tendency in proton multiplicity to have a larger rate of events with one proton with respect to events without protons, while the extracted result has the opposite behaviour.

Finally, in the bin with two or more protons in the final state, the result prefers the SF with 2p2h over the case without 2p2h. Here it can also be seen that the two implementations of LFG give very different predictions, thereby demonstrating the importance of 1p1h modelling also for the events with multiple protons.

It is also interesting to compare the effect of FSI and 2p2h in these distributions, as shown in Figs.~\ref{fig:MultiDiff_0p_FSIComp}-\ref{fig:MultiDiff_Np_FSIComp} in appendix~\ref{app:multidim}, where SF with and without 2p2h is compared together with different strengths of FSI. A larger FSI strength tends to redistribute events between the bins of proton multiplicity: larger FSI increases the rate of events without protons above the momentum threshold and decreases the number of events with one or more protons above it. On the other hand, 2p2h tends to increase the cross section for all proton multiplicities. Since the measurement of the shape of the cross section in proton multiplicity is well known (the uncertainties are dominated by effects that fully correlate all three bins, mostly due to the flux) these results may offer an interesting capability to separate the effects of proton FSI and 2p2h. However, more robust predictions of the outgoing proton kinematics in 1p1h and 2p2h events after FSI would be needed in order to exploit the proton multiplicity measurement to expose a possible significant 2p2h excess in the result. It is particularly striking that no model is able to simultaneously describe accurately the events with and without above-threshold protons in the final state, the LFG being more in agreement with the result in former and the SF being in better agreement in the latter.

The results are also compared in Figs.~\ref{fig:MultiDiff_0p_rfgComp}-\ref{fig:MultiDiff_Np_rfgComp} with RFG models as implemented in NuWro 11q and NEUT 5.3.2.2 and with the RFG with Bodek-Ritchie corrections as implemented in GENIE 2.12.4. GENIE overestimates the cross section in most of the phase space, except for backward-going muons. It also reproduces the same trend as the extracted result in the multiplicity plot, showing more events without a final state proton above the threshold.  
The other RFG implementations in NuWro and NEUT behave similarly to the NuWro LFG model.

It also is interesting to consider the impact of the RES pion-production contribution to the signal (where the pion is absorbed inside the nucleus). In general NEUT 5.3.2.2 predicts that the RES contribution to the cross section is always less than about 5\% except when there is one above-threshold proton in the final state and $\cos(\theta_{\mu})>0.8$, where it reaches around 15\%, and when there is more than one proton in the final state where the cross section is dominated almost equally be RES and 2p2h interactions. Although there is large theoretical uncertainty on this RES contribution (notably from nuclear effects in pion production and in pion FSI), within most bins unrealistically large changes would be required to alter the interpretation of the results.

In general, the interpretation of the aforementioned discrepancies between the result of the multidifferential analysis and different simulations is not straightforward since the measured cross section is affected by multiple initial state and final state nuclear effects which cannot be easily separated in the momentum and angular kinematic distributions. The STV are expressly designed in order to unambiguously distinguish the impact of different nuclear effects, their measurement therefore offers a more transparent interpretation of such discrepancies.
 
%The good agreement with NuWro SF+2p2h in Fig.~\ref{fig:STV_NuWro}, with respect to the clear overestimation of NEUT vXXXX SF+2p2h 
%in Fig.~\ref{fig:STV_NEUT6D}, shows the importance of the details in the different SF implementations in the two MC.

%\textcolor{red}{XL: I think we can remove all $\dphit$ plots, or put them in Appendix}

The comparison of the STV distributions to different CCQE models, as implemented in NuWro, are shown in the left plots of Fig.~\ref{fig:STV_NuWro}. Given the definition of $\dptv$ (Eq.~\ref{eq:dptconvolution}), the dominant contribution below the Fermi surface ($\dpt \sim 230$ MeV) is CCQE with limited FSI strength (as can be seen in the left plots of Fig.~\ref{fig:STV_NEUT6D}), thus the Fermi motion determines the bulk structure of $\dpt$, thereby allowing it to act as a probe of the initial-state nucleon. The measured $\dpt$ distribution strongly disagrees with the RFG prediction: the prominent imprint of the cliff at the Fermi surface, a characteristic of RFG, is firmly disfavoured by this result.

% SD: I don't think this works when the shape is in such good agreemnt with the LFG models. 

%It is also found that the LFG prediction overestimates this soft region of the $\dpt$ distribution, which within an LFG model is dominated by nucleons with small initial momenta at a large radius inside the nucleus where the medium density is low, therefore the discrepancy of the LFG model with the result in this region indicates a mismodeling of the nucleon dynamics close to the nucleus surface.

It is interesting to note that both Fermi gas models (LFG and RFG) exhibit similar excesses over the result, but at different kinematic regions. Indeed, considering shape-only comparisons of data to various simulations in the right plots of Fig.~\ref{fig:STV_NuWro}, it can be seen that the LFG predictions well describe the differential distribution, but are plagued by an overestimation of the overall cross section, even if RPA corrections are applied. Such a normalisation discrepancy could come from a general overly large CCQE cross section or weak proton FSI keeping too many protons above signal threshold. The latter is particularly further supported in the proton multiplicity plot in Fig.~\ref{fig:MultiDiff_Np}, where an increase in proton FSI would migrate events from the 0 proton bin to the 1 proton bin thereby bringing the prediction into better agreement with the results. The fact that the GiBUU (LFG) CCQE prediction in Fig.~\ref{fig:STV_GENIE}, which largely differs from the NuWro and NEUT LFG models through its FSI modelling, seems to provide a normalisation in good agreement with the results adds additional evidence that the normalisation discrepancy seen in the NEUT and NuWro LFG normalisations is at least partially related to FSI modelling. 

In general it seems that the nucleon dynamics for $\dpt\lesssim400$~MeV, are better described by SF than Fermi gas models. The consistency between SF and the result at $\dpt\sim300$~MeV suggest that the nucleon-nucleon correlations captured by SF are required. Future measurements of the STV with higher statistics may allow further exploration of the nature of such correlations. 

Above $\sim400$~MeV, $\dpt$ is driven by nucleon-nucleon correlations and FSI effects, so it is not surprising that the predictions from the Fermi gas and SF models become more similar. The SF model in this region, as it is implemented in the simulations, is not fully consistent since a 2p2h contribution computed for an LFG model is added on top of the CCQE SF. The SF model without a 2p2h contribution is also shown for comparison: within the hard tail of $\dpt$ and $\dphit$ the result clearly indicates the need for additional strength, consistent with that from a 2p2h contribution, beyond the nucleon-nucleon correlations already included in the SF model.

Both RFG and LFG models have consistent predictions regarding the total cross section and the $\dalphat$ distribution, which represents to good approximation the direction of the initial nucleon momentum $\pni$. Furthermore, the distributions of $\dalphat$ show a significant difference between Fermi gas and SF models in the shape. In fact, in NuWro predictions the discrepancy at low $\dalphat$ between Fermi gas models and the data is caused by RPA (see the left plots of Fig.~\ref{fig:STV_NuWroRPAComp}), without which the shape would be consistent.

The left hand plots of Fig.~\ref{fig:STV_NEUT6D} show the comparison of STV with NEUT 5.3.2.2 model, demonstrating in more detail how the 2p2h contribution is clearly located in the tails of $\dpt$ and $\dphit$ , where the agreement with the data is good and the 2p2h contribution seems essential. It also highlights the CCQE dominance in the bulk of the $\dpt$ distribution, where the model tends to overestimate the data.  

The distribution of $\dpt$ beyond 400~MeV and the shape of $\dalphat$ are also sensitive to intra-nucleus momentum exchange such as 2p2h, as already discussed, as well as FSI. As can be seen in right plots of Fig.~\ref{fig:STV_NEUT6D}, in order to bring the NEUT 5.3.2.2 model in to agreement with the data, the proton FSI strength must be increased by reducing the mean free path between re-interactions inside the nucleus by a a factor of two. Although it is challenging to draw firm conclusions from external measurements of electron-nucleus scattering, this appears to be around the maximum plausible variation based on the data~\cite{Dutta:2012ii}. Moreover, in the present semi-classical model of FSI implemented in all the simulations, the FSI mainly affect the probability of observing the outgoing proton (here defined as proton momentum above 450 MeV, see Tab.~\ref{tab:phaseSpace}) thus changing the integrated cross section. Only small modifications to the shape of the STV distributions are visible. As can be seen in Fig.~\ref{fig:STV_NEUT6D}, as the FSI strength increases, both $\dpt$ and $\dphit$ spectra become harder---with depletion and enhancement in regions of small and large imbalances, respectively---as is expected from the intra-nucleus momentum transfer during FSI. Nevertheless, the enhancement in this particular FSI model is much smaller than that caused by the presence of 2p2h in the region of high transverse kinematic imbalance, so it is far too small to invalidate the evidence for a 2p2h contribution.

The GENIE predictions in the left plots of Fig.~\ref{fig:STV_GENIE} strongly overestimate the data in the collinear regions where the proton  momentum aligns with the three-momentum transfer in the transverse plane (see $\vec{p}^\textrm{p}_\textrm{T}$ and $\vec{q}_\textrm{T}$ in Fig.~\ref{fig:STVdiagram}), i.e. in regions where $\dpt\to 0$, $\dphit\to 0$, and  $\dalphat\to 0$ and 180 degrees. Such overprediction originates from the elastic interaction of GENIE's widely used hA FSI model~\cite{Lu:2015tcr}. Moreover, compared to other 2p2h models, the empirical MEC model in GENIE features a much stronger enhancement in regions of large imbalances, where the overall predictions clearly overestimate the data.
%Further comparisons to GIBUU and GENIE MC, divided by interactions, are shown in the Appendix. 

The GiBUU predictions (the right plots of Fig.~\ref{fig:STV_GENIE}) provide an integrated cross section in good agreement with the data but it is characterised by one of the hardest $\dpt$ and $\dphit$ distributions of all the simulations, as can be seen in right plot Fig.~\ref{fig:STV_NuWro}, which is in disagreement with the measured shape of the observables. However, it should be noted that there is theoretical motivation to reduce GiBUU's 2p2h model strength by a factor of two~\cite{Gallmeister:2016dnq} and an exploration of the results presented here within a more recent version of GiBUU with such reduced 2p2h has recently shown much better agreement~\cite{Dolan:2018sbb}.  

Fig.~\ref{fig:STV_GENIE} and the left hand plots of Fig.~\ref{fig:STV_NEUT6D} also demonstrate that the contribution from RES pion-production within the STV restricted phase-space is universally small relative to the 2p2h contributions (except perhaps in the final bins of $\dpt$ and $\dphit$). For this reason even relatively large changes in the prediction for the RES contribution to the result would not invalidate the conclusions regarding 2p2h. 

Finally Figs.\ref{fig:inf_model_mom}-\ref{fig:inf_model_ang} show the results of the cross section measurement as a function of the proton inferred kinematics  ($\Delta p_p, \Delta\theta_p, |\Delta \textbf{p}_p|$, as in Eq.~\ref{eq:inferred}), compared to different models. The most precise measurements come from the region with largest statistics: $0.0<$ cos $\theta<0.8$, $p_\mu>250$ MeV, which corresponds to the region of intermediate $Q^2$. Similarly to what was observed in the STV analysis, this region is best described by the SF model and in the high $|\Delta \overrightarrow{p}_p|$ tail a net preference for the presence of 2p2h contribution is visible. Such indication is independent of the strength of FSI effects, as shown in Figs.\ref{fig:inf_fsi_mom}-\ref{fig:inf_fsi_ang}, where there is also a small preference for larger FSI.

The other regions of muon kinematics are not all consistently well described by any of the models, as can be seen by the high $\chi^2$ values. Depending on the kinematic region and on the observable considered, the LFG or SF may better describe the data. It is also interesting to note that both the very forward-going and low momentum muon kinematic bins suggest a multitude of regions of $\Delta p_p$ and $|\Delta \textbf{p}_p|$ which are largely dominated by the 2p2h contribution and largely independent of FSI variation, as can be seen in figures Figs.\ref{fig:inf_model_mom}-\ref{fig:inf_model_tmom} and Figs.\ref{fig:inf_fsi_mom}-\ref{fig:inf_fsi_tmom} respectively. However whether the result favours these large 2p2h contributions depends on the kinematic bin: for example both the NEUT and NuWro SF and 2p2h predictions largely agree with the result in the $0.8<$ cos $\theta<1.0$, $p_\mu>750$ MeV bin, but are quite different in the $0.8<$ cos $\theta<1.0$, $250 < p_\mu < 750$ MeV bin. This difference is understood to stem from the substantial RES pion-production contribution to these bins, which differs considerably in NuWro and NEUT. These bins therefore offer a powerful probe of the CCnonQE contribution, but specific conclusions are difficult due to the large uncertainties in both 2p2h and nuclear effects in RES interactions, making these results complementary to the multi-differential and STV analyses that mostly have a very small RES contribution.

In appendix~\ref{app:infk} a comparison to RFG models is reported which, consistently with what is observed in the STV analysis, generally gives worse agreement with the result than either LFG or SF. In particular the results clearly show that both the NEUT 5.3.2.2 RFG+RPA model and the GENIE 2.12.4 BRRFG model, which are similar to the nominal models used in recent T2K~\cite{Abe:2017vif} and NO$\nu$A~\cite{Ahmad:2001an} oscillation analyses respectively, do not describe the result well.  This conclusion is supported by also considering the poor agreement between the multi-differential results and these models seen in appendix~\ref{app:multidim}, and even more so considering the aforementioned strong contention between the STV analysis results and the RFG and empirical MEC models.

\section{Conclusions \label{sec:conclusions}}

This paper has presented the measurement with ND280 data of muon and proton multi-differential 
cross section, as well as their kinematic correlations,
for charged-current neutrino-nucleus scattering without pion in the final state.
The muon-proton correlations in the final state are measured through two new sets of variables never
used before as observables in neutrino cross section measurements: the STV and the proton inferred kinematics.
The analysis selection separates events without protons, with 1 proton or with more than 1 proton in the final state above a 500 MeV momentum threshold, thus enabling also the measurement of proton multiplicity.
Particular care has been taken to minimise the model-dependence of the measurement
in efficiency corrections, background subtraction and cross section evaluation,
thus enabling an unbiased and large set of model comparisons with the results.
For the first time in neutrino scattering measurements the concept of data-driven regularisation is introduced to achieve a result that is easy to interpret but contains minimal bias. Overall the results offer a powerful new probe of nuclear effects and that their exploration of kinematic imbalances facilitates a method of separating, at least partially, the different contributions of a CC0$\pi$ measurement.
Since prediction power of proton kinematics in neutrino-interaction simulations is still poor, it remains challenging to draw firm quantitative conclusions. Nevertheless, an extensive comparison with generator predictions has allowed interesting qualitative conclusions to be drawn. As briefly summarised below, the three analyses suggest similar conclusions.

The RFG model is able to describe only a very limited region of phase space (and only when there is no above-threshold proton in the final state) and is categorically disfavoured when considering the result in $\dpt$. 
The LFG prediction shows slightly better agreement with data than RFG when considering interactions with above-threshold protons,
especially considering the distribution of STV, but it still overestimates the soft part of the STV spectrum.
A more consistent LFG implementation, such as the one in NEUT 5.4.0, gives improved
results. It provides better agreement with both the STV distributions and in the region without above-threshold protons in the final state and with large muon angle, where no other model is able to describe the result.
For the events with one or more above-threshold protons in the final state, the best description
of the data is given by the SF model. Beyond the nucleon-nucleon correlations already included in SF, 
a clear requirement for a 2p2h contribution is visible in the hard tail of STV distributions
($\dpt$ and $\dphit$) and of the $|\Delta \overrightarrow{p}_p|$
distribution. The requirement for a large 2p2h contribution to the result remains even with dramatic variations of the semi-classical FSI models available in the simulations.
On the other hand, GENIE's ``empirical MEC'' 2p2h model appears to substantially
overemphasize the hard tail of the STV. The prominent features of the STV predicted by GENIE's hA FSI model (driven by its elastic component) are in very poor agreement with the results. 

%A less significant indication of 2p2h in data can also be seen in the 
%multi-differential cross section with one proton in the final state.
The results with one proton in the final state, when compared to the SF model,
suggest the need for stronger FSI effects, at the limit to what is allowed by external data
of proton-nucleus scattering.
The measurement of proton multiplicity can in principle disentangle 2p2h from FSI effects, with
the former increasing the cross section in all bins of proton
multiplicity while the latter redistributing events between different bins.
Currently the primary limitation is the absence of a model able to properly describe
both the events with and without above-threshold protons in the final state.

The measurement of neutrino-nucleus interactions with a pionless final state with protons
clearly shows the potential to provide an even more detailed characterisation of nuclear effects in neutrino-nucleus scattering in the future. To this aim, larger statistics are needed, alongside more robust predictions of outgoing proton kinematics in 2p2h and FSI models.

\section*{Acknowledgements}
We thank the J-PARC staff for superb accelerator performance. We thank the 
CERN NA61/SHINE Collaboration for providing valuable particle production data.
We acknowledge the support of MEXT, Japan; 
NSERC (Grant No. SAPPJ-2014-00031), NRC and CFI, Canada;
CEA and CNRS/IN2P3, France;
DFG, Germany; 
INFN, Italy;
National Science Centre (NCN) and Ministry of Science and Higher Education, Poland;
RSF, RFBR, and MES, Russia; 
MINECO and ERDF funds, Spain;
SNSF and SERI, Switzerland;
STFC, UK; and 
DOE, USA.
We also thank CERN for the UA1/NOMAD magnet, 
DESY for the HERA-B magnet mover system, 
NII for SINET4, 
the WestGrid and SciNet consortia in Compute Canada, 
and GridPP in the United Kingdom.
In addition, participation of individual researchers and institutions has been further 
supported by funds from ERC (FP7), H2020 Grant No. RISE-GA644294-JENNIFER, EU; 
JSPS, Japan; 
Royal Society, UK; 
the Alfred P. Sloan Foundation and the DOE Early Career program, USA.

\appendix
\section{Further comparisons to models \label{app:furtherModelComp}}

\subsection{Multi-differential cross section measurement \label{app:multidim}}

\begin{figure*}[h!]
\begin{center}
  \end{center}
 \includegraphics[width=1.0\textwidth]{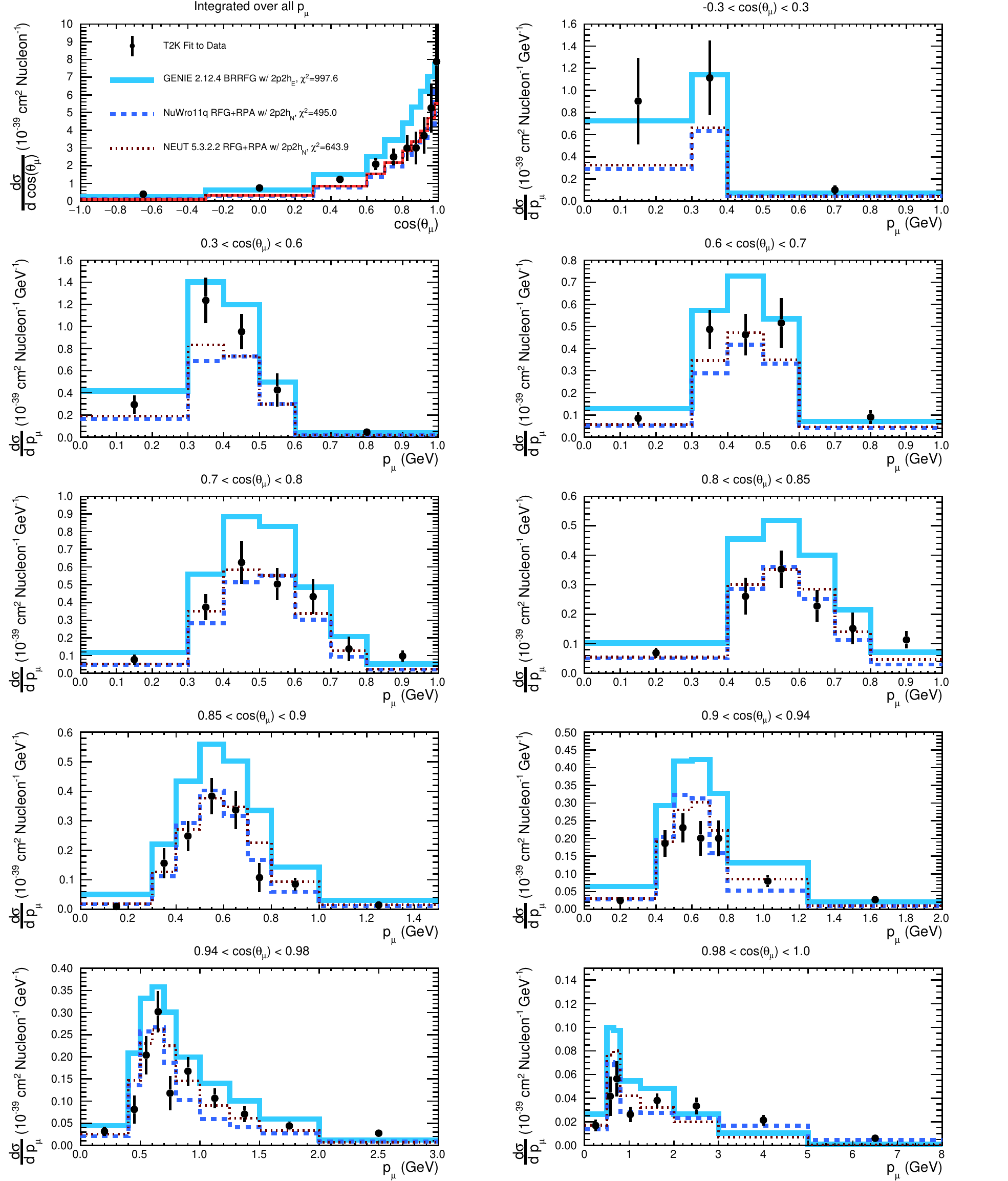} 
\caption{The extracted differential cross section as a function of the muon kinematics in the sample without any protons (with momenta above 500 MeV) compared to GENIE 2.12.4, NuWro 11q and NEUT 5.3.2.2  predictions which utilise an RFG or RFG+RPA nuclear model. GENIE's RFG model also includes the empirical correction from Bodek and Ritchie (BRRFG)~\cite{Bodek:1981wr}. The NEUT and GENIE predictions shown here are similar to those used as a starting point for the T2K and NO$\nu$A experiment's oscillation analyses respectively. More details of these models can be found in Sec.~\ref{sec:simulation}. Note that the last bin in each momentum plot is shortened for improved readability.}
\label{fig:MultiDiff_0p_rfgComp}
\end{figure*}

\begin{figure*}[h!]
\begin{center}
 \includegraphics[width=\textwidth]{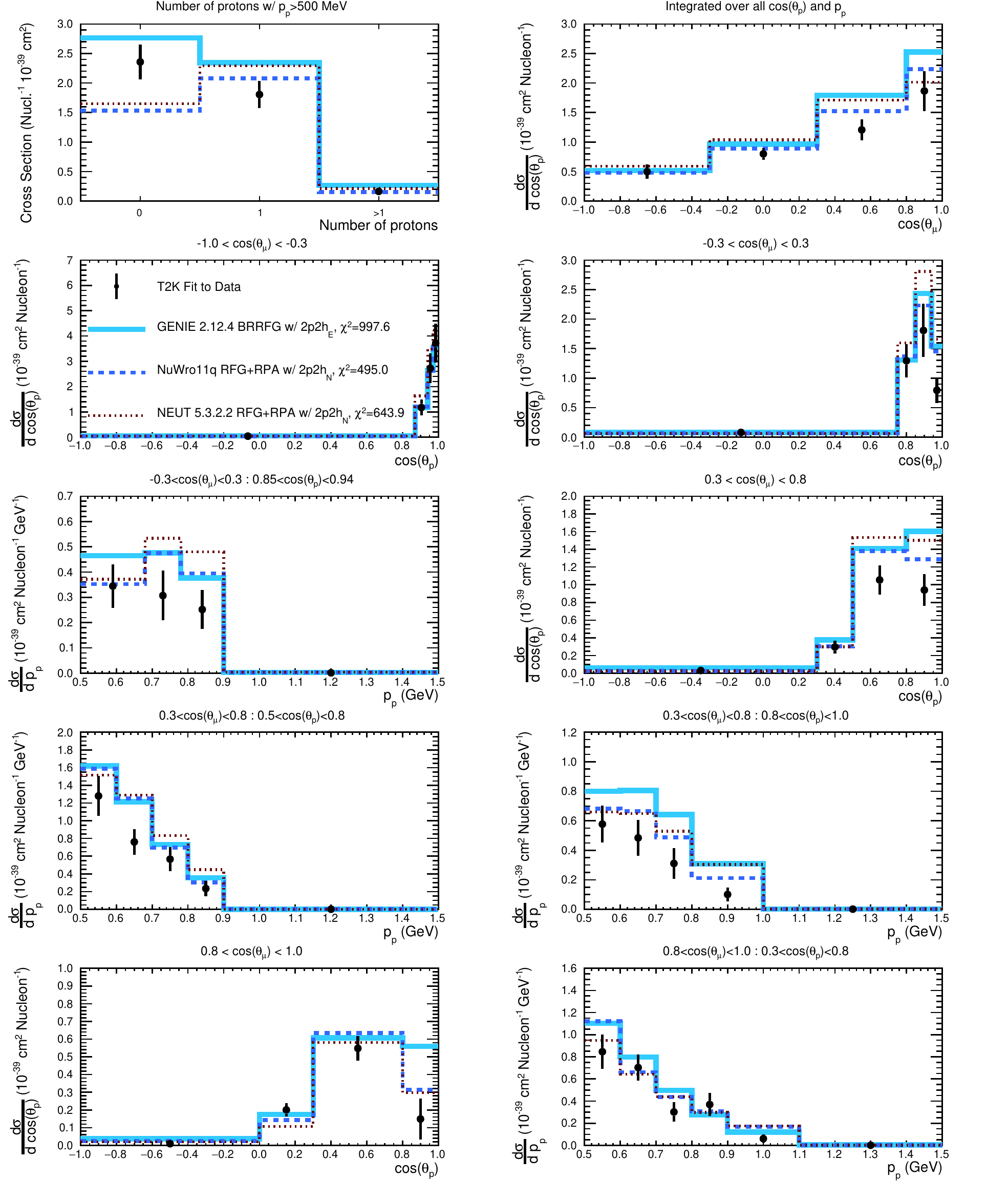} 
  \end{center}
\caption{The extracted differential cross section as a function of the proton multiplicity (top left) and of the proton and muon kinematics in the sample with exactly one proton (with momentum above 500 MeV) compared to GENIE 2.12.4, NuWro 11q and NEUT 5.3.2.2  predictions which utilise an RFG or RFG+RPA nuclear model. GENIE's RFG model also includes the empirical correction from Bodek and Ritchie (BRRFG)~\cite{Bodek:1981wr}. The NEUT and GENIE predictions shown here are similar to those used as a starting point for the T2K and NO$\nu$A experiment's oscillation analyses respectively. The 2p2h subscript indicates whether it is an implementation of the Nieves et. al. model of Ref.~\cite{Nieves:2012} (N) or the GENIE empirical 2p2h model (E). The same models are also compared to the cross section as a function of proton multiplicity. More details of these models can be found in Sec.~\ref{sec:simulation}. Note that the last bin in each momentum plot is shortened for improved readability. }
\label{fig:MultiDiff_Np_rfgComp}
\end{figure*}

\begin{figure*}[h!]
\begin{center}
  \end{center}
 \includegraphics[width=1.0\textwidth]{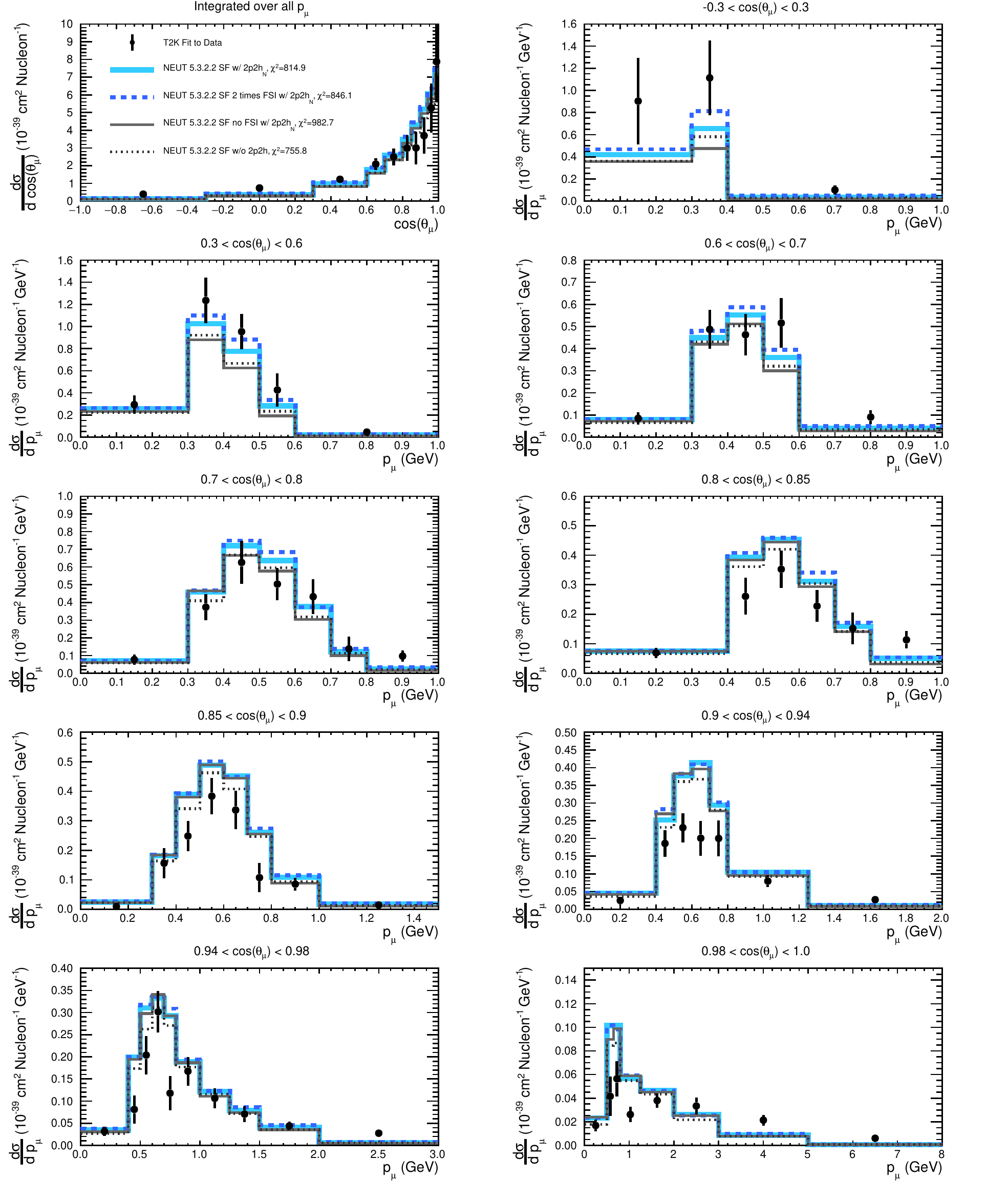} 
\caption{The extracted differential cross section as a function of the muon kinematics in the sample without any protons (with momenta above 500 MeV) compared to NEUT 5.3.2.2 using the SF nuclear model with and without a 2p2h prediction and with zero or doubled FSI strength (achieved with alterations to the mean free path of FSI). 2p2h$_N$ indicates the 2p2h model is an implementation of the Nieves et. al. model of Ref.~\cite{Nieves:2012}. More details of these models can be found in Sec.~\ref{sec:simulation}. Note that the last bin in each momentum plot is shortened for improved readability.}
\label{fig:MultiDiff_0p_FSIComp}
\end{figure*}

\begin{figure*}[h!]
\begin{center}
 \includegraphics[width=\textwidth]{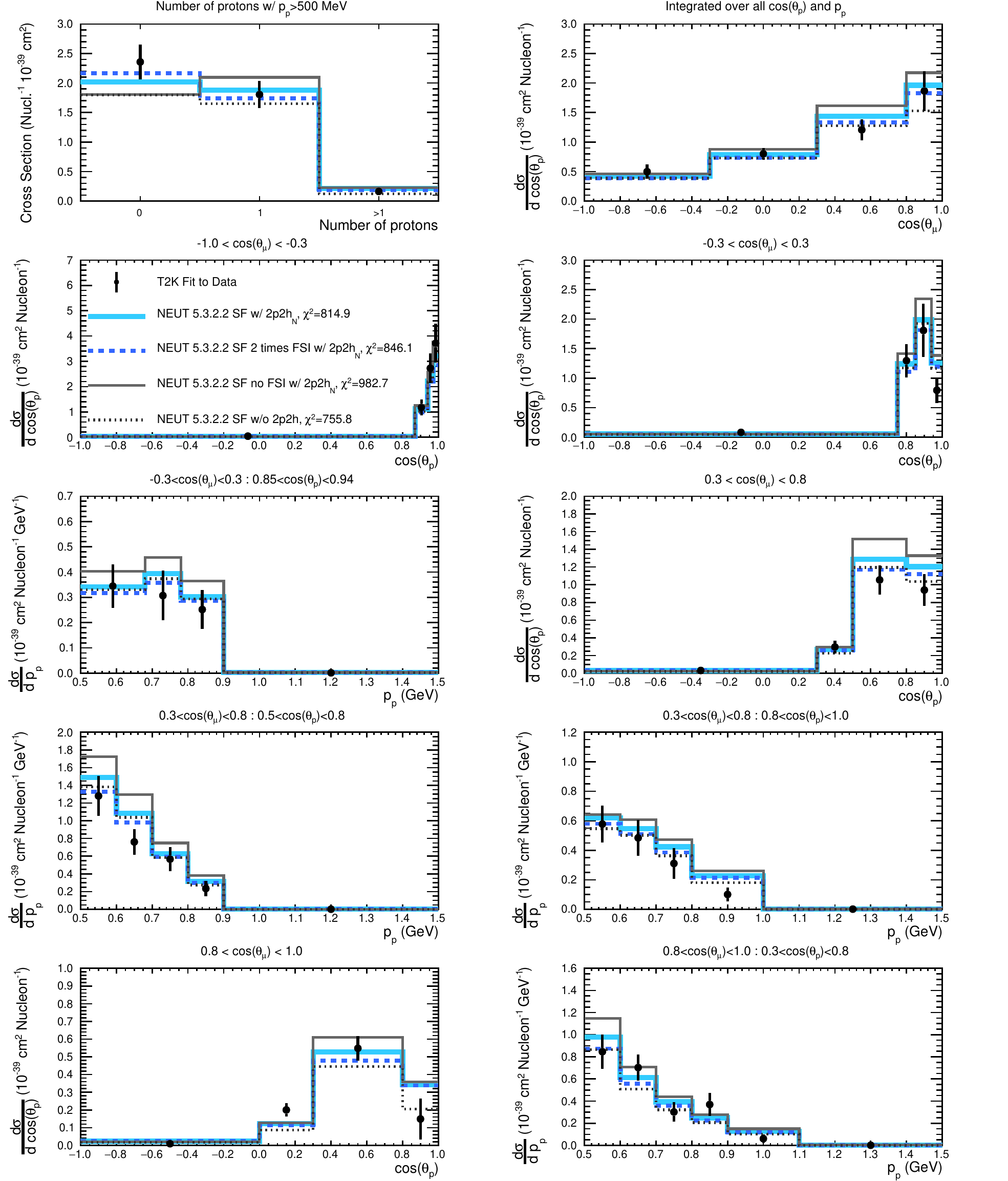} 
  \end{center}
\caption{The extracted differential cross section as a function of the proton multiplicity (top left) and of the proton and muon kinematics in the sample with exactly one proton (with momentum above 500 MeV) compared to NEUT 5.3.2.2 using a Benhar Spectral Function nuclear model with and without a 2p2h prediction (based on the Nieves model) and with zero or doubled FSI strength (achieved with alterations to the mean free path of FSI). The same models are also compared to the cross section as a function of proton multiplicity. 2p2h$_N$ indicates the 2p2h model is an implementation of the Nieves et. al. model of Ref.~\cite{Nieves:2012}. More details of these models can be found in Sec.~\ref{sec:simulation}. Note that the last bin in each momentum plot is shortened for improved readability. }
\label{fig:MultiDiff_Np_FSIComp}
\end{figure*}

As discussed in Sec.~\ref{sec:multidiff} this appendix provides additional comparisons of the multidifferential analysis results to various model predictions. Figures~\ref{fig:MultiDiff_0p_rfgComp} and~\ref{fig:MultiDiff_Np_rfgComp} compare the results to various models that use an RFG/RFG+RPA nuclear model. Figures~\ref{fig:MultiDiff_0p_FSIComp} and~\ref{fig:MultiDiff_Np_FSIComp} assess the impact of 2p2h and of FSI strength alterations on the comparison of NEUT predictions to the results.

\subsection{STV measurement \label{app:stv}}

\begin{figure*}
\begin{center}
\includegraphics[width=0.49\textwidth]{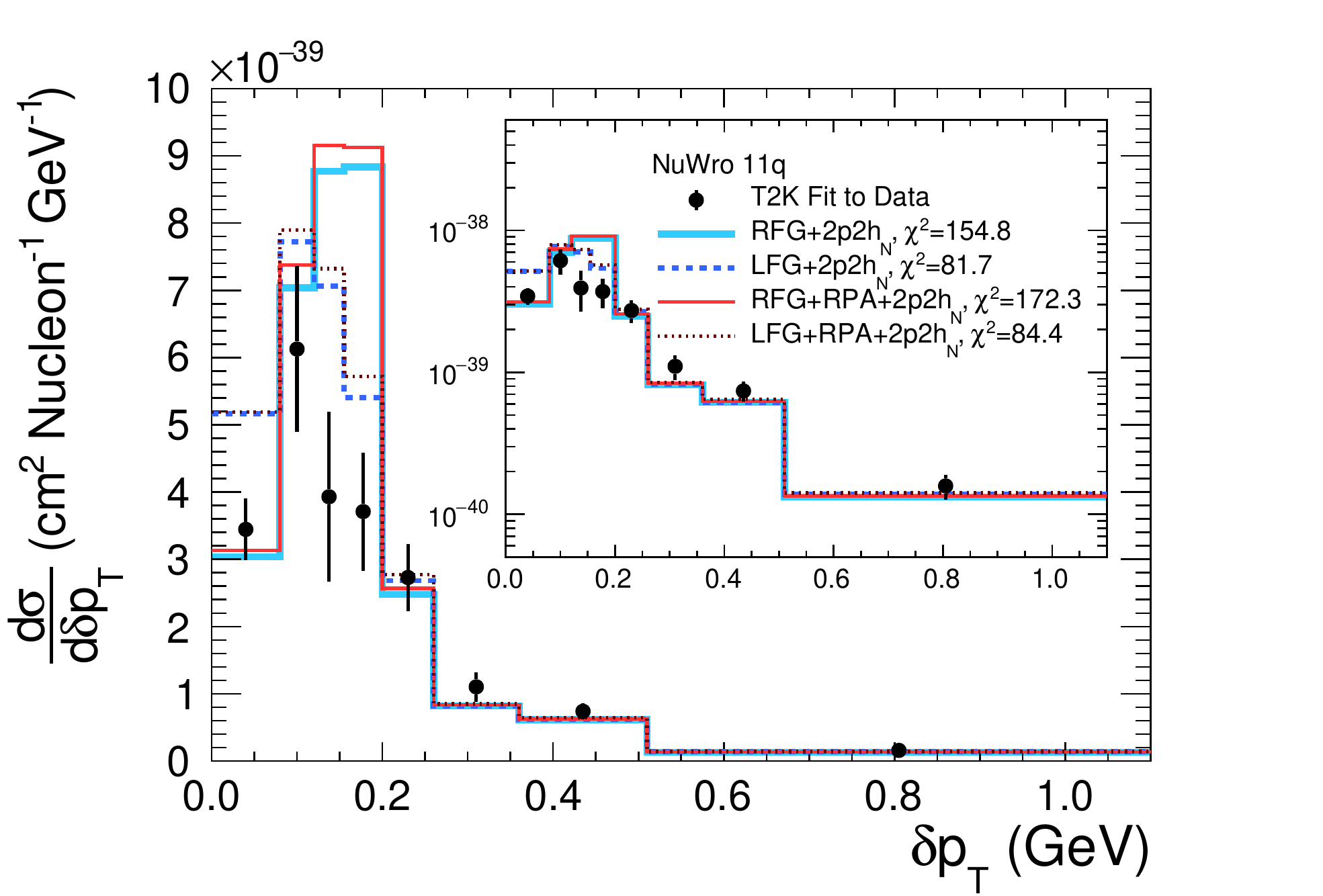}
\includegraphics[width=0.49\textwidth]{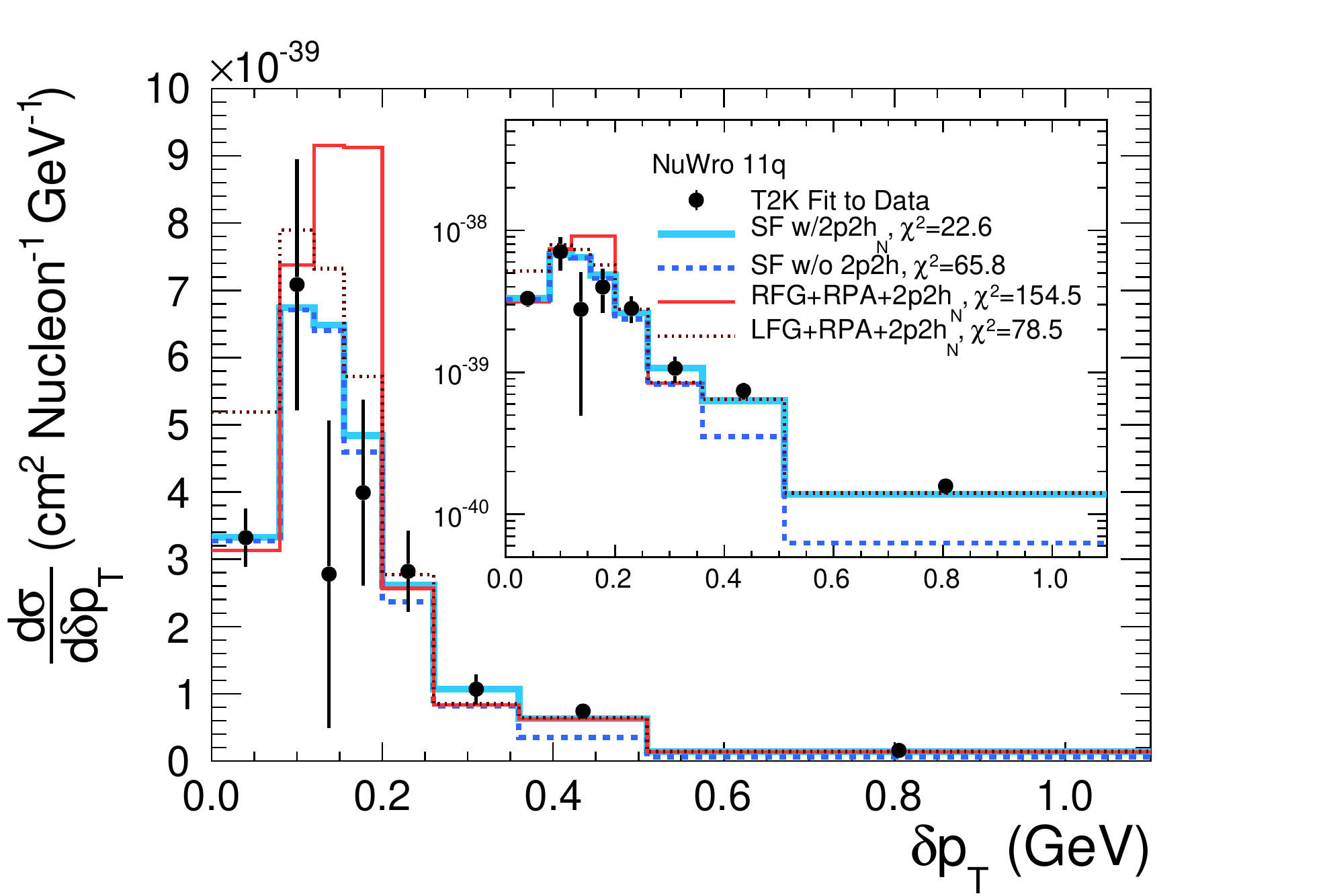}

\includegraphics[width=0.49\textwidth]{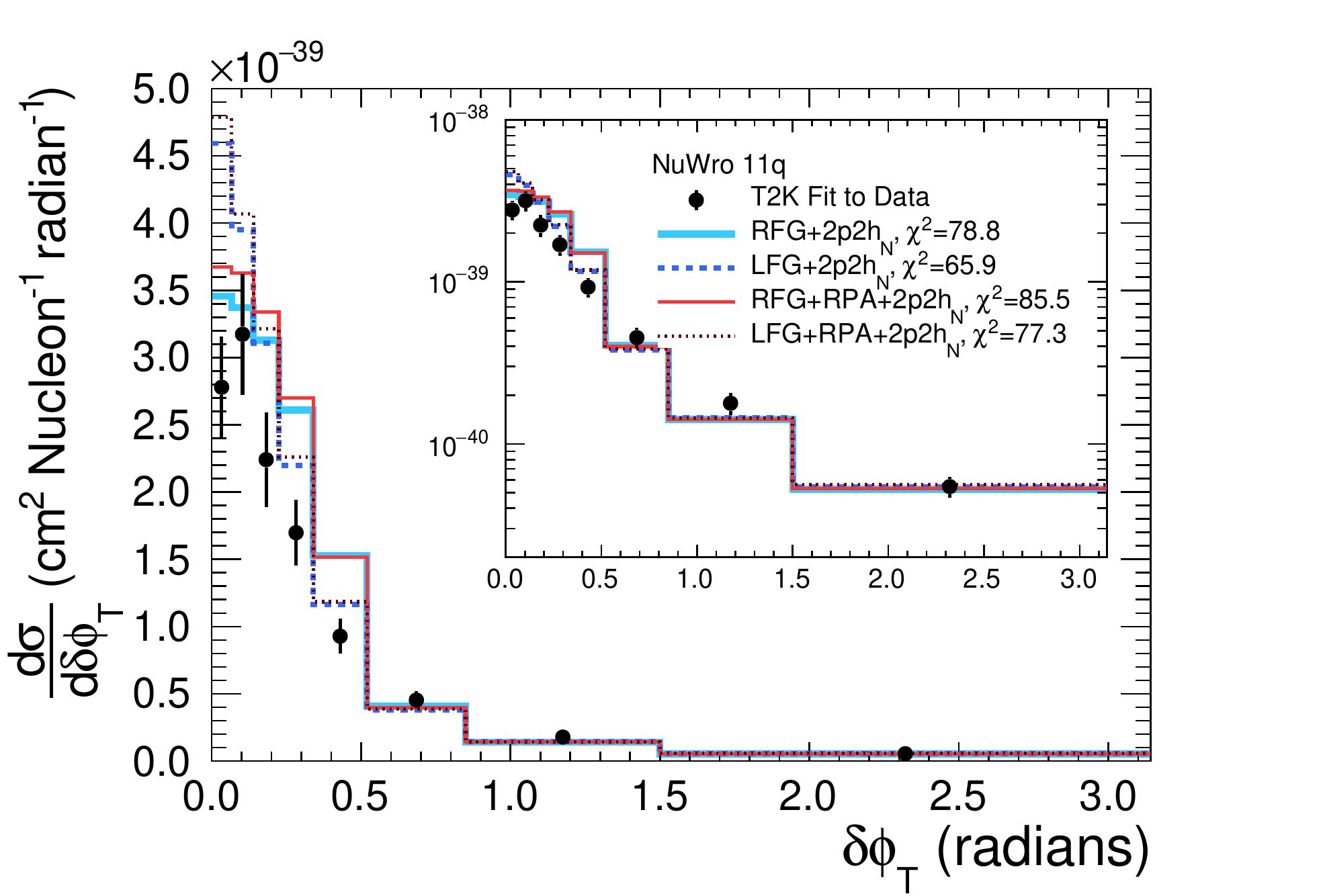}
\includegraphics[width=0.49\textwidth]{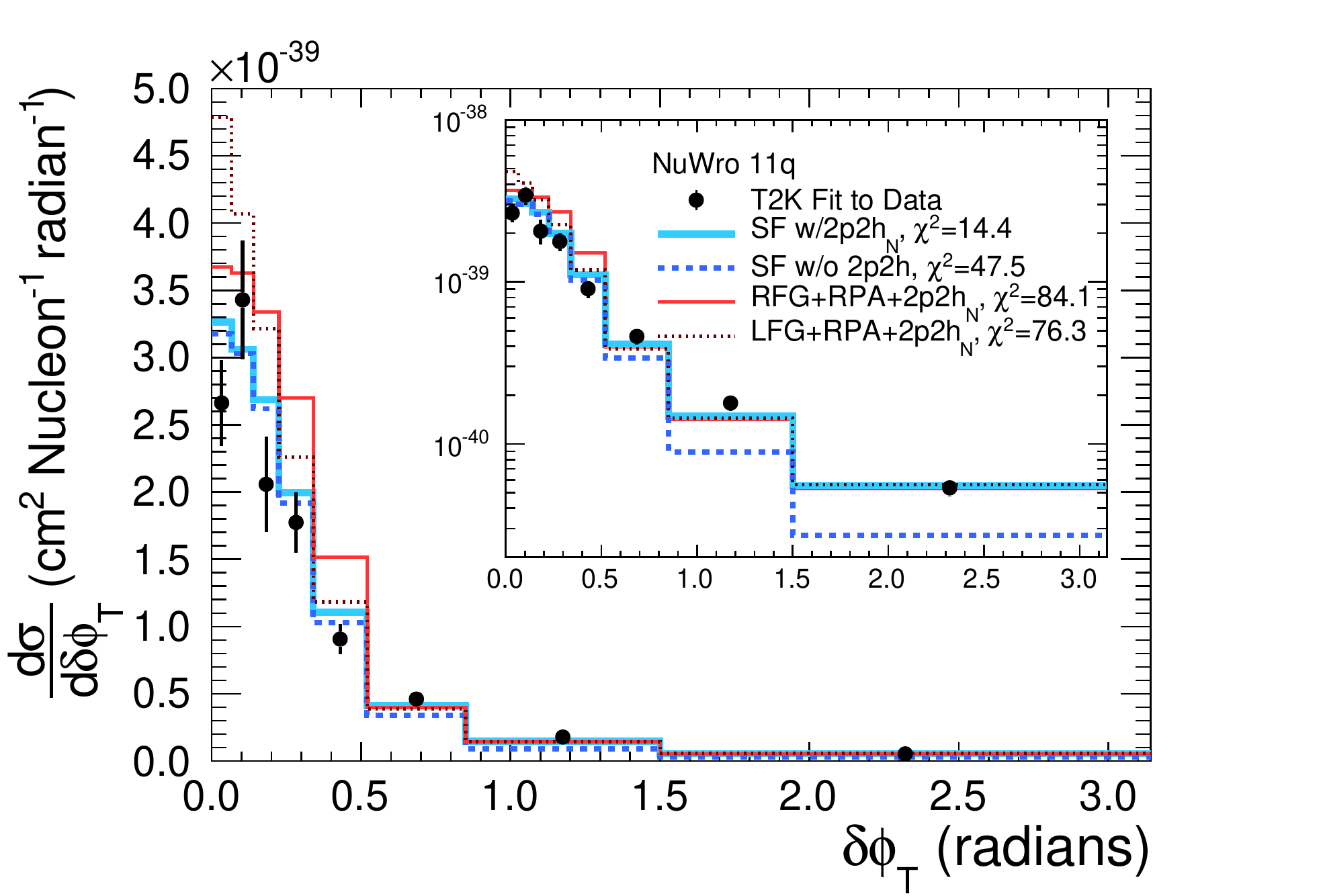}

\includegraphics[width=0.49\textwidth]{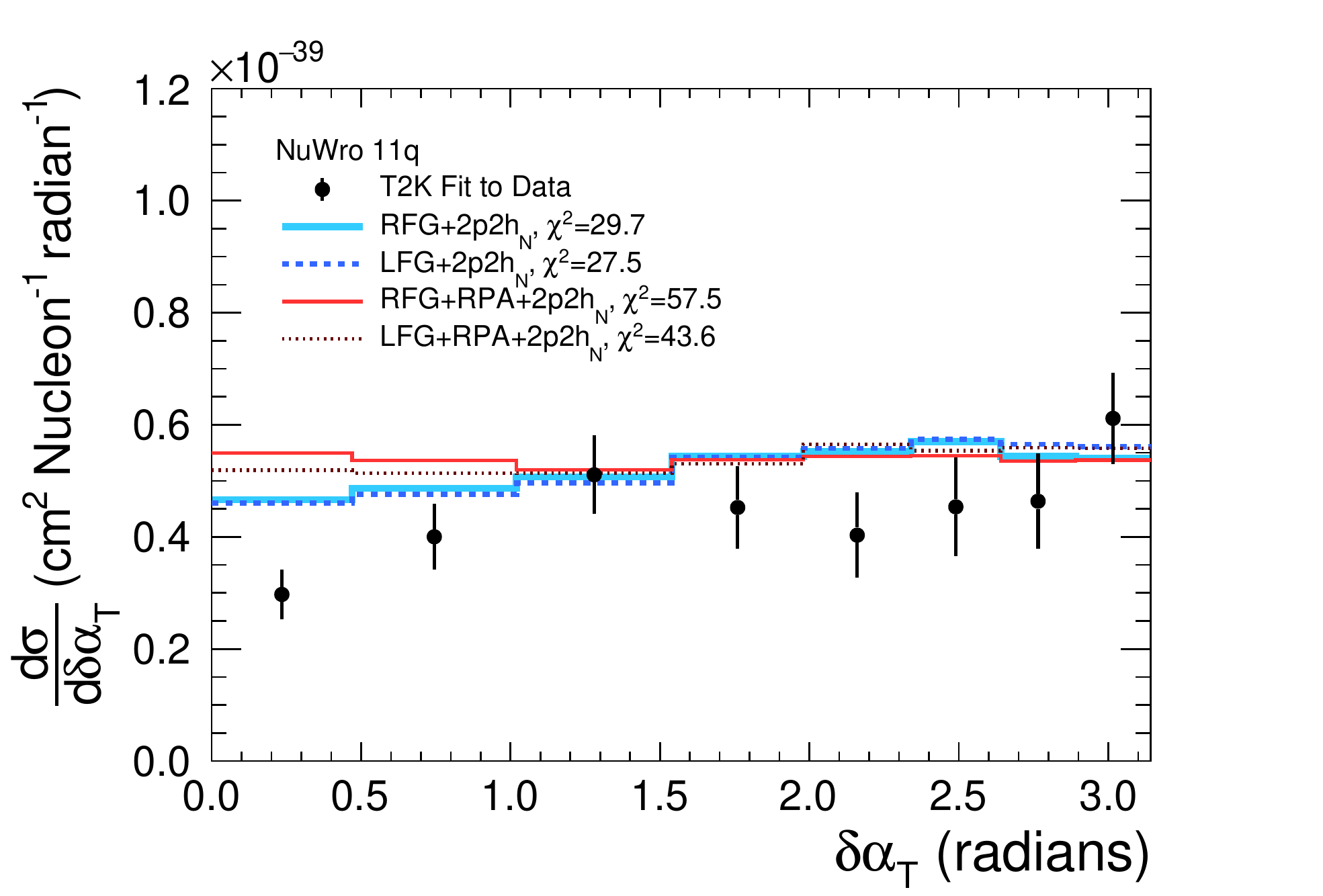}
\includegraphics[width=0.49\textwidth]{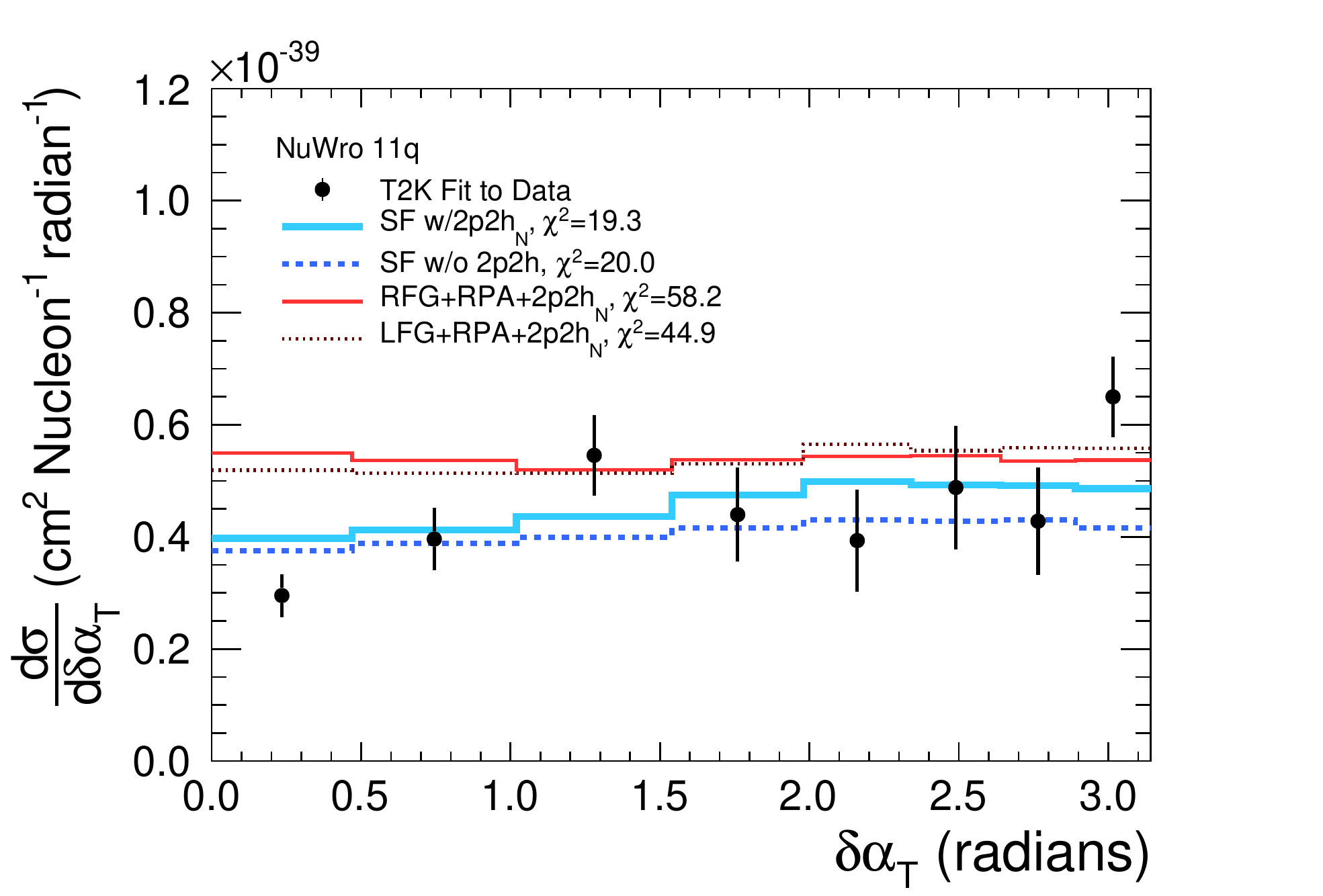}
\end{center}
\caption{The plots on the left show the extracted differential cross section as a function of the single transverse variables compared to different initial state models in the NuWro 11q simulation with and without RPA. The plots on the right show the extracted unregularised differential cross section as a function of the single transverse variables compared to different initial state models in the NuWro 11q simulation. 2p2h$_N$ indicates the Nieves et. al. model of Ref.~\cite{Nieves:2012}. More details of these models can be found in Sec.~\ref{sec:simulation}. The inlays show the same comparisons on a logarithmic scale. }
\label{fig:STV_NuWroRPAComp}
\end{figure*}

As discussed in Sec.~\ref{sec:STV} this appendix provides additional comparisons of the single transverse analysis results to assess the impact of RPA on the data-simulation comparisons and also to identify the role of regularisation in the cross section extraction procedure.  This is shown in Fig.~\ref{fig:STV_NuWroRPAComp} which first shows NuWro 11q predictions with and without RPA compared to the same regularised results. Alongside this, the figure also shows the predictions of different nuclear models, already shown in Fig.~\ref{fig:STV_NuWro}, compared to the unregularised results. In this final figure it should be noted that, while the nominal result is different than the regularised result presented in Sec.~\ref{sec:STV}, the physics conclusions remain the same with a similar goodness of fit.

%\begin{figure*}
%\begin{center}
%\end{center}
%\caption{The extracted unregularised differential cross section as a function of the single transverse variables compared to different initial state models in the NuWro 11q simulation (all with $M_A^{QE}=1.0$ GeV) }
%\label{fig:STV_NoReg_NuWroModelComp}
%\end{figure*}

\subsection{Inferred proton kinematics measurement \label{app:infk}}

\begin{figure*}[h!]
\begin{center}
 \includegraphics[width=\textwidth]{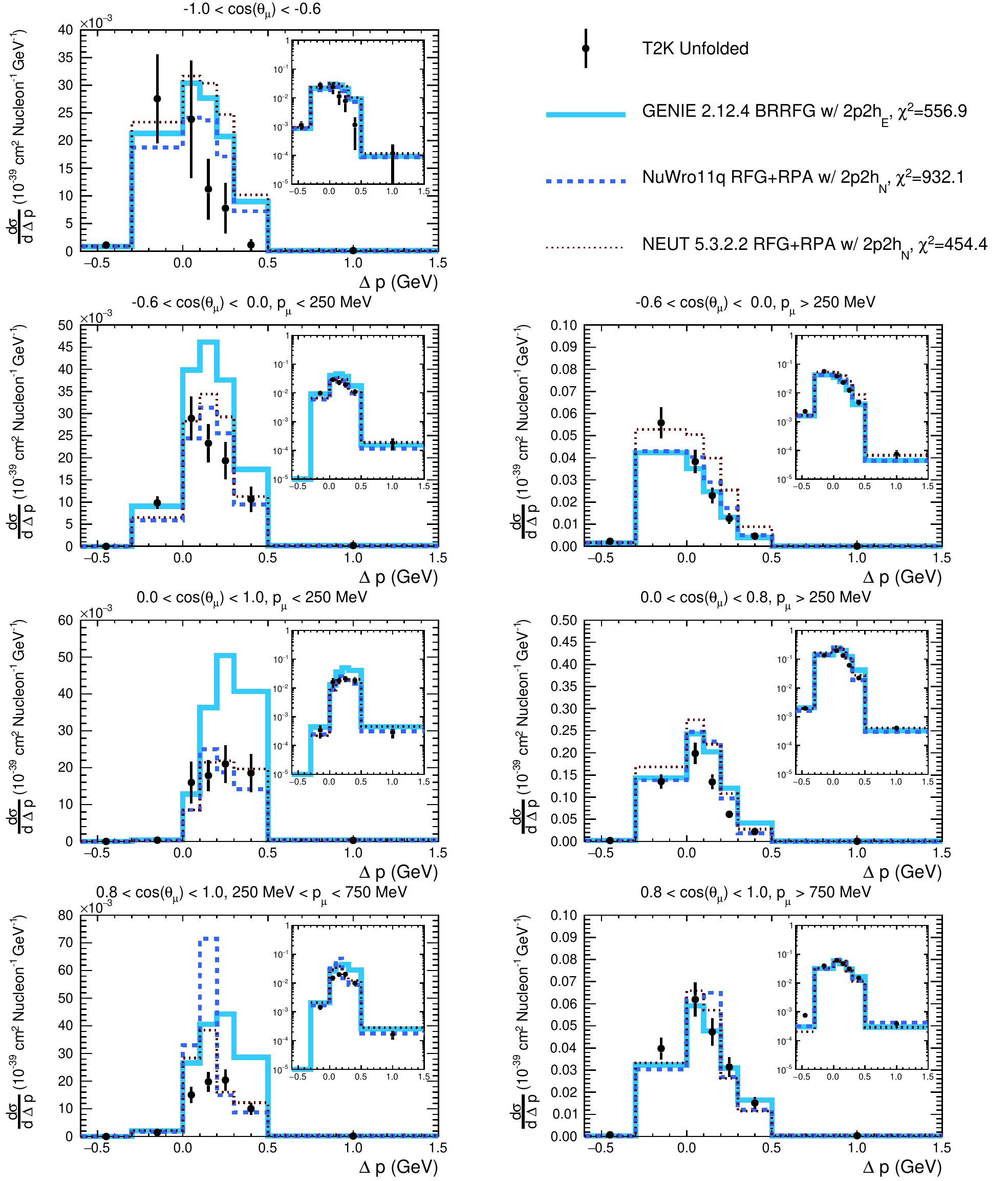} 
  \end{center}
\caption{The extracted differential cross section as a function of the inferred and true proton momentum difference in different muon kinematic bins, within a restricted proton kinematic phase space, compared to GENIE 2.12.4, NuWro 11q and NEUT 5.3.2.2  predictions which utilise an RFG or RFG+RPA nuclear model. GENIE's RFG model also includes the empirical correction from Bodek and Ritchie (BRRFG)~\cite{Bodek:1981wr}. The NEUT and GENIE predictions shown here are similar to those used as a starting point for the T2K and NO$\nu$A experiment's oscillation analyses respectively. More details of these models can be found in Sec.~\ref{sec:simulation}. The 2p2h subscript indicates whether it is an implementation of the Nieves et. al. model of Ref.~\cite{Nieves:2012} (N) or the GENIE empirical 2p2h model (E). More details of these models can be found in Sec.~\ref{sec:simulation}. Note that the first and last bin in each plot is shortened for improved readability. The inlays show the same comparisons on a logarithmic scale.}
\label{fig:inf_rfg_mom}
\end{figure*}

\begin{figure*}[h!]
\begin{center}
 \includegraphics[width=\textwidth]{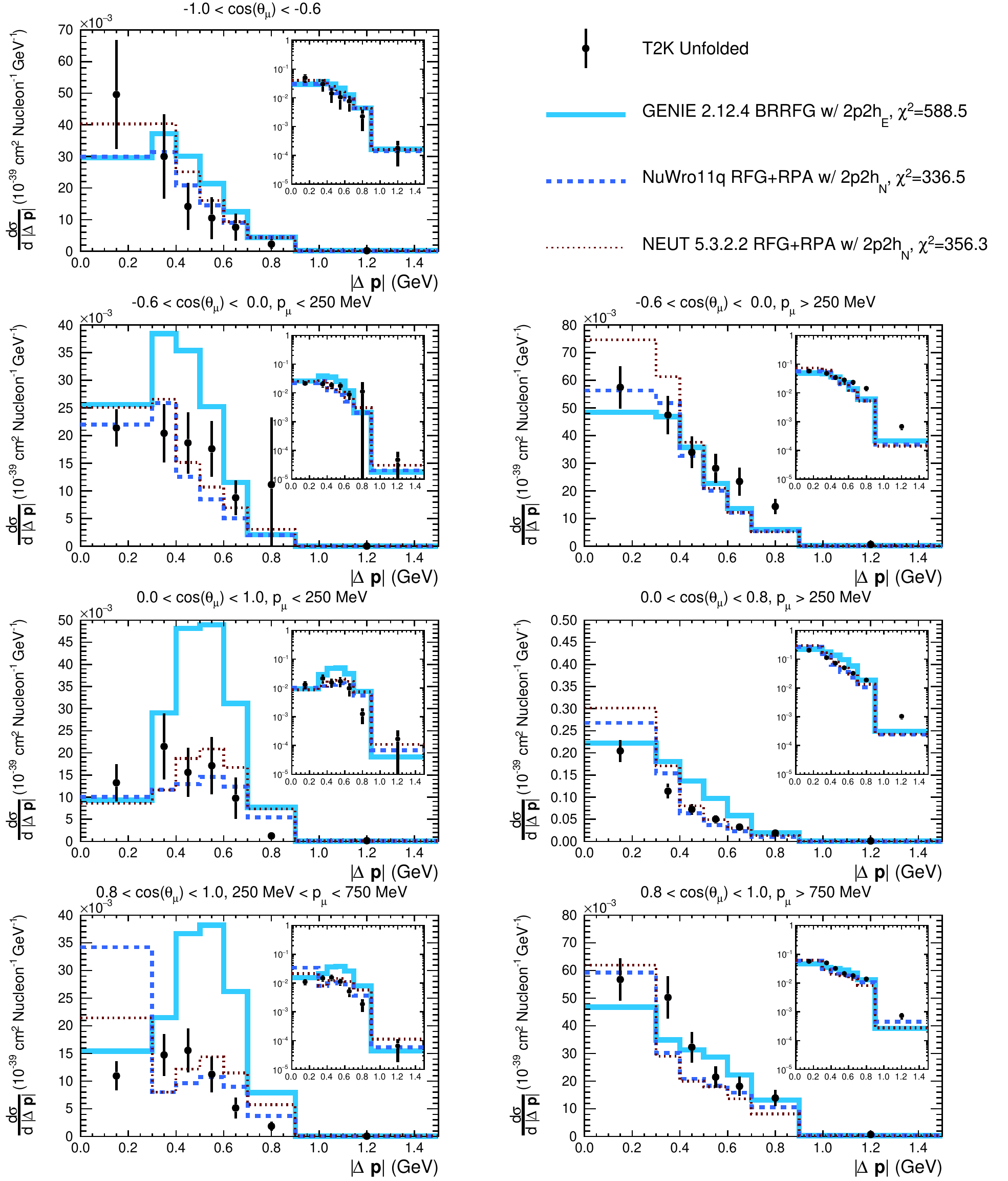} 
  \end{center}
\caption{The extracted differential cross section as a function of the modulus of the inferred and true proton three-momentum difference in different muon kinematic bins, within a restricted proton kinematic phase space, compared to GENIE 2.12.4, NuWro 11q and NEUT 5.3.2.2  predictions which utilise an RFG or RFG+RPA nuclear model. GENIE's RFG model also includes the empirical correction from Bodek and Ritchie (BRRFG)~\cite{Bodek:1981wr}. The NEUT and GENIE predictions shown here are similar to those used as a starting point for the T2K and NO$\nu$A experiment's oscillation analyses respectively. The 2p2h subscript indicates whether it is an implementation of the Nieves et. al. model of Ref.~\cite{Nieves:2012} (N) or the GENIE empirical 2p2h model (E). More details of these models can be found in Sec.~\ref{sec:simulation}. Note that the last bin in each plot is shortened for improved readability. The inlays show the same comparisons on a logarithmic scale.}
\label{fig:inf_rfg_tmom}
\end{figure*}

\begin{figure*}[h!]
\begin{center}
 \includegraphics[width=\textwidth]{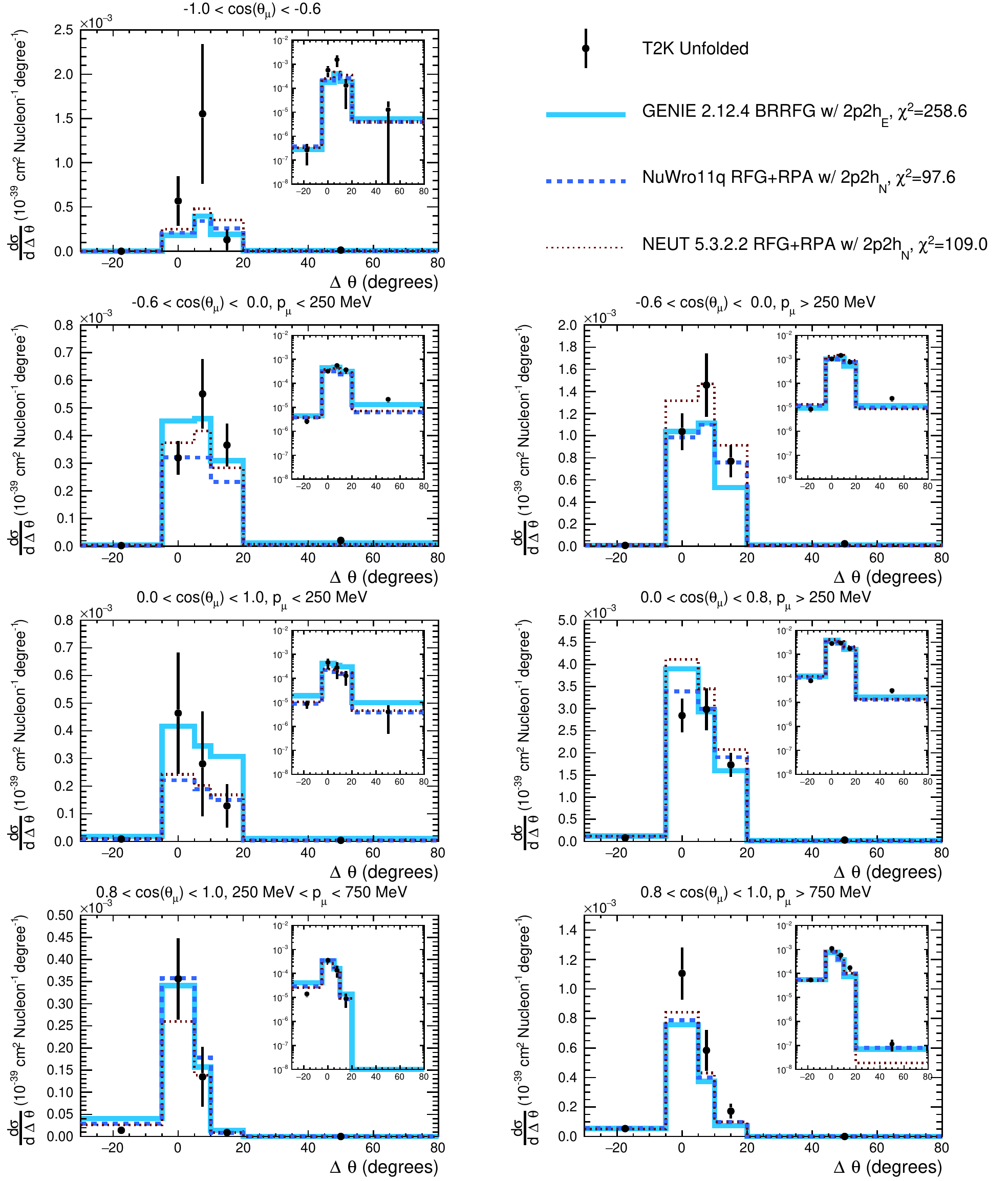} 
  \end{center}
\caption{The extracted differential cross section as a function of the inferred and true proton outgoing-angle difference in different muon kinematic bins, within a restricted proton kinematic phase space, compared to GENIE 2.12.4, NuWro 11q and NEUT 5.3.2.2  predictions which utilise an RFG or RFG+RPA nuclear model. GENIE's RFG model also includes the empirical correction from Bodek and Ritchie (BRRFG)~\cite{Bodek:1981wr}. The NEUT and GENIE predictions shown here are similar to those used as a starting point for the T2K and NO$\nu$A experiment's oscillation analyses respectively. The 2p2h subscript indicates whether it is an implementation of the Nieves et. al. model of Ref.~\cite{Nieves:2012} (N) or the GENIE empirical 2p2h model (E). More details of these models can be found in Sec.~\ref{sec:simulation}. Note that the first and last bin in each plot is shortened for improved readability. The inlays show the same comparisons on a logarithmic scale.}
\label{fig:inf_rfg_ang}
\end{figure*}

As discussed in Sec.~\ref{sec:inferred} this appendix provides additional comparisons of the proton inferred kinematics analysis results to RFG nuclear models. This is shown in Fig.~\ref{fig:inf_rfg_mom}-\ref{fig:inf_rfg_ang}, which compare the extracted results to RFG predictions from NEUT 5.3.2.2, NuWro 11q and GENIE 2.12.4.

\section{$\chi^2$ comparisons to the STV results with and without regularisation \label{app:unregchi2}}

As discussed in Sec.~\ref{subsec:mlfitting}, the regularisation of cross-section extraction methods allows results that are easy to interpret at the cost of some bias. Although the regularised STV results presented in Sec.~\ref{sec:STV} minimise this using a data-driven method, the unregularised results are also produced to guarantee no unfolding bias. To demonstrate that the application of regularisation does not alter the physical interpretation of the results, Tab.~\ref{tab:dptchi2} to ~\ref{tab:datchi2} show a summary of the $\chi^2$ agreement between the various models considered in Sec.~\ref{sec:STV} and both the regularised and the unregularised results. The shape-only $\chi^2$ is also shown for each model as this does not suffer from Peelle's Pertinent Puzzle discussed in Sec.~\ref{sec:discussion} and so may be the most useful quantitative metric for comparing model agreement. This shape-only $\chi^2$ is formed by numerically decomposing the full covariance matrix into a shape and normalisation component. Occasionally it was found that it was necessary to add a small component to the diagonal of the shape-only covariance matrix to make the matrix invertible, thereby slightly overestimating the error (this is responsible for the two instances of seeing a shape-only $\chi^2$ greater than the full $\chi^2$ in Tab.~\ref{tab:datchi2}). 

Overall the $\chi^2$ tables show very little difference between the regularised and unregularised result, demonstrating that the cautious data-driven choice of regularisation strength means that only a very small bias is added. The tables echo the conclusions drawn in Sec.~\ref{sec:discussion}: a 2p2h contribution and a non-RFG nuclear model is shown to be a hallmark of the lower $\chi^2$ in the comparisons. 

\begin{table*}[h]
\begin{center}
%\footnotesize
\centering
\begin{tabular}{|l|c|c|c|c|}
\hline
   & \multicolumn{2}{|c|}{Full} & \multicolumn{2}{|c|}{Shape Only} \\
\hline
 Generator & No Reg. & Nom. Reg. & No Reg. & Nom. Reg. \\
\hline
 NEUT 5.4.0 (LFG$_N$+2p2h$_N$) & 31.6 & 30.4 & 3.38 & 2.60 \\ 
 NEUT 5.3.2.2 (SF+2p2h$_N$+2$\times$FSI) & 15.9 & 14.8 & 11.0 & 10.1 \\ 
 NEUT 5.3.2.2 (SF+2p2h$_N$) & 31.9 & 30.3 & 16.6 & 15.5 \\ 
 NuWro 11q (SF+2p2h$_N$) & 22.6 & 23.1 & 16.8 & 15.6 \\ 
 NuWro 11q (LFG+2p2h$_N$) & 81.5 & 81.7 & 39.0 & 15.6 \\ 
 NuWro 11q (LFG+RPA+2p2h$_N$) & 78.5 & 84.4 & 39.9 & 36.3 \\ 
 NEUT 5.3.2.2 (SF+2p2h$_N$+No FSI) & 114 & 112 & 42.9 & 41.4 \\ 
 GENIE 2.12.4 (RFG+2p2h$_E$) & 92.9 & 92.4 & 47.9 & 47.7 \\ 
 NuWro 11q (SF w/o 2p2h) & 65.8 & 68.7 & 55.4 & 54.8 \\ 
 NEUT 5.3.2.2 (SF w/o 2p2h) & 93.3 & 91.5 & 61.2 & 59.6 \\ 
 GiBUU 2016 (LFG+2p2h$_G$) & 77.0 & 78.9 & 66.1 & 59.6 \\ 
 NuWro 11q (RFG+2p2h$_N$) & 150 & 155 & 67.2 & 69.0 \\ 
 NuWro 11q (RFG+RPA+2p2h$_N$) & 155 & 172 & 68.6 & 70.4 \\ 
 GENIE 2.12.4 (RFG w/o 2p2h) & 94.6 & 97.8 & 74.1 & 76.2 \\ 
\hline
\end{tabular}
\caption{\label{tab:dptchi2} The full and shape-only $\chi^2$ comparisons to the $\delta p_T$ result with nominal and no regularisation. The table is ordered by the size of the no-regularisation shape-only $\chi^2$. More details of these models can be found in Sec.~\ref{sec:simulation}.}
\end{center}
\end{table*}

\begin{table*}[h]
\begin{center}
%\footnotesize
\centering
\begin{tabular}{|l|c|c|c|c|}
\hline
   & \multicolumn{2}{|c|}{Full} & \multicolumn{2}{|c|}{Shape Only} \\
\hline
 Generator & No Reg. & Nom. Reg. & No Reg. & Nom. Reg. \\
\hline
 NEUT 5.4.0 (LFG$_N$+2p2h$_N$) & 39.0 & 36.7 & 7.55 & 6.40 \\ 
 NEUT 5.3.2.2 (SF+2p2h$_N$+2$\times$FSI) & 9.95 & 8.70 & 7.71 & 6.57 \\ 
 NEUT 5.3.2.2 (SF+2p2h$_N$) & 18.4 & 17.0 & 9.59 & 8.45 \\ 
 NuWro 11q (SF+2p2h$_N$) & 14.4 & 13,5 & 10.8 & 9.70 \\ 
 NuWro 11q (LFG+2p2h$_N$) & 66.8 & 65.9 & 29.7.0 & 29.0 \\ 
 NEUT 5.3.2.2 (SF+2p2h$_N$+No FSI) & 81.5 & 81.4 & 30.5 & 30.1 \\ 
 NuWro 11q (LFG+RPA+2p2h$_N$) & 76.3 & 77.3 & 32.1 & 31.3 \\ 
 NuWro 11q (RFG+RPA+2p2h$_N$) & 84.7 & 85.5 & 40.1 & 39.4 \\ 
 NuWro 11q (SF w/o 2p2h) & 47.5 & 48.9 & 42.1 & 42.3 \\ 
 NuWro 11q (RFG+2p2h$_N$) & 79.3 & 78.8 & 42.6 & 42.0 \\ 
 NEUT 5.3.2.2 (SF w/o 2p2h) & 60.6 & 61.0 & 43.7 & 43.8 \\ 
 GiBUU 2016 (LFG+2p2h$_G$) & 43.4 & 44.1 & 45.6 & 46.2 \\ 
 GENIE 2.12.4 (RFG+2p2h$_E$) & 208 & 211 & 114 & 115 \\ 
 GENIE 2.12.4 (RFG w/o 2p2h) & 192 & 193 & 128 & 128 \\ 
\hline
\end{tabular}
\caption{\label{tab:dphitchi2} The full and shape-only $\chi^2$ comparisons to the $\delta \phi_T$ result with nominal and no regularisation. The table is ordered by the size of the no-regularisation shape-only $\chi^2$. More details of these models can be found in Sec.~\ref{sec:simulation}.}
\end{center}
\end{table*}

\begin{table*}[h]
\begin{center}
%\footnotesize
\centering
\begin{tabular}{|l|c|c|c|c|}
\hline
   & \multicolumn{2}{|c|}{Full} & \multicolumn{2}{|c|}{Shape Only} \\
\hline
 Generator & No Reg. & Nom. Reg. & No Reg. & Nom. Reg. \\
\hline
 NEUT 5.3.2.2 (SF+2p2h$_N$+2$\times$FSI) &17.7&  15.8&  16.3&  14.2 \\
 NuWro 11q (SF+2p2h$_N$) &19.3&  18.0&  18.6&  16.6 \\
 NEUT 5.3.2.2 (SF+2p2h$_N$) &24.8&  23.0&  18.8&  16.8 \\
 NuWro 11q (LFG+2p2h$_N$)&29.6&  27.5&  19.0&  16.9 \\
 NuWro 11q (RFG+2p2h$_N$)&31.6&  29.7&  20.7&  18.7 \\
 NEUT 5.3.2.2 (SF w/o 2p2h) &21.0&  19.5&  21.7&  19.6 \\
 NEUT 5.4.0 (LFG$_N$+2p2h$_N$) &63.0&  60.7&  22.8&  20.8 \\
 NuWro 11q (SF w/o 2p2h)&20.0&  18.9&  23.4&  21.4 \\
 NEUT 5.3.2.2 (SF+2p2h$_N$+No FSI)&49.9&  48.2&  28.3&  26.3 \\
 NuWro 11q (LFG+RPA+2p2h$_N$) &44.9&  43.6&  28.6&  26.3 \\
 GiBUU 2016 (LFG+2p2h$_G$) &41.3&  40.2&  35.5&  33.7 \\
 NuWro 11q (RFG+RPA+2p2h$_N$) &58.2&  57.5&  38.1&  35.8 \\
 GENIE 2.12.4 (RFG+2p2h$_E$) &88.5&  90.2&  40.1&  39.6 \\
 GENIE 2.12.4 (RFG w/o 2p2h) &38.6&  72.0&  62.6&  64.1 \\
\hline
\end{tabular}
\caption{\label{tab:datchi2} The full and shape-only $\chi^2$ comparisons to the $\delta \alpha_T$ result with nominal and no regularisation. The table is ordered by the size of the no-regularisation shape-only $\chi^2$. More details of these models can be found in Sec.~\ref{sec:simulation}.}
\end{center}
\end{table*}

\clearpage
\bibliography{main}

\end{document}